\documentclass{2022AMSE_preprint}

\usepackage{bm}
\usepackage{multirow}
\usepackage{placeins}
\usepackage{float}
\makeatletter
\renewcommand{\@biblabel}[1]{[#1]}
\makeatother

\begin{document}

\ensubject{Solid Mechanics}

\ArticleType{RESEARCH PAPER}
\Year{2026}

\title{Voxel-based block variational quantum linear solver: a hybrid quantum--classical method for static analysis of solids}{Voxel-based block variational quantum linear solver: a hybrid quantum--classical method for static analysis of solids}

\author{Feng Wu}{}
\author{Chen Li}{}
\author{Li Zhu}{}
\author{Yuxiang Yang}{}
\author{Xu Guo}{guoxu@dlut.edu.cn}

\AuthorMark{Xu Guo}
\AuthorCitation{F. Wu, C. Li, L. Zhu, Y. Yang, and X. Guo}

\address{State Key Laboratory of Structural Analysis, Optimization and CAE Software for Industrial Equipment,\\School of Mechanics and Aerospace Engineering, Dalian University of Technology, Dalian 116024, Liaoning, P. R. China}

\abstract{In solid mechanics, finite element discretization of large-scale static problems produces large sparse linear systems whose solution requires substantial computation time and memory. The variational quantum linear solver (VQLS) offers a hybrid quantum--classical route, but its use in quantum finite element analysis is limited by the decomposition of nonunitary matrices, barren plateaus, and the measurement cost of expectation values. We propose a voxel-based block variational quantum linear solver (Voxel-BVQLS) that combines structured matrix decomposition, the principle of minimum potential energy, and batched quantum tests. First, we construct an LCU decomposition of the stiffness matrix from the recursive block-banded structure of voxel-grid finite element matrices, with the number of unitary terms bounded independently of the problem size. Second, we optimize the ansatz parameters using a minimum-potential-energy objective in place of a conventional VQLS loss function, thereby mitigating barren plateaus in the studied problems. Third, we introduce a block-Hadamard test whose circuit directly estimates weighted sums of multiple inner products, reducing the number of circuit configurations required per iteration. We assessed the proposed method in noiseless classical simulations using three examples. These examples show that the method reduces both the number of unitary terms in the LCU decomposition and the number of iterations required to converge, while still yielding solutions of finite accuracy. Voxel-BVQLS thus provides a structured hybrid quantum--classical framework for quantum finite element analysis on regular grids.}


\setlength{\textheight}{23.6cm}
\thispagestyle{empty}

\maketitle
\renewcommand{\thefootnote}{\fnsymbol{footnote}}
\setcounter{footnote}{1}
\footnotetext{Corresponding author: Xu Guo. E-mail: \href{mailto:guoxu@dlut.edu.cn}{guoxu@dlut.edu.cn}}
\renewcommand{\thefootnote}{\arabic{footnote}}
\setcounter{footnote}{0}
\setlength{\parindent}{1em}

\vspace{-1mm}

\section{Introduction}

Advances in computing power have continually driven numerical analysis in solid mechanics \cite{Gao2012,Mosby2016}. The finite element method (FEM) has evolved alongside classical computers and has become one of the most important numerical methods in solid mechanics \cite{Turner1956,Clough1960,Wu2013InterBelt}. FEM partitions the problem domain into elements, constructs element stiffness equations, and assembles a global equilibrium system. Fine discretization of very large structures often produces large sparse linear systems \cite{Bangerth2011,Koric2014}. As the number of degrees of freedom grows to millions or beyond, solution time, memory use, and parallel communication overhead typically increase substantially\cite{Wu2024Discrepancy,Wu2023ADDP}. At billions of degrees of freedom, large-scale parallel computing resources are usually required, which imposes a significant computational bottleneck.  New computing approaches that alleviate the cost of classical finite element solvers are therefore of considerable interest.

Quantum computing offers a different way to process information in scientific computation \cite{Feynman1982,Lloyd1996}. Superposition, entanglement, and interference allow quantum algorithms to manipulate information in high-dimensional state spaces. These features may offer computational advantages for particular tasks \cite{Kim2023,Preskill2018}. Applications in quantum chemistry and quantum simulation have shown promise \cite{Peruzzo2014,Cao2019,McArdle2020}. At the intersection of scientific computing and mechanics, quantum algorithms have been studied for partial differential equations \cite{AuYeung2024}, materials and structural analysis \cite{Balducci2022}, and fluid mechanics \cite{MengYang2023,MengYang2024,Meng2024Processor}. Solid mechanics, however, often involves heterogeneous materials, complex geometries, and varied boundary conditions \cite{Wu2026VBQC,Zhang2010Heterogeneous}. The entries and sparsity patterns of the resulting stiffness matrices depend on the particular problem, complicating uniform quantum encodings and circuit implementations. Moreover, most governing equations in solid mechanics, especially static equilibrium equations, do not directly take the form of a unitary Schr\"odinger evolution suitable for quantum simulation \cite{XuHu2026Potential,Jin2024Schrodingerization}. Quantum computing for solid mechanics therefore remains at an early stage. Recent studies have constructed quantum Hamiltonians for solid dynamics through energy-conservation mappings \cite{Xu2026Elasto}, or developed decomposition-free variational solvers for structural statics \cite{Xu2025}. Most results nevertheless concern equation mappings or small-scale proof-of-principle demonstrations. End-to-end quantum speedup for general solid mechanics problems has not been established. Methods tailored to the structure of solid mechanics problems are therefore needed.

The main computational tasks in FEM include solving large sparse linear systems and eigenvalue problems. Both are closely related to quantum linear algebra. Embedding quantum subroutines in finite element workflows is therefore a route for research at the interface of quantum computing and solid mechanics. Here, the quantum finite element method denotes a hybrid quantum--classical framework. Geometry modeling, mesh discretization, element calculations, and global assembly remain classical, while a quantum or hybrid algorithm solves the discrete algebraic system or estimates selected physical quantities \cite{Raisuddin2022}. In the gate model, two representative routes for finite element linear systems are quantum linear systems algorithms and the variational quantum linear solver (VQLS). The HHL algorithm is an example of the former. It uses quantum phase estimation and controlled rotations to prepare a solution state proportional to $\mathbf{K}^{-1}\mathbf{f}$ \cite{Harrow2009}. Under ideal assumptions, HHL-type algorithms may offer exponential speedup. However, a finite element complexity analysis found that the advantage over classical algorithms is generally at most polynomial \cite{Montanaro2016}. HHL-type methods are thus better suited to estimating a small number of solution-state observables. 

VQLS casts a linear system as a hybrid quantum--classical optimization problem \cite{Cerezo2021Review,BravoPrieto2023}. Its core idea is to prepare a normalized trial solution state with the ansatz circuit $\mathbf{V}(\boldsymbol{\alpha})$ and to iteratively update its parameters with a classical optimizer, so that the state approaches the exact solution. VQLS can use shallower circuits than phase-estimation-based quantum linear systems algorithms, making it a leading approach in recent quantum finite element studies. Improvements have focused on objective functions, matrix encodings, and ansatz design \cite{Ying2023,Sato2021,Patil2022,PellowJarman2021}. Variational quantum methods have already been applied to several finite element and related partial differential equation problems. Trahan et al.\ \cite{Trahan2023} used Pauli decomposition to study discrete finite element systems for the one-dimensional steady Poisson, transient heat conduction, and wave equations. Ali and Kabel \cite{Ali2023} studied variational solutions of the Poisson equation on a noiseless simulator and a quantum device. They reported substantial limitations on current hardware. Arora et al.\ \cite{Arora2025} examined a quantum finite element workflow for steady heat conduction with different elements, material properties, and boundary conditions.  Liu et al.\ \cite{Liu2024} combined FEM with a variational quantum eigensolver to analyze the natural vibrations of trusses, beams, and continua. Although these studies demonstrate the potential of VQLS for shallow circuits and problem-aware design, VQLS still faces three interrelated challenges in solid mechanics. First, trainability of parameterized circuits is not guaranteed. Certain random ansatz circuits and global objective functions can exhibit barren plateaus \cite{McClean2018,Cerezo2021BP}. Second, a quantum representation of the stiffness matrix may contain many unitary terms. An $n$-qubit system has $4^n$ Pauli strings. Without exploiting matrix structure, the nonzero terms and their controlled implementations may incur substantial circuit and measurement costs \cite{Childs2012,Wu2025Voxel,Chakraborty2024}. A common global VQLS loss contains $\mathbf{K}^{\mathrm{T}}\mathbf{K}$, so representing $\mathbf{K}$ by $M$ terms in an LCU decomposition leads to $O(M^2)$ inner products in the direct expansion, each of which also requires enough measurements to control statistical error \cite{BravoPrieto2023,Trahan2023}.

To address these challenges for real symmetric positive-definite finite element stiffness systems, we propose the voxel-based block variational quantum linear solver (Voxel-BVQLS). Voxel-BVQLS exploits regular voxel structure to reduce both the number of LCU unitary terms and the circuit configurations needed to evaluate the minimum-potential-energy objective and its gradient. The main contributions are as follows.

 (1) We develop an LCU decomposition for stiffness matrices on voxel grids. It exploits the recursive block-banded structure of the finite element matrix together with cyclic permutation matrices and Euler's formula, and its number of unitary terms is bounded independently of the problem dimension.

 (2) We construct a variational objective from the principle of minimum potential energy. For a constrained symmetric positive-definite stiffness matrix $\mathbf{K}$, the quadratic form associated with linear-static total potential energy has the unique minimizer $\mathbf{K}^{-1}\mathbf{f}$ over the full displacement space, and the resulting objective can mitigate barren plateaus in the studied settings.

 (3) We propose a block-Hadamard test and use it to construct the block variational solver. Quantum multiplexors and a block-partitioning strategy allow one circuit to estimate weighted sums of multiple inner products, which reduces the number of circuit configurations required per iteration.

 (4) We test the method in noiseless classical simulations on a two-dimensional truss, a plane-stress model, and a three-dimensional steady-state heat-conduction model with one degree of freedom per node. All three examples produce approximate displacement or temperature fields, and the structured LCU uses fewer unitary terms than the Pauli decomposition.

The remainder of this paper is organized as follows. Section 2 reviews quantum computing fundamentals, LCU decomposition, VQLS, and its main challenges in solid mechanics. Section 3 presents the voxel-grid stiffness-matrix LCU decomposition and the minimum-potential-energy formulation. Section 4 develops the block-Hadamard test and Voxel-BVQLS. Section 5 reports three numerical examples. Section 6 summarizes the conclusions and limitations.

\section{Quantum finite element method and VQLS}

Quantum finite element analysis combines FEM with quantum computing. Its basic approach retains established classical steps for geometry modeling, mesh discretization, element calculations, and global assembly. The computationally demanding algebraic solve is then expressed in a form suitable for a quantum computer. After discretization and imposition of the required boundary conditions, the finite element equilibrium equation is

\begin{equation}\mathbf{K}\mathbf{u}=\mathbf{f},\label{eq:1}\end{equation}
where $\mathbf{K}\in\mathbb{R}^{N\times N}$ is the finite element system matrix after boundary conditions have been imposed. The vectors $\mathbf{u},\mathbf{f}\in\mathbb{R}^{N}$ are the unknown displacement and load vectors, respectively. Here, $N$ is the number of unknown degrees of freedom.

In the quantum finite element workflow used here, the finite element equations are assembled classically and solved by a hybrid quantum--classical algorithm. On the quantum side, the VQLS encodes the right-hand-side vector as a quantum state, represents the system matrix as an LCU, and prepares a trial solution with a parameterized quantum circuit $\mathbf{V}(\boldsymbol{\alpha})$. Quantum measurements estimate the objective function. On the classical side, an optimizer updates $\boldsymbol{\alpha}$ from the measurement results. The two sides alternate until a stopping criterion is met. 

This section is organized as follows: Section 2.1 introduces quantum amplitude encoding and basic gates, Section 2.2 presents matrix LCU decomposition, Section 2.3 reviews the VQLS framework including the ansatz, loss functions, and the Hadamard test, and Section 2.4 discusses the main challenges of applying VQLS to solid mechanics and outlines the proposed solutions.

\subsection{Quantum amplitude encoding and basic quantum gates}

To solve Eq.~\eqref{eq:1} with VQLS, the right-hand-side vector must first be encoded as a quantum state. The basic information unit of a quantum computer is the qubit. A qubit has two computational basis states, $|0\rangle$ and $|1\rangle$, corresponding to classical bit values 0 and 1. Unlike a classical bit, a qubit may occupy a linear combination of these basis states: $|\mathbf{a}\rangle=a_0|0\rangle+a_1|1\rangle$. The complex numbers $a_0$ and $a_1$ are probability amplitudes. Measurement yields $|0\rangle$ and $|1\rangle$ with probabilities $|a_0|^2$ and $|a_1|^2$, respectively. Hence, $|a_0|^2+|a_1|^2=1$.

An $n$-qubit system has $2^n$ computational basis states, and a general state can be expanded as $|\mathbf{a}\rangle=\sum_{i=0}^{2^n-1}a_i|i\rangle$. Here $|i\rangle$ denotes the $n$-bit basis state corresponding to integer $i$. A quantum state corresponds to a normalized complex vector of length $2^n$. Conversely, any nonzero vector $\mathbf{x}\in\mathbb{C}^{2^n}$ can be normalized and written as

\begin{equation}|\mathbf{x}\rangle=\frac{1}{\|\mathbf{x}\|_2}\sum_{i=0}^{2^n-1}x_i|i\rangle.\label{eq:2}\end{equation}

Equation~\eqref{eq:2} defines amplitude encoding. The normalized quantum states of the displacement and load vectors in Eq.~\eqref{eq:1} are

\begin{align}& \left| \mathbf{f} \right\rangle =\frac{1}{{{\left\| \mathbf{f} \right\|}_{2}}}\sum\limits_{i=0}^{{{2}^{n}}-1}{{{f}_{i}}\left| i \right\rangle },\ \ \ \mathbf{f}={{\left( {{f}_{0}},\ {{f}_{1}},\ \cdots ,\ {{f}_{N-1}} \right)}^{\mathrm{T}}}\notag \\ & \left| \mathbf{u} \right\rangle =\frac{1}{{{\left\| \mathbf{u} \right\|}_{2}}}\sum\limits_{i=0}^{{{2}^{n}}-1}{{{u}_{i}}\left| i \right\rangle },\ \ \ \mathbf{u}={{\left( {{u}_{0}},\ {{u}_{1}},\ \cdots ,\ {{u}_{N-1}} \right)}^{\mathrm{T}}}.\label{eq:3}\end{align}
The computational basis state $|i\rangle$ is labeled by the $n$-bit binary representation of integer $i$.

Classical logic gates change bit values, whereas quantum gates change quantum states in the gate model. A deterministic quantum gate acting on a closed system is represented by a unitary matrix. Unitarity preserves the norm of the state. A quantum register usually starts in the readily prepared state $|0\rangle^{\otimes n}=|0\rangle|0\rangle\cdots|0\rangle$. For any target pure state, a unitary matrix $\mathbf{F}$ exists such that

\begin{equation}\left| \mathbf{f} \right\rangle =\mathbf{F}{{\left| 0 \right\rangle }^{\otimes n}},\label{eq:4}\end{equation}
where $\mathbf{F}$ is an $N\times N$ unitary matrix. Equation~\eqref{eq:4} shows that the load vector can be represented as a quantum state. Any unitary transformation can be built from a universal set of elementary gates. Table~\ref{tab:1} lists the single-qubit and two-qubit gates used below, along with their circuit symbols.

\begin{table}[H]
\centering
 \caption{Single-qubit and two-qubit gates used in this work}\label{tab:1}
\renewcommand{\arraystretch}{1.2}
\setlength{\extrarowheight}{2pt}
\begin{tabular}{>{\centering\arraybackslash}m{0.17\linewidth}>{\centering\arraybackslash}m{0.14\linewidth}>{\raggedright\arraybackslash}m{0.42\linewidth}>{\centering\arraybackslash}m{0.15\linewidth}}
\toprule
 & Basic gate & Unitary matrix & Quantum circuit \\
\midrule
\multirow{4}{*}[-32pt]{Single-qubit gates} & Hadamard gate &
$\displaystyle\mathbf{H}=\frac{1}{\sqrt{2}}\left[ \begin{matrix} 1 & 1 \\ 1 & -1 \\ \end{matrix} \right]$ &
\includegraphics[width=2.2cm]{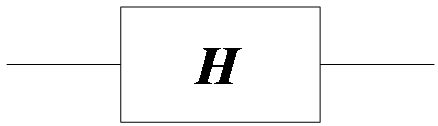} \\ \addlinespace[4pt]
& Pauli-X gate &
$\displaystyle\mathbf{X}=\left[ \begin{matrix} 0 & 1 \\ 1 & 0 \\ \end{matrix} \right]$ &
\includegraphics[width=2.2cm]{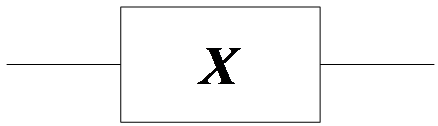} \\ \addlinespace[4pt]
& $R_y$ rotation gate &
$\displaystyle{{\mathbf{R}}_{y}}\left( \theta \right)=\left[ \begin{matrix} \cos {\theta }/{2}\; & -\sin {\theta }/{2}\; \\ \sin {\theta }/{2}\; & \cos {\theta }/{2}\; \\ \end{matrix} \right]$ &
\includegraphics[width=2.2cm]{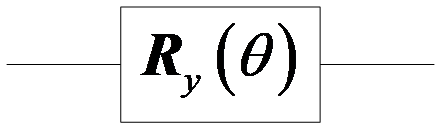} \\ \addlinespace[4pt]
& $R_z$ rotation gate &
$\displaystyle{{\mathbf{R}}_{z}}\left( \theta \right)=\left[ \begin{matrix} {{e}^{-\text{i}{\theta }/{2}\;}} & 0 \\ 0 & {{e}^{\text{i}{\theta }/{2}\;}} \\ \end{matrix} \right]$ &
\includegraphics[width=2.2cm]{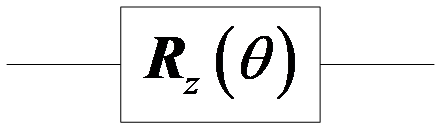} \\ \addlinespace[4pt]
\multirow{3}{*}[-35pt]{Two-qubit gates} & SWAP gate
& $\left[ \begin{matrix} 1 & {} & {} & {} \\ {} & {} & 1 & {} \\ {} & 1 & {} & {} \\ {} & {} & {} & 1 \\ \end{matrix} \right]$ &
\includegraphics[width=2.2cm]{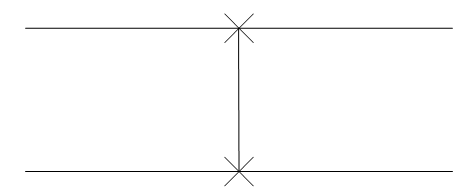} \\ \addlinespace[4pt]
& Open-controlled $U$ gate & $\left[ \begin{matrix} \mathbf{U} & {} \\ {} & {{\mathbf{I}}_{2\times 2}} \\ \end{matrix} \right]$ &
\includegraphics[width=2.2cm]{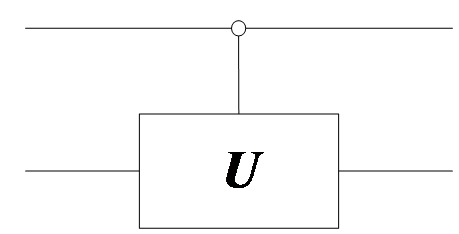} \\ \addlinespace[4pt]
& Controlled-$U$ gate & $\left[ \begin{matrix} {{\mathbf{I}}_{2\times 2}} & {} \\ {} & \mathbf{U} \\ \end{matrix} \right]$ &
\includegraphics[width=2.2cm]{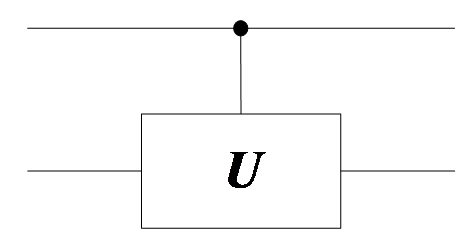} \\ \addlinespace[6pt]
\bottomrule
\end{tabular}
\end{table}

The ansatz circuits also use the CNOT gate, which is the controlled-$U$ gate with $\mathbf{U}=\mathbf{X}$. All subsequent circuits are constructed from the gates listed in Table~\ref{tab:1}.

\subsection{LCU decomposition}

In the gate model, deterministic coherent evolution is implemented by unitary operators. The constrained finite element stiffness matrix $\mathbf{K}$ is real symmetric positive definite but generally does not satisfy $\mathbf{K}^{\mathrm{H}}\mathbf{K}=\mathbf{I}$. It therefore cannot act directly as a quantum gate. To represent $\mathbf{K}$ in a quantum circuit, we write it as a weighted sum of unitary matrices:

\begin{equation}\mathbf{K}=\sum\limits_{i=0}^{M-1}{{{a}_{i}}{{\mathbf{K}}_{i}}},\label{eq:5}\end{equation}
where $\mathbf{K}_i$ is unitary and $a_i\in\mathbb{C}$ is its coefficient. Equation~\eqref{eq:5} expresses a nonunitary matrix as a linear combination of unitaries. We call this an LCU decomposition of the stiffness matrix. Each $a_i\mathbf{K}_i$ is a unitary term, and $M$ is the number of terms. A common LCU construction uses Pauli decomposition. The single-qubit Pauli matrices are

\begin{equation}{{\boldsymbol{\sigma}}_{0}}=\left[ \begin{matrix} 1 & 0 \\ 0 & 1 \\ \end{matrix} \right],\ \ {{\boldsymbol{\sigma}}_{1}}=\left[ \begin{matrix} 0 & 1 \\ 1 & 0 \\ \end{matrix} \right],\ \ {{\boldsymbol{\sigma}}_{2}}=\left[ \begin{matrix} 0 & -\text{i} \\ \text{i} & 0 \\ \end{matrix} \right],\ \ {{\boldsymbol{\sigma}}_{3}}=\left[ \begin{matrix} 1 & 0 \\ 0 & -1 \\ \end{matrix} \right].\label{eq:6}\end{equation}
Thus $\boldsymbol{\sigma}_0=\mathbf{I}$, $\boldsymbol{\sigma}_1=\mathbf{X}$, $\boldsymbol{\sigma}_2=\mathbf{Y}$, and $\boldsymbol{\sigma}_3=\mathbf{Z}$. The $i$-th $n$-qubit Pauli string is

\begin{equation}{{\mathbf{K}}_{i}}={{\boldsymbol{\sigma}}_{{{i}_{0}}}}\otimes {{\boldsymbol{\sigma}}_{{{i}_{1}}}}\otimes \cdots \otimes {{\boldsymbol{\sigma}}_{{{i}_{n-1}}}}.\label{eq:7}\end{equation}
Each position in the tensor product corresponds to one qubit. To enumerate the Pauli strings with a single integer, write $i$ in base four with $n$ digits:

\begin{equation}i=\sum\limits_{t=0}^{n-1}{{{i}_{t}}\times {{4}^{n-1-t}}}={{i}_{0}}\times {{4}^{n-1}}+{{i}_{1}}\times {{4}^{n-2}}+\cdots +{{i}_{n-1}}\times {{4}^{0}},\label{eq:8}\end{equation}
where $i_t\in\{0,1,2,3\}$. There are $4^n=N^2$ candidate Pauli strings, which satisfy the trace-orthogonality relation $\operatorname{tr}(\mathbf{K}_i\mathbf{K}_j)=2^n\delta_{ij}$. The coefficients in the full expansion are therefore

\begin{equation}{{a}_{i}}=\frac{1}{{{2}^{n}}}\text{tr}\left( \mathbf{K}{{\mathbf{K}}_{i}} \right).\label{eq:9}\end{equation}

Equation~\eqref{eq:9} gives the coefficients by projecting the stiffness matrix onto the Pauli strings. Several efficient Pauli-decomposition algorithms have been proposed, including tensorized Pauli decomposition (TPD) \cite{Hantzko2024} and MSPD \cite{MSPD2026}. MSPD exploits matrix sparsity and uses a fast Hadamard transform to accelerate the decomposition. It is among the most efficient reported approaches.

\subsection{Overview of VQLS}

Building on quantum amplitude encoding and LCU decomposition, we review the VQLS framework. It has three components: an ansatz circuit that prepares a trial solution, a loss function that measures solution quality, and a quantum measurement procedure that estimates the loss. The following subsections introduce each component.

\subsubsection{Ansatz}

\begin{figure*}[htbp]
\centering
\includegraphics[width=14.00cm]{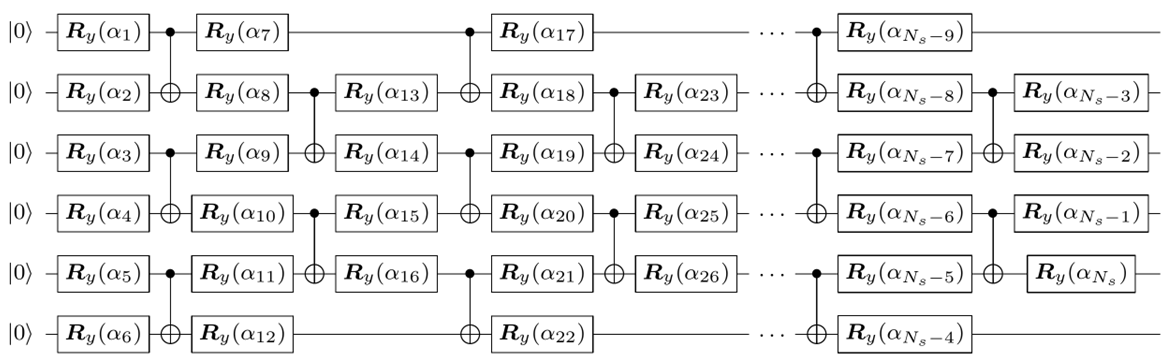}
 \caption{Real-amplitude ansatz circuit with alternating $\mathbf{R}_{y}$ and CNOT gates}\label{fig:1}
\end{figure*}

We first describe the ansatz used to approximate the unknown displacement vector in Eq.~\eqref{eq:1}. Equation~\eqref{eq:3} defines the target displacement state $|\mathbf{u}\rangle$. VQLS uses a parameterized unitary $\mathbf{V}(\boldsymbol{\alpha})$ to prepare a normalized trial state

\begin{equation}\left| \mathbf{u}\left( \boldsymbol{\alpha} \right) \right\rangle =\mathbf{V}\left( \boldsymbol{\alpha} \right){{\left| 0 \right\rangle }^{\otimes n}}=\mathbf{V}\left( \boldsymbol{\alpha} \right){{\mathbf{z}}_{n}}.\label{eq:10}\end{equation}
This state approximates $|\mathbf{u}\rangle$. The circuit implementing $\mathbf{V}(\boldsymbol{\alpha})$ is the ansatz circuit, also called a parameterized quantum circuit (PQC). We denote the vector corresponding to $|0\rangle^{\otimes n}$ by $\mathbf{z}_n=(1,0,\ldots,0)^{\mathrm{T}}\in\mathbb{R}^{N}$. Figure~\ref{fig:1} shows a representative ansatz with $N_{\alpha}$ parameters. Alternating $\mathbf{R}_y$ and CNOT gates act on $|0\rangle^{\otimes 6}$ and encode a displacement vector of dimension $2^6$.

The $\mathbf{R}_y$ gates in Fig.~\ref{fig:1} adjust the computational-basis amplitudes, while the CNOT gates create correlations between qubits. Additional rotation and entangling layers can increase ansatz expressivity. They also increase parameter count, circuit depth, and optimization difficulty. Ansatz design must therefore balance expressivity against trainability.

The state $|\mathbf{u}(\boldsymbol{\alpha})\rangle$ is normalized and specifies only the direction of the displacement vector. A classical scalar $c$ sets its magnitude. From Eq.~\eqref{eq:10}, the displacement vector is

\begin{equation}\mathbf{u}=c\mathbf{V}\left( \boldsymbol{\alpha} \right){{\mathbf{z}}_{n}},\ \ \ c={{\left\| \mathbf{u} \right\|}_{2}}.\label{eq:11}\end{equation}
Substituting Eq.~\eqref{eq:11} into Eq.~\eqref{eq:1} gives

\begin{equation}\mathbf{K}c\mathbf{V}\left( \boldsymbol{\alpha} \right){{\mathbf{z}}_{n}}=\mathbf{f}.\label{eq:12}\end{equation}
Its quantum-state form is

\begin{equation}c\mathbf{K}\left| \mathbf{u}\left( \boldsymbol{\alpha} \right) \right\rangle ={{\left\| \mathbf{f} \right\|}_{2}}\left| \mathbf{f} \right\rangle.\label{eq:13}\end{equation}

The unknowns are thus $c$ and $\boldsymbol{\alpha}$. Once $\boldsymbol{\alpha}$ is determined, $c$ follows. The main task is therefore to determine $\boldsymbol{\alpha}$.

\subsubsection{Loss functions}

To determine $\boldsymbol{\alpha}$, VQLS casts the linear-system solve as an optimization problem. A loss function measures trial-solution quality, and a classical optimizer repeatedly updates the parameters to reduce that loss. The loss functions below are evaluated at the ansatz state $|\mathbf{u}(\boldsymbol{\alpha})\rangle$. We omit its argument $\boldsymbol{\alpha}$ in the following formulas for brevity. Common choices are:

 (1) The global loss \cite{BravoPrieto2023} measures how far $\mathbf{K}|\mathbf{u}\rangle$ departs from the direction of $|\mathbf{f}\rangle$:

\begin{equation}{{\hat{L}}_{G}}=\left\langle \mathbf{u} \right|{{\mathbf{H}}_{G}}\left| \mathbf{u} \right\rangle ,\ \ \ {{\mathbf{H}}_{G}}={{\mathbf{K}}^{\mathrm{T}}}\left( \mathbf{I}-\left| \mathbf{f} \right\rangle \left\langle \mathbf{f} \right| \right)\mathbf{K}.\label{eq:14}\end{equation}
Its normalized form is

\begin{equation}{{L}_{G}}=\frac{{{{\hat{L}}}_{G}}}{\left\langle \mathbf{u} \right|{{\mathbf{K}}^{\mathrm{T}}}\mathbf{K}\left| \mathbf{u} \right\rangle }=1-\frac{\left\langle \mathbf{u} \right|{{\mathbf{K}}^{\mathrm{T}}}\left| \mathbf{f} \right\rangle \left\langle \mathbf{f} \right|\mathbf{K}\left| \mathbf{u} \right\rangle }{\left\langle \mathbf{u} \right|{{\mathbf{K}}^{\mathrm{T}}}\mathbf{K}\left| \mathbf{u} \right\rangle }.\label{eq:15}\end{equation}

 (2) The local loss \cite{BravoPrieto2023} measures this departure separately on each qubit:

\begin{align}& {{{\hat{L}}}_{L}}=\left\langle \mathbf{u} \right|{{\mathbf{H}}_{L}}\left| \mathbf{u} \right\rangle ,\ \ \ {{\mathbf{H}}_{L}}={{\mathbf{K}}^{\mathrm{T}}}\left( \mathbf{I}-\mathbf{FP}{{\mathbf{F}}^{\mathrm{H}}} \right)\mathbf{K}\notag \\ & \mathbf{P}=\frac{1}{n}\sum\limits_{j=1}^{n}{\mathbf{I}_{2}^{\otimes \left( j-1 \right)}\otimes \mathbf{\tau }\otimes \mathbf{I}_{2}^{\otimes \left( n-j \right)}},\ \ \ \mathbf{\tau }=\left| 0 \right\rangle \left\langle 0 \right|=\left[ \begin{matrix} 1 & 0 \\ 0 & 0 \\ \end{matrix} \right],\label{eq:16}\end{align}
where $\mathbf{F}$ is the unitary defined in Eq.~\eqref{eq:4}, with $\mathbf{F}|0\rangle^{\otimes n}=|\mathbf{f}\rangle$, and $\mathbf{F}^{\mathrm{H}}$ is its conjugate transpose. The operator $\mathbf{P}$ is the weighted average of $n$ single-qubit projectors $\boldsymbol{\tau}=|0\rangle\langle 0|$, which measures how far $\mathbf{F}^{\mathrm{H}}\mathbf{K}|\mathbf{u}\rangle$ departs from $|0\rangle^{\otimes n}$. Its normalized form is

\begin{equation}{{L}_{L}}=\frac{{{{\hat{L}}}_{L}}}{\left\langle \mathbf{u} \right|{{\mathbf{K}}^{\mathrm{T}}}\mathbf{K}\left| \mathbf{u} \right\rangle }.\label{eq:17}\end{equation}

 (3) The residual loss \cite{Huang2021Regression} directly measures the residual of Eq.~\eqref{eq:1}:

\begin{equation}{{L}_{R}}=\left\| c\mathbf{K}\left| \mathbf{u} \right\rangle -{{\left\| \mathbf{f} \right\|}_{2}}\left| \mathbf{f} \right\rangle \right\|_{2}^{2},\label{eq:18}\end{equation}
where $c$ is the scaling factor defined in Eq.~\eqref{eq:11}. The normalized global and local losses both lie in $[0,1]$, and the residual loss is the squared $L_2$ norm of the residual of Eq.~\eqref{eq:1}.

\begin{figure*}[htbp]
\centering
\includegraphics[width=12.00cm]{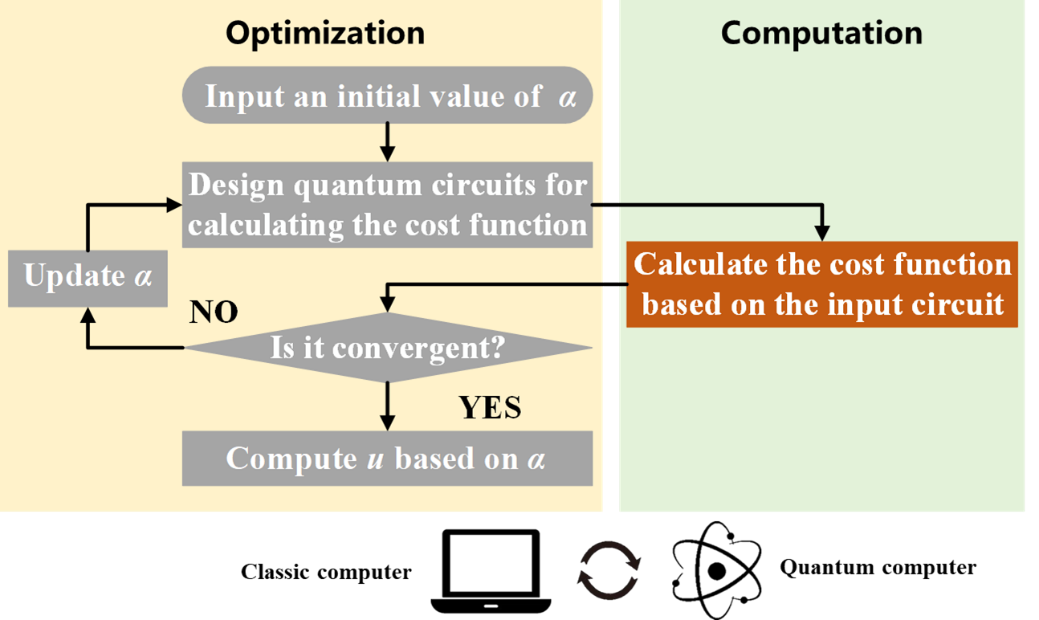}
\caption{Hybrid quantum--classical optimization loop of VQLS}\label{fig:2}
\end{figure*}

If $\mathbf{K}$ is invertible and the ansatz can represent the normalized exact solution, the global minimum of each loss is zero. In principle, optimizing $\boldsymbol{\alpha}$ to reach that minimum yields the solution. Estimating the loss and its gradient is the main computational expense during optimization. Figure~\ref{fig:2} shows the VQLS workflow: the quantum processor estimates the objective, and the classical optimizer updates the parameters.

\subsubsection{Hadamard test}

The loss functions contain vector inner products and operator expectation values. A Hadamard test maps the real part of a unitary expectation value or of an overlap between two quantum states to the measurement probability of an ancilla qubit. Equations~\eqref{eq:14}--\eqref{eq:18} involve $\langle\mathbf{u}|\mathbf{K}^{\mathrm{T}}\mathbf{K}|\mathbf{u}\rangle$, $\langle\mathbf{f}|\mathbf{K}|\mathbf{u}\rangle$, and $\langle\mathbf{u}|\mathbf{K}^{\mathrm{T}}\mathbf{F}\mathbf{P}\mathbf{F}^{\mathrm{H}}\mathbf{K}|\mathbf{u}\rangle$. Explicitly forming $\mathbf{K}^{\mathrm{T}}\mathbf{K}$ on a classical computer requires $O(N^2)$ storage and arithmetic in the stated setting, whereas the corresponding scalars for sparse $\mathbf{K}$ can be evaluated at a cost proportional to the number of nonzeros. In the quantum formulation, substituting Eq.~\eqref{eq:5} into the first term gives

\begin{equation}\left\langle \mathbf{u} \right|{{\mathbf{K}}^{\mathrm{T}}}\mathbf{K}\left| \mathbf{u} \right\rangle =\sum\limits_{i=0}^{M-1}{\sum\limits_{j=0}^{M-1}{a_{i}^{*}{{a}_{j}}\left\langle \mathbf{u} \right|{{\mathbf{U}}_{i,j}}\left| \mathbf{u} \right\rangle }},\ \ \ {{\mathbf{U}}_{i,j}}=\mathbf{K}_{i}^{\mathrm{H}}{{\mathbf{K}}_{j}}.\label{eq:19}\end{equation}

\begin{figure*}[htbp]
\centering
\includegraphics[width=9.83cm]{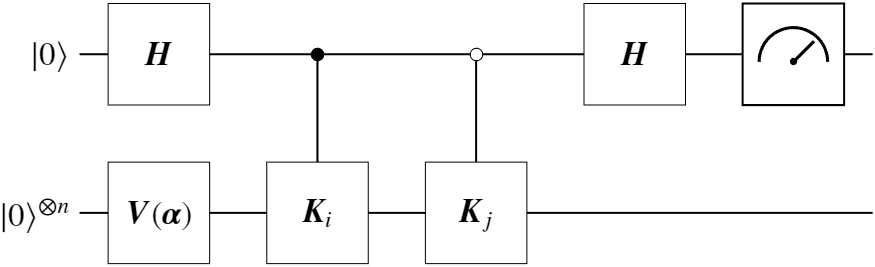}
 \caption{Hadamard-test circuit for estimating $\operatorname{Re}\langle\mathbf{u}|\mathbf{K}_{i}^{\mathrm{H}}\mathbf{K}_{j}|\mathbf{u}\rangle$}\label{fig:3}
\end{figure*}

Because the product $\mathbf{U}_{i,j}$ of two unitary matrices is unitary, evaluating the loss reduces to estimating its expectation value in $|\mathbf{u}\rangle$. Figure~\ref{fig:3} shows the corresponding Hadamard-test circuit.

Run the circuit in Fig.~\ref{fig:3} and denote the probability of measuring the ancilla in $|0\rangle$ by $p_{i,j}$. Then

\begin{equation}\operatorname{Re}\left( \left\langle \mathbf{u} \right|{{\mathbf{K}}_{i}}^{\mathrm{H}}{{\mathbf{K}}_{j}}\left| \mathbf{u} \right\rangle \right)=2{{p}_{i,j}}-1.\label{eq:20}\end{equation}

Likewise, Eq.~\eqref{eq:5} gives

\begin{equation}\left\langle \mathbf{f} \right|\mathbf{K}\left| \mathbf{u} \right\rangle =\sum\limits_{i=0}^{M-1}{{{a}_{i}}\left\langle \mathbf{f} \right|{{\mathbf{K}}_{i}}\left| \mathbf{u} \right\rangle }.\label{eq:21}\end{equation}

If the probability of measuring the ancilla in $|0\rangle$ is $\beta_i$, then

\begin{equation}\operatorname{Re}\left( \left\langle \mathbf{f} \right|{{\mathbf{K}}_{i}}\left| \mathbf{u} \right\rangle \right)=2{{\beta }_{i}}-1.\label{eq:22}\end{equation}

These probabilities are estimated from finitely many repeated measurements. For independent Bernoulli samples, the standard error is $\sqrt{p(1-p)/S}=O(S^{-1/2})$, where $p$ is the true probability of the outcome and $S$ is the number of shots.

The matrix $\mathbf{K}$, vector $\mathbf{f}$, and ansatz amplitudes are real in this work, so the final objectives are real. Individual unitary terms may still have complex expectation values, but their imaginary parts cancel in the weighted sum. General complex-valued problems would require a phase gate on the ancilla to estimate the imaginary part. Note also that $\mathbf{K}|\mathbf{u}\rangle$ is generally unnormalized and cannot be prepared directly as the output of an ordinary unitary gate.

\subsection{Three challenges for VQLS}

When combining classical finite element discretization with VQLS, comparing only linear-system dimensions is insufficient: one must also account for data encoding, matrix representation, quantum measurement, and classical optimization. Based on Eqs.~\eqref{eq:5}--\eqref{eq:22}, we focus on three challenges.

 (1) Term-by-term objective evaluation is costly. Global, local, and residual losses commonly contain $\mathbf{K}^{\mathrm{T}}\mathbf{K}$. If $\mathbf{K}$ has $M$ unitary terms, the direct double expansion in Eq.~\eqref{eq:19} contains $M^2$ ordered inner products. Conjugate symmetry reduces the prefactor but not the $O(M^2)$ scaling.

 (2) Generic matrix representations do not exploit the structure of finite element stiffness matrices. Pauli decomposition applies to any matrix, but an $n$-qubit system has $4^n=N^2$ candidate Pauli strings. The number of nonzero terms depends on stiffness-matrix sparsity, degree-of-freedom ordering, and coefficient structure. Using mechanical structure may reduce the number of unitary terms in the LCU decomposition.

 (3) VQLS convergence remains a challenge. The losses in Eqs.~\eqref{eq:14}--\eqref{eq:18} are constructed from different error measures, and with many ansatz parameters their optimization can stall or fail to find a solution. In particular, the loss may become very small while the displacement error remains large.

To address these challenges, Section 3.1 constructs an LCU decomposition of the stiffness matrix using regular voxel-grid adjacency, Section 3.2 develops a variational objective with one action of $\mathbf{K}$ from the principle of minimum potential energy and relates its gradient to displacement error, and Section 4 encodes multiple unitary terms and parameter-shifted states in index registers through a block-Hadamard test, reducing the number of circuit configurations. Together, these steps define a voxel-based quantum finite element framework intended to lower quantum computational cost and improve convergence.

\section{Quantum finite element method on voxel grids}

Section 2 presented the general VQLS procedure. Applying VQLS to finite elements requires two further steps specific to mechanics. First, we exploit stiffness-matrix structure to reduce the number of unitary terms in the LCU decomposition. Second, we choose a variational objective consistent with the equilibrium equations. The Pauli decomposition in Section 2.2 applies to any $N\times N$ matrix but has up to $4^n=N^2$ candidate Pauli strings. Finite element stiffness matrices are assembled from local element matrices according to nodal connectivity. Their nonzero patterns therefore reflect mesh adjacency. We construct the quantum representation from the mesh and degree-of-freedom ordering rather than treating the stiffness matrix as an arbitrary dense matrix.

Voxel grids discretize a geometric domain with regularly arranged elements of identical topology. They provide uniform element connectivity, regular degree-of-freedom ordering, and convenient parallel assembly \cite{Schneider2022Voxel,Wu2026VBQC}. For finite elements with local support, lexicographic numbering along the coordinate directions yields a banded stiffness matrix in one dimension. In two and three dimensions, the matrices admit recursive block-banded forms. This structure directly links geometric discretization to quantum operator decomposition. Local coupling between adjacent nodes produces a finite number of block-diagonal offsets. Cyclic permutation matrices handle these offsets, and the remaining blocks are decomposed into unitaries.

We use this structure to construct an LCU decomposition. The upper bound on the number of unitary terms depends on the spatial dimension and the degrees of freedom per node, but not on the node count along each coordinate direction. Section 3.1 derives the decomposition for voxel finite elements. Section 3.2 develops a variational objective from the principle of minimum potential energy in solid mechanics.

\subsection{LCU decomposition of stiffness matrices on voxel grids}

This section derives LCU decompositions for real symmetric stiffness matrices on one-, two-, and three-dimensional voxel grids. We begin in one dimension, where local coupling directly gives a banded matrix. The recursive block structure then extends the result to two and three dimensions. Each decomposition has three steps. A cyclic permutation moves off-diagonal blocks onto the main diagonal. Singular value decomposition is applied to the $2\times2$ real blocks. Euler's formula then expresses each nonnegative diagonal matrix as a sum of two diagonal unitaries. The resulting unitary terms contain cyclic permutations, $2\times2$ orthogonal blocks, and diagonal phases. Remark 2 and Appendices 1 and 2 describe their quantum circuits and resource assumptions.

\subsubsection{LCU decomposition for a one-dimensional voxel grid}

\paragraph{(a) One degree of freedom per node}

Consider a one-dimensional finite element problem with $N_x=2^{n_x}$ nodes along the $x$ direction, where $n_x$ is the number of qubits needed for the node index. We first consider one degree of freedom per node, as in elementary heat-conduction and one-dimensional string-vibration problems. The finite element stiffness matrix $\mathbf{K}$ is

\begin{equation}\mathbf{K}=\left[ \begin{matrix} {{k}_{0,0}} & {{k}_{0,1}} & {} & {} & {} \\ {{k}_{1,0}} & {{k}_{1,1}} & {{k}_{1,2}} & {} & {} \\ {} & {{k}_{2,1}} & {{k}_{2,2}} & \ddots & {} \\ {} & {} & \ddots & \ddots & {{k}_{{{N}_{x}}-2,{{N}_{x}}-1}} \\ {} & {} & {} & {{k}_{{{N}_{x}}-1,{{N}_{x}}-2}} & {{k}_{{{N}_{x}}-1,{{N}_{x}}-1}} \\ \end{matrix} \right].\label{eq:23}\end{equation}

To obtain an LCU decomposition, split this stiffness matrix into three matrices:

\begin{align}& {{\mathbf{K}}_{1}}=\left[ \begin{matrix} {{k}_{0,0}} & {} & {} & {} & {} \\ {} & {{k}_{1,1}} & {} & {} & {} \\ {} & {} & {{k}_{2,2}} & {} & {} \\ {} & {} & {} & \ddots & {} \\ {} & {} & {} & {} & {{k}_{{{N}_{x}}-1,{{N}_{x}}-1}} \\ \end{matrix} \right]=\underset{n=0}{\overset{{{N}_{x}}-1}{\mathop{\oplus }}}\,{{k}_{n,n}}\notag \\ & {{\mathbf{K}}_{2}}=\left[ \begin{matrix} {} & {{k}_{0,1}} & {} & {} & {} \\ {} & {} & {{k}_{1,2}} & {} & {} \\ {} & {} & {} & \ddots & {} \\ {} & {} & {} & {} & {{k}_{{{N}_{x}}-2,{{N}_{x}}-1}} \\ {} & {} & {} & {} & {} \\ \end{matrix} \right]=\left[ \underset{n=0}{\overset{{{N}_{x}}-2}{\mathop{\oplus }}}\,{{k}_{n,n+1}}\oplus 0 \right]\mathbf{S}_{{{N}_{x}}}^{{}}\notag \\ & {{\mathbf{K}}_{3}}=\left[ \begin{matrix} {} & {} & {} & {} & {} \\ {{k}_{1,0}} & {} & {} & {} & {} \\ {} & {{k}_{2,1}} & {} & {} & {} \\ {} & {} & \ddots & {} & {} \\ {} & {} & {} & {{k}_{{{N}_{x}}-1,{{N}_{x}}-2}} & {} \\ \end{matrix} \right]=\mathbf{S}_{{{N}_{x}}}^{\mathrm{T}}\left[ \underset{n=0}{\overset{{{N}_{x}}-2}{\mathop{\oplus }}}\,{{k}_{n+1,n}}\oplus 0 \right],\label{eq:24}\end{align}
where $\mathbf{S}_{N_x}=\left[ \begin{matrix} \mathbf{0} & \mathbf{I}_{N_x-1} \\ 1 & \mathbf{0} \\ \end{matrix} \right]$ is a cyclic permutation matrix. It is unitary and its quantum circuit and resource assumptions are given in Remark 2.

Because a product of unitaries is unitary, only the following three diagonal matrices require further decomposition:
\begin{align*}
\mathbf{D}_1&=\underset{n=0}{\overset{{{N}_{x}}-1}{\mathop{\oplus }}}\,{{k}_{n,n}},\\
\mathbf{D}_2&=\left(\underset{n=0}{\overset{{{N}_{x}}-2}{\mathop{\oplus }}}\,{{k}_{n,n+1}}\right)\oplus 0,\\
\mathbf{D}_3&=\left(\underset{n=0}{\overset{{{N}_{x}}-2}{\mathop{\oplus }}}\,{{k}_{n+1,n}}\right)\oplus 0.
\end{align*}

For a unified derivation, denote each matrix by $\bigoplus_{n=0}^{N_x-1}\hat{k}_n$. Divide each diagonal entry by the largest absolute entry so that it lies in $[-1,1]$. Euler's formula then expresses it as the average of two conjugate phases. Indeed, any real $a$ with $0\le |a|\le 1$ can be written as

\begin{equation}a=\frac{1}{2}{{e}^{\text{i}\phi }}+\frac{1}{2}{{e}^{-\text{i}\phi }},\ \ \ \phi =\arccos \left( a \right)\in \left[ 0,\ \pi \right].\label{eq:25}\end{equation}

By Eq.~\eqref{eq:25},

\begin{equation}\underset{n=0}{\overset{{{N}_{x}}-1}{\mathop{\oplus }}}\,{{\hat{k}}_{n}}=\frac{{{k}_{\max }}}{2}\underset{n=0}{\overset{{{N}_{x}}-1}{\mathop{\oplus }}}\,{{e}^{\text{i}{{\phi }_{n}}}}+\frac{{{k}_{\max }}}{2}\underset{n=0}{\overset{{{N}_{x}}-1}{\mathop{\oplus }}}\,{{e}^{-\text{i}{{\phi }_{n}}}},\label{eq:26}\end{equation}
where

\begin{equation}{{k}_{\max }}=\underset{0\le n\le {{N}_{x}}-1}{\mathop{\max }}\,\left( \left| {{{\hat{k}}}_{n}} \right| \right),\ \ \ {{\phi }_{n}}=\arccos \left( \frac{{{{\hat{k}}}_{n}}}{{{k}_{\max }}} \right).\label{eq:27}\end{equation}
If $k_{\max}=0$, the diagonal matrix is zero and should be omitted from the LCU decomposition. Equation~\eqref{eq:27} applies only when $k_{\max}>0$.

Applying Eq.~\eqref{eq:27} to the three diagonal matrices gives the following expression. If $\kappa_1$ or $\kappa_2$ is zero, the corresponding zero matrix is omitted from the LCU decomposition.

\begin{align}& \underset{n=0}{\overset{{{N}_{x}}-1}{\mathop{\oplus }}}\,{{k}_{n,n}}=\frac{{{\kappa }_{1}}}{2}\left( \underset{n=0}{\overset{{{N}_{x}}-1}{\mathop{\oplus }}}\,{{e}^{\text{i}{{\phi }_{n}}}}+\underset{n=0}{\overset{{{N}_{x}}-1}{\mathop{\oplus }}}\,{{e}^{-\text{i}{{\phi }_{n}}}} \right)\notag \\ & \underset{n=0}{\overset{{{N}_{x}}-2}{\mathop{\oplus }}}\,{{k}_{n,n+1}}\oplus 0=\underset{n=0}{\overset{{{N}_{x}}-2}{\mathop{\oplus }}}\,{{k}_{n+1,n}}\oplus 0=\frac{{{\kappa }_{2}}}{2}\left[ \left(\underset{n=0}{\overset{{{N}_{x}}-2}{\mathop{\oplus }}}\,{{e}^{\text{i}{{\psi }_{n}}}}\right)\oplus 1+\left(\underset{n=0}{\overset{{{N}_{x}}-2}{\mathop{\oplus }}}\,{{e}^{-\text{i}{{\psi }_{n}}}}\right)\oplus \left( -1 \right) \right],\label{eq:28}\end{align}
where

\begin{align}& {{\kappa }_{1}}=\underset{0\le n\le {{N}_{x}}-1}{\mathop{\max }}\,\left| {{k}_{n,n}} \right|,\ \ \ {{\phi }_{n}}=\arccos \frac{{{k}_{n,n}}}{{{\kappa }_{1}}}\notag \\ & {{\kappa }_{2}}=\underset{0\le n\le {{N}_{x}}-2}{\mathop{\max }}\,\left| {{k}_{n,n+1}} \right|,\ \ \ {{\psi }_{n}}=\arccos \frac{{{k}_{n,n+1}}}{{{\kappa }_{2}}}.\label{eq:29}\end{align}

Substituting Eq.~\eqref{eq:28} into Eq.~\eqref{eq:24} gives

\begin{align}& \mathbf{K}=\frac{{{\kappa }_{1}}}{2}\left( \underset{n=0}{\overset{{{N}_{x}}-1}{\mathop{\oplus }}}\,{{e}^{\text{i}{{\phi }_{n}}}}+\underset{n=0}{\overset{{{N}_{x}}-1}{\mathop{\oplus }}}\,{{e}^{-\text{i}{{\phi }_{n}}}} \right)+\frac{{{\kappa }_{2}}}{2}\left[ \left(\underset{n=0}{\overset{{{N}_{x}}-2}{\mathop{\oplus }}}\,{{e}^{\text{i}{{\psi }_{n}}}}\right)\oplus 1+\left(\underset{n=0}{\overset{{{N}_{x}}-2}{\mathop{\oplus }}}\,{{e}^{-\text{i}{{\psi }_{n}}}}\right)\oplus \left( -1 \right) \right]\mathbf{S}_{{{N}_{x}}}^{{}}\notag \\ & \ \ \ \ +\frac{{{\kappa }_{2}}}{2}\mathbf{S}_{{{N}_{x}}}^{\mathrm{T}}\left[ \left(\underset{n=0}{\overset{{{N}_{x}}-2}{\mathop{\oplus }}}\,{{e}^{\text{i}{{\psi }_{n}}}}\right)\oplus 1+\left(\underset{n=0}{\overset{{{N}_{x}}-2}{\mathop{\oplus }}}\,{{e}^{-\text{i}{{\psi }_{n}}}}\right)\oplus \left( -1 \right) \right].\label{eq:30}\end{align}
{\emergencystretch=4em

Thus, the one-dimensional stiffness matrix is a sum of at most six unitary terms. The stated implementation time for the cyclic permutation matrix $\mathbf{S}_{N_x}$ is $O(\mathrm{poly}(\log N))$, and the other unitaries are diagonal and can be implemented with quantum multiplexors \cite{Shende2006Multiplexor,Soudackov2026QFlux}, also with the stated $O(\mathrm{poly}(\log N))$ time complexity. See Remark 2 and Appendix 2 for the associated assumptions.
\par}

\paragraph{(b) Two degrees of freedom per node}

Next, consider two degrees of freedom per node, such as the deflection and rotation at each node of an Euler--Bernoulli beam element. For $N_x=2^{n_x}$ nodes along $x$ and $d=2$ degrees of freedom per node, the total number of degrees of freedom is $N=dN_x$, and $\mathbf{K}$ has the following block-tridiagonal form:

\begin{equation}\mathbf{K}=\left[ \begin{matrix} {{\mathbf{k}}_{0,0}} & {{\mathbf{k}}_{0,1}} & {} & {} & {} \\ {{\mathbf{k}}_{1,0}} & {{\mathbf{k}}_{1,1}} & {{\mathbf{k}}_{1,2}} & {} & {} \\ {} & {{\mathbf{k}}_{2,1}} & {{\mathbf{k}}_{2,2}} & \ddots & {} \\ {} & {} & \ddots & \ddots & {{\mathbf{k}}_{{{N}_{x}}-2,{{N}_{x}}-1}} \\ {} & {} & {} & {{\mathbf{k}}_{{{N}_{x}}-1,{{N}_{x}}-2}} & {{\mathbf{k}}_{{{N}_{x}}-1,{{N}_{x}}-1}} \\ \end{matrix} \right],\label{eq:31}\end{equation}
where each submatrix $\mathbf{k}_{i,j}$ is $2\times 2$. This block structure facilitates quantum encoding: $n_x$ qubits encode the node index, and another $\log_2 d$ qubits encode the degrees of freedom at each node (one qubit for $d=2$). For the LCU decomposition, write $\mathbf{K}$ as

\begin{equation}\mathbf{K}={{\mathbf{K}}_{1}}+{{\mathbf{K}}_{2}}+{{\mathbf{K}}_{3}},\label{eq:32}\end{equation}
where

\begin{align}& {{\mathbf{K}}_{1}}=\left[ \begin{matrix} {{\mathbf{k}}_{0,0}} & {} & {} & {} & {} \\ {} & {{\mathbf{k}}_{1,1}} & {} & {} & {} \\ {} & {} & {{\mathbf{k}}_{2,2}} & {} & {} \\ {} & {} & {} & \ddots & {} \\ {} & {} & {} & {} & {{\mathbf{k}}_{{{N}_{x}}-1,{{N}_{x}}-1}} \\ \end{matrix} \right]=\underset{n=0}{\overset{{{N}_{x}}-1}{\mathop{\oplus }}}\,{{\mathbf{k}}_{n,n}}\notag \\ & {{\mathbf{K}}_{2}}=\left[ \begin{matrix} {} & {{\mathbf{k}}_{0,1}} & {} & {} & {} \\ {} & {} & {{\mathbf{k}}_{1,2}} & {} & {} \\ {} & {} & {} & \ddots & {} \\ {} & {} & {} & {} & {{\mathbf{k}}_{{{N}_{x}}-2,{{N}_{x}}-1}} \\ {} & {} & {} & {} & {} \\ \end{matrix} \right]=\left[ \underset{n=0}{\overset{{{N}_{x}}-2}{\mathop{\oplus }}}\,{{\mathbf{k}}_{n,n+1}}\oplus {{\mathbf{0}}_{2}} \right]\left[ \mathbf{S}_{{{N}_{x}}}^{{}}\otimes {{\mathbf{I}}_{d}} \right]\notag \\ & {{\mathbf{K}}_{3}}=\left[ \begin{matrix} {} & {} & {} & {} & {} \\ {{\mathbf{k}}_{1,0}} & {} & {} & {} & {} \\ {} & {{\mathbf{k}}_{2,1}} & {} & {} & {} \\ {} & {} & \ddots & {} & {} \\ {} & {} & {} & {{\mathbf{k}}_{{{N}_{x}}-1,{{N}_{x}}-2}} & {} \\ \end{matrix} \right]=\left[ \mathbf{S}_{{{N}_{x}}}^{\mathrm{T}}\otimes {{\mathbf{I}}_{d}} \right]\left[ \underset{n=0}{\overset{{{N}_{x}}-2}{\mathop{\oplus }}}\,{{\mathbf{k}}_{n+1,n}}\oplus {{\mathbf{0}}_{2}} \right],\label{eq:33}\end{align}
where $\mathbf{0}_{2}$ and $\mathbf{I}_{d}$ are the $2\times2$ zero matrix and the $d\times d$ identity matrix, respectively. Since $\mathbf{S}_{N_x}$ is unitary, only three block-diagonal matrices with $2\times2$ blocks need decomposition: $\bigoplus_{n=0}^{N_x-1}\mathbf{k}_{n,n}$, $(\bigoplus_{n=0}^{N_x-2}\mathbf{k}_{n,n+1})\oplus\mathbf{0}_{2}$, and $(\bigoplus_{n=0}^{N_x-2}\mathbf{k}_{n+1,n})\oplus\mathbf{0}_{2}$. Denote any of them by $\bigoplus_{n=0}^{N_x-1}\widehat{\mathbf{k}}_n$. The singular value decomposition of each block is

\begin{equation}{{\mathbf{\hat{k}}}_{n}}={{\mathbf{U}}_{n}}{{\Sigma}_{n}}\mathbf{V}_{n}^{\mathrm{T}},\ \ \ {{\Sigma}_{n}}=\left[ \begin{matrix} {{\sigma }_{1,n}} & {} \\ {} & {{\sigma }_{2,n}} \\ \end{matrix} \right].\label{eq:34}\end{equation}

The formulas for $\mathbf{U}_n$, $\mathbf{V}_n$, and the singular values $\sigma_{1,n}\ge\sigma_{2,n}\ge0$ are given in Lemma 1 of Appendix 1. Substituting Eq.~\eqref{eq:34} into $\bigoplus_{n=0}^{N_x-1}\hat{\mathbf{k}}_n$ gives

\begin{equation}\underset{n=0}{\overset{{{N}_{x}}-1}{\mathop{\oplus }}}\,{{\mathbf{\hat{k}}}_{n}}=\left( \underset{n=0}{\overset{{{N}_{x}}-1}{\mathop{\oplus }}}\,{{\mathbf{U}}_{n}} \right)\left( \underset{n=0}{\overset{{{N}_{x}}-1}{\mathop{\oplus }}}\,{{\Sigma}_{n}} \right)\left( \underset{n=0}{\overset{{{N}_{x}}-1}{\mathop{\oplus }}}\,\mathbf{V}_{n}^{\mathrm{T}} \right).\label{eq:35}\end{equation}

Both $\bigoplus_{n=0}^{N_x-1}\mathbf{U}_n$ and $\bigoplus_{n=0}^{N_x-1}\mathbf{V}_n^{\mathrm T}$ are unitary, so completing the LCU decomposition of $\bigoplus_{n=0}^{N_x-1}\hat{\mathbf{k}}_n$ requires only the decomposition of $\bigoplus_{n=0}^{N_x-1}\operatorname{diag}(\sigma_{1,n},\sigma_{2,n})$. By Eq.~\eqref{eq:25}, this matrix can be written as

\begin{equation}\underset{n=0}{\overset{{{N}_{x}}-1}{\mathop{\oplus }}}\,\left[ \begin{matrix} {{\sigma }_{1,n}} & {} \\ {} & {{\sigma }_{2,n}} \\ \end{matrix} \right]=\frac{{{\sigma }_{\max }}}{2}\underset{n=0}{\overset{{{N}_{x}}-1}{\mathop{\oplus }}}\,\left[ \begin{matrix} {{e}^{\text{i}{{\phi }_{1,n}}}} & {} \\ {} & {{e}^{\text{i}{{\phi }_{2,n}}}} \\ \end{matrix} \right]+\frac{{{\sigma }_{\max }}}{2}\underset{n=0}{\overset{{{N}_{x}}-1}{\mathop{\oplus }}}\,\left[ \begin{matrix} {{e}^{-\text{i}{{\phi }_{1,n}}}} & {} \\ {} & {{e}^{-\text{i}{{\phi }_{2,n}}}} \\ \end{matrix} \right],\label{eq:36}\end{equation}
where

\begin{equation}{{\sigma }_{\max }}=\underset{0\le n\le {{N}_{x}}-1}{\mathop{\max }}\,\left\{ {{\sigma }_{1,n}},{{\sigma }_{2,n}} \right\},\ \ \ {{\phi }_{1,n}}=\arccos \frac{{{\sigma }_{1,n}}}{{{\sigma }_{\max }}},\ \ \ {{\phi }_{2,n}}=\arccos \frac{{{\sigma }_{2,n}}}{{{\sigma }_{\max }}}.\label{eq:37}\end{equation}
If $\sigma_{\max}=0$, all $2\times2$ blocks vanish and this component can be omitted, whereas the following expression assumes $\sigma_{\max}>0$.
Substituting Eq.~\eqref{eq:36} into Eq.~\eqref{eq:35} completes the LCU decomposition of $\bigoplus_{n=0}^{N_x-1}\hat{\mathbf{k}}_n$:

\begin{align}& \underset{n=0}{\overset{{{N}_{x}}-1}{\mathop{\oplus }}}\,{{\mathbf{\hat{k}}}_{n}}=\frac{{{\sigma }_{\max }}}{2}\left( \underset{n=0}{\overset{{{N}_{x}}-1}{\mathop{\oplus }}}\,{{\mathbf{U}}_{n}} \right)\left( \underset{n=0}{\overset{{{N}_{x}}-1}{\mathop{\oplus }}}\,\left[ \begin{matrix} {{e}^{\text{i}{{\phi }_{1,n}}}} & {} \\ {} & {{e}^{\text{i}{{\phi }_{2,n}}}} \\ \end{matrix} \right] \right)\left( \underset{n=0}{\overset{{{N}_{x}}-1}{\mathop{\oplus }}}\,\mathbf{V}_{n}^{\mathrm{T}} \right)\notag \\ & \ \ \ \ \ \ \ \ \ +\frac{{{\sigma }_{\max }}}{2}\left( \underset{n=0}{\overset{{{N}_{x}}-1}{\mathop{\oplus }}}\,{{\mathbf{U}}_{n}} \right)\left( \underset{n=0}{\overset{{{N}_{x}}-1}{\mathop{\oplus }}}\,\left[ \begin{matrix} {{e}^{-\text{i}{{\phi }_{1,n}}}} & {} \\ {} & {{e}^{-\text{i}{{\phi }_{2,n}}}} \\ \end{matrix} \right] \right)\left( \underset{n=0}{\overset{{{N}_{x}}-1}{\mathop{\oplus }}}\,\mathbf{V}_{n}^{\mathrm{T}} \right).\label{eq:38}\end{align}
Thus, any block-diagonal matrix $\bigoplus_{n=0}^{N_x-1}\hat{\mathbf{k}}_n$ with $2\times2$ blocks is a linear combination of two unitaries. Applying the same steps as in Eqs.~\eqref{eq:34}--\eqref{eq:38} to $\mathbf{K}_1$, $\mathbf{K}_2$, and $\mathbf{K}_3$ gives two unitaries for each, so the one-dimensional voxel-grid stiffness matrix requires at most six unitary terms for $d=2$.

The unitaries in Eq.~\eqref{eq:38} are block-diagonal matrices with $2\times2$ unitary blocks. In particular, $\mathbf{U}_n$ and $\mathbf{V}_n$ are $2\times2$ orthogonal matrices and can be represented by elementary rotation gates, as detailed in Appendix 1.

\paragraph{(c) Four degrees of freedom per node}

Finally, consider $d=4$, for which $N=4N_x$. A model with three degrees of freedom per node can be extended by adding one auxiliary degree of freedom with zero load and a positive diagonal entry $\gamma$. This produces a $4\times4$ nodal block, preserves the solution on the original three degrees of freedom, and avoids a zero eigenvalue.

\begin{figure*}[htbp]
\centering
\includegraphics[width=14.00cm]{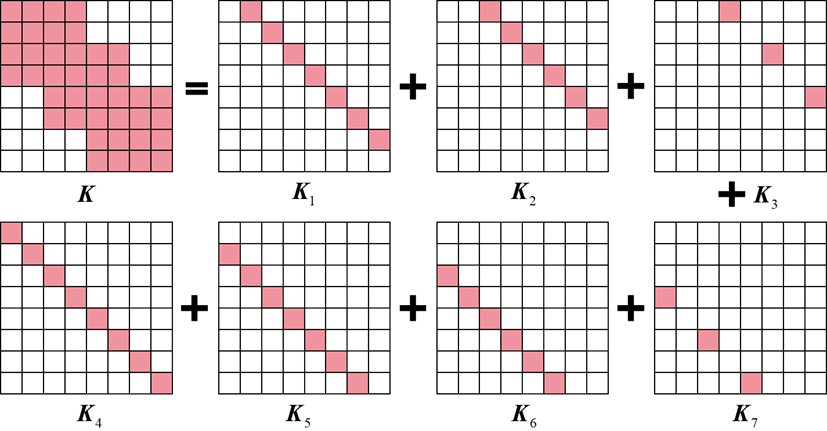}
\caption{Rearrangement of a block-tridiagonal matrix with $4\times4$ blocks into a block-heptadiagonal matrix with $2\times2$ blocks}\label{fig:4}
\end{figure*}

With four degrees of freedom per node, $\mathbf{K}$ remains the block-tridiagonal matrix in Eq.~\eqref{eq:31}, but each block is $4\times4$. As shown in Fig.~\ref{fig:4}, rearrange it as the sum of seven matrices with $2\times2$ blocks:

\begin{equation}\mathbf{K}={{\mathbf{K}}_{1}}+{{\mathbf{K}}_{2}}+{{\mathbf{K}}_{3}}+{{\mathbf{K}}_{4}}+{{\mathbf{K}}_{5}}+{{\mathbf{K}}_{6}}+{{\mathbf{K}}_{7}},\label{eq:39}\end{equation}
where

\begin{align}& {{\mathbf{K}}_{1}}=\underset{n=0}{\overset{2{{N}_{x}}-1}{\mathop{\oplus }}}\,\left[ \begin{matrix} {{k}_{2n,2n}} & {{k}_{2n,2n+1}} \\ {{k}_{2n+1,2n}} & {{k}_{2n+1,2n+1}} \\ \end{matrix} \right],\text{ }\notag \\ & {{\mathbf{K}}_{2}}=\left[ \underset{n=0}{\overset{2{{N}_{x}}-2}{\mathop{\oplus }}}\,\left[ \begin{matrix} {{k}_{2n,2n+2}} & {{k}_{2n,2n+3}} \\ {{k}_{2n+1,2n+2}} & {{k}_{2n+1,2n+3}} \\ \end{matrix} \right]\oplus {{\mathbf{0}}_{2}} \right]\left[ \mathbf{S}_{2{{N}_{x}}}^{{}}\otimes {{\mathbf{I}}_{2}} \right],\notag \\ & {{\mathbf{K}}_{3}}=\left[ \mathbf{S}_{2{{N}_{x}}}^{\mathrm{T}}\otimes {{\mathbf{I}}_{2}} \right]\left[ \underset{n=0}{\overset{2{{N}_{x}}-2}{\mathop{\oplus }}}\,\left[ \begin{matrix} {{k}_{2n+2,2n}} & {{k}_{2n+2,2n+1}} \\ {{k}_{2n+3,2n}} & {{k}_{2n+3,2n+1}} \\ \end{matrix} \right]\oplus {{\mathbf{0}}_{2}} \right],\notag \\ & {{\mathbf{K}}_{4}}=\left[ \underset{n=0}{\overset{2{{N}_{x}}-3}{\mathop{\oplus }}}\,\left[ \begin{matrix} {{k}_{2n,2n+4}} & {{k}_{2n,2n+5}} \\ {{k}_{2n+1,2n+4}} & {{k}_{2n+1,2n+5}} \\ \end{matrix} \right]\oplus {{\mathbf{0}}_{4}} \right]\left[ \mathbf{S}_{2{{N}_{x}}}^{2}\otimes {{\mathbf{I}}_{2}} \right],\notag \\ & {{\mathbf{K}}_{5}}=\left[ {{\left( \mathbf{S}_{2{{N}_{x}}}^{\mathrm{T}} \right)}^{2}}\otimes {{\mathbf{I}}_{2}} \right]\left[ \underset{n=0}{\overset{2{{N}_{x}}-3}{\mathop{\oplus }}}\,\left[ \begin{matrix} {{k}_{2n+4,2n}} & {{k}_{2n+4,2n+1}} \\ {{k}_{2n+5,2n}} & {{k}_{2n+5,2n+1}} \\ \end{matrix} \right]\oplus {{\mathbf{0}}_{4}} \right],\notag \\ & {{\mathbf{K}}_{6}}=\left[ \underset{n=0}{\overset{2{{N}_{x}}-4}{\mathop{\oplus }}}\,\left[ \begin{matrix} {{k}_{2n,2n+6}} & {{k}_{2n,2n+7}} \\ {{k}_{2n+1,2n+6}} & {{k}_{2n+1,2n+7}} \\ \end{matrix} \right]\oplus {{\mathbf{0}}_{6}} \right]\left[ \mathbf{S}_{2{{N}_{x}}}^{3}\otimes {{\mathbf{I}}_{2}} \right],\notag \\ & {{\mathbf{K}}_{7}}=\left[ {{\left( \mathbf{S}_{2{{N}_{x}}}^{\mathrm{T}} \right)}^{3}}\otimes {{\mathbf{I}}_{2}} \right]\left[ \underset{n=0}{\overset{2{{N}_{x}}-4}{\mathop{\oplus }}}\,\left[ \begin{matrix} {{k}_{2n+6,2n}} & {{k}_{2n+6,2n+1}} \\ {{k}_{2n+7,2n}} & {{k}_{2n+7,2n+1}} \\ \end{matrix} \right]\oplus {{\mathbf{0}}_{6}} \right],\label{eq:40}\end{align}
where $\mathbf{S}_{2N_x}=\left[ \begin{matrix} {} & \mathbf{I}_{2N_x-1} \\ 1 & {} \\ \end{matrix} \right]$ is a cyclic permutation matrix. The matrices $\mathbf{0}_2$, $\mathbf{0}_4$, and $\mathbf{0}_6$ are zero matrices of sizes $2\times2$, $4\times4$, and $6\times6$, respectively. As in the $d=2$ case, each $\mathbf{K}_i$ is a linear combination of two unitaries. Thus, for $d=4$, the one-dimensional voxel-grid stiffness matrix has an LCU decomposition with 14 unitary terms.

\subsubsection{LCU decomposition for two- and three-dimensional voxel grids}

\begin{figure*}[htbp]
\centering
\includegraphics[width=14.66cm]{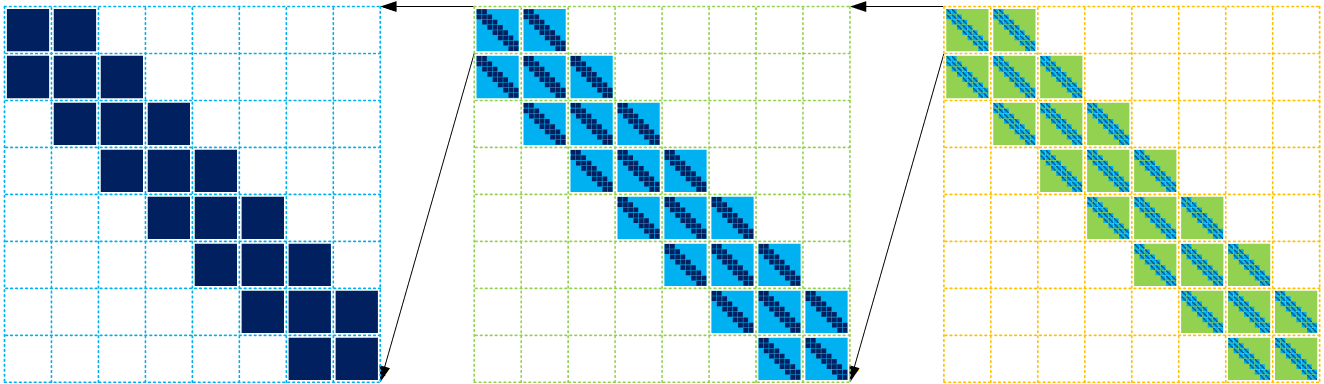}
\caption{Recursive block-banded structure of voxel-grid stiffness matrices}\label{fig:5}
\end{figure*}

A stiffness matrix assembled on a voxel grid has a self-similar block structure, as shown in Fig.~\ref{fig:5}: the three-dimensional matrix can be viewed as a block-tridiagonal matrix of two-dimensional voxel-grid stiffness matrices, and each two-dimensional matrix is likewise block tridiagonal with one-dimensional voxel-grid stiffness matrices as blocks. This property extends the decomposition in Section 3.1.1 to two and three dimensions.

\paragraph{(a) Two-dimensional case}

\begin{figure*}[htbp]
\centering
\includegraphics[width=14.99cm]{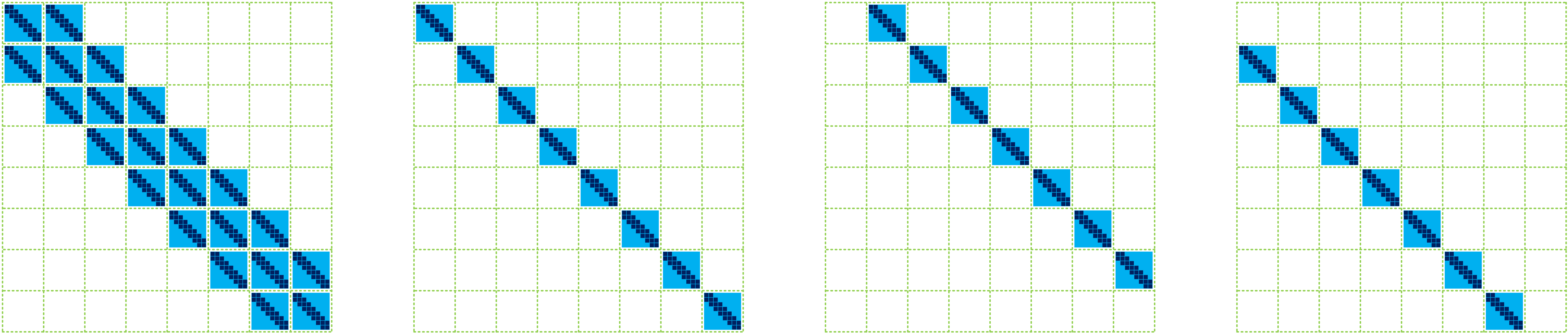}
\caption{Three-part decomposition of a two-dimensional voxel-grid stiffness matrix}\label{fig:6}
\end{figure*}

For a two-dimensional voxel-grid stiffness matrix $\mathbf{K}$, the block-tridiagonal structure gives

\begin{equation}\mathbf{K}=\mathbf{K}_{1}^{\left( 1 \right)}+\mathbf{K}_{2}^{\left( 1 \right)}+\mathbf{K}_{3}^{\left( 1 \right)},\label{eq:41}\end{equation}
where $\mathbf{K}_{1}^{(1)}$ contains the main-diagonal blocks of $\mathbf{K}$, whereas $\mathbf{K}_{2}^{(1)}$ and $\mathbf{K}_{3}^{(1)}$ contain the blocks immediately above and below the main diagonal, respectively. As in Eqs.~\eqref{eq:24} and \eqref{eq:33}, the latter two matrices are products of block-diagonal matrices and cyclic permutation matrices:

\begin{equation}\mathbf{K}_{2}^{\left( 1 \right)}=\left[ \underset{n=0}{\overset{{{N}_{y}}-2}{\mathop{\oplus }}}\,{{\mathbf{B}}_{n,n+1}}\oplus {{\mathbf{0}}_{d{{N}_{x}}}} \right]\left( {{\mathbf{S}}_{{{N}_{y}}}}\otimes {{\mathbf{I}}_{d{{N}_{x}}}} \right),\ \ \ \mathbf{K}_{3}^{\left( 1 \right)}={{\left( {{\mathbf{S}}_{{{N}_{y}}}}\otimes {{\mathbf{I}}_{d{{N}_{x}}}} \right)}^{\mathrm{T}}}\left[ \underset{n=0}{\overset{{{N}_{y}}-2}{\mathop{\oplus }}}\,{{\mathbf{B}}_{n+1,n}}\oplus {{\mathbf{0}}_{d{{N}_{x}}}} \right],\label{eq:42}\end{equation}
where $N_y=2^{n_y}$ is the number of nodes along $y$, and $\mathbf{B}_{n,n+1}$ and $\mathbf{B}_{n+1,n}$ are $dN_x\times dN_x$ coupling blocks between adjacent nodal rows. The main-diagonal and coupling blocks remain block banded along $x$, so the decomposition in Section 3.1.1 applies. Figure~\ref{fig:6} illustrates the two-dimensional splitting.

First, split the block-tridiagonal matrix into three parts along $y$. Then apply the one-dimensional decomposition along $x$ to each part. The $y$ direction produces three classes of block-diagonal offsets. Each class yields at most six unitary terms for $d=1$ or 2, or 14 terms for $d=4$. Thus, the two-dimensional stiffness matrix requires at most $3\times6=18$ terms for $d=1$ or 2, and $3\times14=42$ terms for $d=4$.

\paragraph{(b) Three-dimensional case}

The three-dimensional decomposition is analogous. First, split the voxel-grid stiffness matrix as

\begin{equation}\mathbf{K}=\mathbf{K}_{1}^{\left( 2 \right)}+\mathbf{K}_{2}^{\left( 2 \right)}+\mathbf{K}_{3}^{\left( 2 \right)},\label{eq:43}\end{equation}
where $\mathbf{K}_{1}^{(2)}$, $\mathbf{K}_{2}^{(2)}$, and $\mathbf{K}_{3}^{(2)}$ contain the main-, upper-, and lower-diagonal blocks, respectively. Each top-level block has the structure of a two-dimensional voxel-grid stiffness matrix in the $x$--$y$ plane. It can therefore be split into three components as in Eq.~\eqref{eq:41}. Each two-dimensional component is finally reduced to a one-dimensional decomposition from Section 3.1.1. Equivalently, the three-dimensional decomposition introduces a three-part split in each of the $z$ and $y$ directions. Its upper bound on unitary terms is $3^2$ times the one-dimensional bound: $9\times6=54$ for $d=1$ or 2, and $9\times14=126$ for $d=4$.

Table~\ref{tab:2} lists the upper bounds on the number of unitary terms in the LCU decomposition of $\mathbf{K}$ for different spatial dimensions and nodal degrees of freedom. Unlike a generic Pauli decomposition, the bound for the proposed voxel-grid decomposition does not increase with the matrix dimension when the spatial dimension and $d$ are fixed. It depends on the spatial dimension and $d$. Consequently, the potential reduction relative to a Pauli decomposition can grow with problem size.

\begin{table}[htbp]
\centering
\caption{Upper bounds on the number of unitary terms for voxel-grid stiffness matrices}\label{tab:2}
\begin{tabular}{lccc}
\toprule
 & One-dimensional
 & Two-dimensional
 & Three-dimensional
 \\
\midrule
$d=1\text{ or }2$ & 6 & 18 & 54 \\
$d=4$ & 14 & 42 & 126 \\
\bottomrule
\end{tabular}
\end{table}

\textbf{Remark 1}
Section 3.1 used three techniques: cyclic permutations to handle block-diagonal offsets, singular value decomposition of real $2\times2$ blocks, and Euler's formula to express a nonnegative diagonal matrix as the sum of two diagonal unitaries. The construction can extend to other matrices with known finite block bandwidth, but the term count then depends on the block bandwidth and block size and does not imply a constant term count for arbitrary unstructured meshes.

\textbf{Remark 2}
The unitary terms obtained here mainly comprise cyclic permutations, $2\times2$ orthogonal blocks, and diagonal unitaries. A cyclic permutation can be implemented with a quantum Fourier transform or a reversible addition circuit. The gate count and depth depend on the gate set and whether an approximate QFT is used. A block-diagonal $R_y$ operator, $\bigoplus_{i=0}^{N-1}\mathbf{R}_y(\theta_i)$, can be expressed by a change of basis as
$\begin{aligned}\left( {{\mathbf{I}}_{N}}\otimes \mathbf{\xi } \right)\underset{i=0}{\overset{N-1}{\mathop{\oplus }}}\,{{\mathbf{R}}_{z}}\left( {{\theta }_{i}} \right){{\left( {{\mathbf{I}}_{N}}\otimes \mathbf{\xi } \right)}^{\mathrm{H}}}\end{aligned}$, where $\mathbf{\xi }=\frac{1}{\sqrt{2}}\left[ \begin{matrix} -\text{i} & \text{i} \\ 1 & 1 \\ \end{matrix} \right]$ is a single-qubit unitary. Thus, implementing the unitaries in this decomposition ultimately depends on efficient implementations of diagonal unitaries. In general, such a unitary has the form $\bigoplus_{i=0}^{2^n-1}e^{\mathrm{i}\theta_i}$. A standard approach uses quantum multiplexors \cite{Shende2006Multiplexor,Soudackov2026QFlux}. Appendix 2 states that even in the worst case, when all $\theta_i$ differ, this operator can be implemented with $O(N)$ qubits and an $O(\log N)$-time quantum circuit under the data-access assumptions stated in Appendix 2.

\subsection{Solving finite element stiffness equations by minimum potential energy}

Section 2 described variational quantum optimization and its convergence challenges. The loss functions in Section 2.3.2 are based on different error measures. With many ansatz parameters, optimization with these loss functions may converge slowly or fail to find a solution. In particular, the loss can become small while the displacement error remains large.
Steepest descent and conjugate gradient are common classical finite element solvers. Both can be viewed as optimization methods. For a symmetric positive-definite stiffness matrix, the principle of minimum potential energy recasts the stiffness equation \eqref{eq:1} as

\begin{equation}\Pi \left( \mathbf{u} \right)=\frac{1}{2}{{\mathbf{u}}^{\mathrm{T}}}\mathbf{K}\mathbf{u}-{{\mathbf{f}}^{\mathrm{T}}}\mathbf{u}.\label{eq:44}\end{equation}

Solvers based on minimum potential energy, particularly conjugate gradient, have been highly successful in classical finite element analysis. This motivates using discrete total potential energy as a variational objective in place of a residual-based VQLS loss. The ansatz parameters are then optimized by minimizing that energy. Previous studies \cite{Sato2021,XuHu2026Potential} showed that, for a given relative displacement-error target, estimating the normalized energy difference can require less precision than estimating a VQLS loss. We therefore adopt a minimum-potential-energy variational algorithm and briefly describe its implementation.

Substituting Eq.~\eqref{eq:11} into Eq.~\eqref{eq:44} gives

\begin{equation}\Pi \left( c,\ \boldsymbol{\alpha} \right)=\frac{{{c}^{2}}}{2}\left\langle \mathbf{u}\left( \boldsymbol{\alpha} \right) \right|\mathbf{K}\left| \mathbf{u}\left( \boldsymbol{\alpha} \right) \right\rangle -c{{\left\| \mathbf{f} \right\|}_{2}}\left\langle \mathbf{f} | \mathbf{u}\left( \boldsymbol{\alpha} \right) \right\rangle.\label{eq:45}\end{equation}
The problem becomes $\min_{c\ge0,\,\boldsymbol{\alpha}\in\Omega_{\alpha}}\Pi(c,\boldsymbol{\alpha})$, where $\Omega_{\alpha}$ is the ansatz-parameter domain.

Several classical optimizers can solve this problem. Gradient-based methods, including steepest descent and sequential quadratic programming (SQP), require the gradient $\left(\partial\Pi/\partial c,\ \partial\Pi/\partial\alpha_1,\ldots,\partial\Pi/\partial\alpha_{N_\alpha}\right)^{\mathrm T}$. It can be calculated analytically as follows:

\begin{equation}\left\{ \begin{aligned} & \frac{\partial \Pi }{\partial c}=c\left\langle \mathbf{u}\left( \boldsymbol{\alpha} \right) \right|\mathbf{K}\left| \mathbf{u}\left( \boldsymbol{\alpha} \right) \right\rangle -{{\left\| \mathbf{f} \right\|}_{2}}\left\langle \mathbf{f} | \mathbf{u}\left( \boldsymbol{\alpha} \right) \right\rangle \\ & \frac{\partial \Pi }{\partial {{\alpha }_{j}}}={{c}^{2}}\left\langle \mathbf{u}\left( \boldsymbol{\alpha} \right) \right|\mathbf{K}\left| {{\mathbf{u}}_{{{\alpha }_{j}}}}\left( \boldsymbol{\alpha} \right) \right\rangle -c{{\left\| \mathbf{f} \right\|}_{2}}\left\langle \mathbf{f} | {{\mathbf{u}}_{{{\alpha }_{j}}}}\left( \boldsymbol{\alpha} \right) \right\rangle ,\ \ \ 1\le j\le {{N}_{\alpha }} \end{aligned} \right.,\label{eq:46}\end{equation}
where

\begin{equation}\left| {{\mathbf{u}}_{{{\alpha }_{j}}}}\left( \boldsymbol{\alpha} \right) \right\rangle =\frac{\text{d}\left| \mathbf{u}\left( \boldsymbol{\alpha} \right) \right\rangle }{\text{d}{{\alpha }_{j}}}=\frac{\text{d}\mathbf{V}\left( \boldsymbol{\alpha} \right)}{\text{d}{{\alpha }_{j}}}{{\left| 0 \right\rangle }^{\otimes n}}={{\mathbf{V}}_{{{\alpha }_{j}}}}{{\left| 0 \right\rangle }^{\otimes n}}.\label{eq:47}\end{equation}

If the ansatz in Fig.~\ref{fig:1} is used, each parameter occurs in exactly one $R_y$ gate. Since $\mathrm d\mathbf{R}_y(\alpha)/\mathrm d\alpha=\mathbf{R}_y(\alpha+\pi)/2$, we obtain

\begin{equation}{{\mathbf{V}}_{{{\alpha }_{j}}}}\left( \boldsymbol{\alpha} \right)=\frac{1}{2}\mathbf{V}\left( {{\boldsymbol{\alpha}}_{j,\pi }} \right),\ \ \ {{\boldsymbol{\alpha}}_{j,\pi }}={{\left( {{\alpha }_{1}},\ {{\alpha }_{2}},\ \cdots ,\ {{\alpha }_{j}}+\pi ,\ \cdots ,\ {{\alpha }_{{{N}_{\alpha }}}} \right)}^{\mathrm{T}}}.\label{eq:48}\end{equation}

Substituting Eq.~\eqref{eq:48} into the derivative of the total potential energy gives

\begin{equation}\left\{ \begin{aligned} & \frac{\partial \Pi }{\partial c}=c\left\langle \mathbf{u}\left( \boldsymbol{\alpha} \right) \right|\mathbf{K}\left| \mathbf{u}\left( \boldsymbol{\alpha} \right) \right\rangle -{{\left\| \mathbf{f} \right\|}_{2}}\left\langle \mathbf{f} | \mathbf{u}\left( \boldsymbol{\alpha} \right) \right\rangle \\ & \frac{\partial \Pi }{\partial {{\alpha }_{j}}}=\frac{1}{2}\left[ {{c}^{2}}\left\langle \mathbf{u}\left( \boldsymbol{\alpha} \right) \right|\mathbf{K}\left| \mathbf{u}\left( {{\boldsymbol{\alpha}}_{j,\pi }} \right) \right\rangle -c{{\left\| \mathbf{f} \right\|}_{2}}\left\langle \mathbf{f} | \mathbf{u}\left( {{\boldsymbol{\alpha}}_{j,\pi }} \right) \right\rangle \right],\ \ \ 1\le j\le {{N}_{\alpha }} \end{aligned} \right..\label{eq:49}\end{equation}

Alternatively, a forward finite difference approximates the gradient:

\begin{equation}\left\{ \begin{aligned} & \frac{\partial \Pi }{\partial c}=c\left\langle \mathbf{u}\left( \boldsymbol{\alpha} \right) \right|\mathbf{K}\left| \mathbf{u}\left( \boldsymbol{\alpha} \right) \right\rangle -{{\left\| \mathbf{f} \right\|}_{2}}\left\langle \mathbf{f} | \mathbf{u}\left( \boldsymbol{\alpha} \right) \right\rangle \\ & \frac{\partial \Pi }{\partial {{\alpha }_{j}}}\approx \frac{\Pi \left( c,\ {{\boldsymbol{\alpha}}_{j,\Delta \alpha }} \right)-\Pi \left( c,\ \boldsymbol{\alpha} \right)}{\Delta \alpha } \end{aligned} \right.,\label{eq:50}\end{equation}
where $\boldsymbol{\alpha}_{j,\Delta\alpha}=(\alpha_1,\ldots,\alpha_j+\Delta\alpha,\ldots,\alpha_{N_\alpha})^{\mathrm T}$.

Treat $\mathbf{s}=(c,\boldsymbol{\alpha}^{\mathrm T})^{\mathrm T}$ as the unknown parameters, and write the displacement vector as $\mathbf{u}(\mathbf{s})$. Define its Jacobian as
\[
\mathbf{J}(\mathbf{s})=\frac{\partial \mathbf{u}(\mathbf{s})}{\partial \mathbf{s}^{\mathrm{T}}}\in\mathbb{R}^{N\times(N_{\alpha}+1)}.
\]
If $\mathbf{J}(\mathbf{s})$ has full row rank, so that its smallest singular value $\sigma_{\min}(\mathbf{J}(\mathbf{s}))>0$, the following error bounds hold:
\begin{equation}
\frac{\left\|\nabla_{\mathbf{s}}\Pi(\mathbf{s})\right\|_{2}}
{\sigma_{\max}\!\left(\mathbf{J}(\mathbf{s})\right)\lambda_{\max}(\mathbf{K})}
\leq
\left\|\mathbf{u}(\mathbf{s})-\mathbf{K}^{-1}\mathbf{f}\right\|_{2}
\leq
\frac{\left\|\nabla_{\mathbf{s}}\Pi(\mathbf{s})\right\|_{2}}
{\sigma_{\min}\!\left(\mathbf{J}(\mathbf{s})\right)\lambda_{\min}(\mathbf{K})},
\label{eq:51}
\end{equation}
where $\lambda_{\min}$ and $\lambda_{\max}$ are the smallest and largest eigenvalues of $\mathbf{K}$, while $\sigma_{\min}$ and $\sigma_{\max}$ are the smallest and largest singular values of $\mathbf{J}$. The bounds show how the gradient and ansatz Jacobian jointly affect displacement error. When the Jacobian has full row rank, the gradient norm can indicate the solution error. Moreover, $\sigma_{\min}(\mathbf{J})$ reflects, to some extent, the sensitivity of the ansatz to parameter changes. A smaller value indicates weaker local sensitivity and can make small gradients and slow convergence more likely. A larger value generally improves local identifiability of the ansatz parameters and can assist optimization.

Minimizing the total potential energy over $\mathbf{s}=(c,\boldsymbol{\alpha}^{\mathrm T})^{\mathrm T}$ mainly requires $\langle\mathbf{u}(\boldsymbol{\alpha})|\mathbf{K}|\mathbf{u}(\boldsymbol{\alpha})\rangle$, $\langle\mathbf{f}|\mathbf{u}(\boldsymbol{\alpha})\rangle$, and $\langle\mathbf{u}(\boldsymbol{\alpha})|\mathbf{K}|\mathbf{u}(\boldsymbol{\alpha}_{j,\pi})\rangle$. The load-state overlap $\langle\mathbf{f}|\mathbf{u}(\boldsymbol{\alpha})\rangle$ can be estimated directly with a Hadamard test. Applying the LCU decomposition in Eq.~\eqref{eq:5} to $\mathbf{K}$ gives
\begin{equation}
\begin{aligned}
\left\langle \mathbf{u}(\boldsymbol{\alpha})\right|\mathbf{K}\left|\mathbf{u}(\boldsymbol{\alpha})\right\rangle
&=\sum_{i=0}^{M-1}a_i\left\langle \mathbf{u}(\boldsymbol{\alpha})\right|\mathbf{K}_i\left|\mathbf{u}(\boldsymbol{\alpha})\right\rangle,\\
\left\langle \mathbf{u}(\boldsymbol{\alpha})\right|\mathbf{K}\left|\mathbf{u}(\boldsymbol{\alpha}_{j,\pi})\right\rangle
&=\sum_{i=0}^{M-1}a_i\left\langle \mathbf{u}(\boldsymbol{\alpha})\right|\mathbf{K}_i\left|\mathbf{u}(\boldsymbol{\alpha}_{j,\pi})\right\rangle.
\end{aligned}
\label{eq:52}
\end{equation}

These terms are estimated by the block-Hadamard circuits in Section 4.

\section{Block-Hadamard quantum test}

Estimating Eq.~\eqref{eq:52} with the Hadamard test from Section 2 requires a separate test for each $\mathbf{K}_i$. Each potential-energy or derivative term therefore needs $O(M)$ controlled-unitary circuit configurations. Evaluating all variational parameters increases this number with $N_\alpha$. This section proposes a block-Hadamard quantum test. A quantum multiplexor coherently selects several $\mathbf{K}_i$ within one circuit. The ancilla measurement probability then contains a weighted sum of overlaps. This enables batched estimation of $\langle\mathbf{u}|\mathbf{K}_i|\mathbf{u}\rangle$ and $\langle\mathbf{u}|\mathbf{K}_i|\mathbf{u}_{\alpha_j}\rangle$.

\subsection{Block-Hadamard tests for potential energy and its derivatives}

Let $\mathbf{V}_0=\mathbf{V}(\boldsymbol{\alpha})$ and $|\mathbf{u}_0\rangle=|\mathbf{u}(\boldsymbol{\alpha})\rangle$. For $j=1,\ldots,N_\alpha$, define the normalized parameter-shifted state
\[\mathbf{V}_j=\mathbf{V}(\boldsymbol{\alpha}_{j,\pi}),\qquad |\mathbf{u}_j\rangle=\mathbf{V}_j|0\rangle^{\otimes n}.\]

By Eq.~\eqref{eq:48}, the derivative state is $|\mathbf{u}_{\alpha_j}\rangle=|\mathbf{u}_j\rangle/2$. The quantum circuit implements only the unitary $\mathbf{V}_j$. The factor $1/2$ is retained in the classical postprocessing of Eq.~\eqref{eq:49}.
The loss functions in Section 2.3.2 typically require overlaps of the form $\langle\mathbf{u}|\mathbf{K}_i^{\mathrm H}\mathbf{K}_j|\mathbf{u}\rangle$, involving $O(M^2)$ circuit calls when evaluated term by term. By contrast, the total-potential-energy objective from Section 3.2 requires only $\langle\mathbf{u}|\mathbf{K}_i|\mathbf{u}\rangle$ and $\langle\mathbf{u}|\mathbf{K}_i|\mathbf{u}_{\alpha_j}\rangle$. Their term-by-term evaluation uses $O(M)$ calls. Measurement estimates also have sampling error of order $O(S^{-1/2})$, where $S$ is the number of shots. More overlaps therefore increase the total number of shots needed for a given precision. We now derive block-Hadamard tests for the two required overlap types. First, consider $\langle\mathbf{u}|\mathbf{K}|\mathbf{u}_j\rangle$:

\begin{equation}\left\langle \mathbf{u} \right|\mathbf{K}\left| {{\mathbf{u}}_{j}} \right\rangle =\sum\limits_{i=0}^{M-1}{{{a}_{i}}\left\langle \mathbf{u} \right|{{\mathbf{K}}_{i}}\left| {{\mathbf{u}}_{j}} \right\rangle }.\label{eq:53}\end{equation}

A term-by-term Hadamard test for the $M$ overlaps on the right-hand side of Eq.~\eqref{eq:53} requires $M$ circuit configurations, each sampled repeatedly. Instead, group the terms into blocks:

\begin{equation}\left\langle \mathbf{u} \right|\mathbf{K}\left| {{\mathbf{u}}_{j}} \right\rangle =\sum\limits_{i=0}^{B-1}{{{a}_{i}}\left\langle \mathbf{u} \right|{{\mathbf{K}}_{i}}\left| {{\mathbf{u}}_{j}} \right\rangle }+\sum\limits_{i=B}^{2B-1}{{{a}_{i}}\left\langle \mathbf{u} \right|{{\mathbf{K}}_{i}}\left| {{\mathbf{u}}_{j}} \right\rangle }+\cdots +\sum\limits_{i=M-B}^{M-1}{{{a}_{i}}\left\langle \mathbf{u} \right|{{\mathbf{K}}_{i}}\left| {{\mathbf{u}}_{j}} \right\rangle },\label{eq:54}\end{equation}
where $M=2^m$ and $B=2^b$.

We now design a circuit that evaluates the first block in Eq.~\eqref{eq:54}, $\sum_{i=0}^{B-1}a_i\langle\mathbf{u}|\mathbf{K}_i|\mathbf{u}_j\rangle$, in one batch. We consider (a) nonnegative real coefficients $a_i$ and (b) general complex coefficients.

\paragraph{(a)
$a_i$ are nonnegative real numbers}

First, suppose all $a_i$ are nonnegative real numbers. Amplitude-encode the coefficients $(\sqrt{a_0},\ldots,\sqrt{a_{B-1}})$ of the first block:

\begin{equation}\left| {{\hat{\mathbf{a}}}_{1}} \right\rangle =\frac{1}{\sqrt{\sum\limits_{i=0}^{B-1}{{a}_{i}}}}\sum\limits_{i=0}^{B-1}{\sqrt{{a}_{i}}\left| i \right\rangle }={{\mathbf{\hat{A}}}_{1}}{{\left| 0 \right\rangle }^{\otimes b}}.\label{eq:55}\end{equation}

\begin{figure*}[htbp]
\centering
\includegraphics[width=10.55cm]{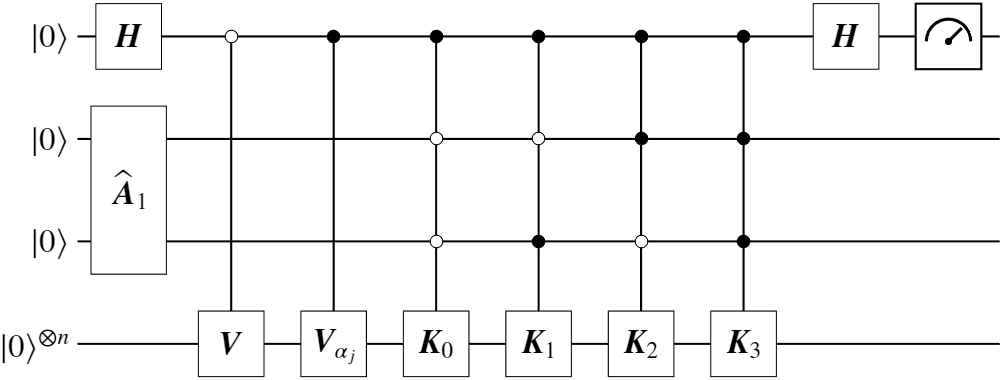}
\caption{Block-Hadamard circuit for the weighted sum in the first block with nonnegative real coefficients ($B=4$)}\label{fig:7}
\end{figure*}

Figure~\ref{fig:7} shows the circuit for $B=4$. Here, $\mathbf{V}_j$ is the state-preparation unitary for the normalized parameter-shifted state, not the nonunitary derivative $\partial\mathbf{V}/\partial\alpha_j$.

The input state is $|\psi\rangle=|0\rangle|0\rangle^{\otimes b}|0\rangle^{\otimes n}$. After the first Hadamard gate, it becomes $|\psi\rangle=(|0\rangle+|1\rangle)|\hat{\mathbf{a}}_1\rangle|0\rangle^{\otimes n}/\sqrt{2}$. The next two layers are controlled. When the first qubit is $|0\rangle$, $\mathbf{V}(\boldsymbol{\alpha})$ acts on the $n$-qubit register. When it is $|1\rangle$, $\mathbf{V}_j$ acts on that register. The state is then

\begin{equation}\left| \psi \right\rangle =\frac{1}{\sqrt{2}}\left| 0 \right\rangle \left| {{\hat{\mathbf{a}}}_{1}} \right\rangle \left| \mathbf{u} \right\rangle +\frac{1}{\sqrt{2}}\left| 1 \right\rangle \left| {{\hat{\mathbf{a}}}_{1}} \right\rangle \left| {{\mathbf{u}}_{j}} \right\rangle.\label{eq:56}\end{equation}

After the quantum multiplexors, the state becomes

\begin{equation}\left| \psi \right\rangle =\frac{1}{\sqrt{2}}\frac{1}{\sqrt{\sum\limits_{i=0}^{B-1}{{a}_{i}}}}\left[ \left| 0 \right\rangle \sum\limits_{i=0}^{B-1}{\sqrt{{a}_{i}}\left| i \right\rangle }\left| \mathbf{u} \right\rangle +\left| 1 \right\rangle \sum\limits_{i=0}^{B-1}{\sqrt{{a}_{i}}\left| i \right\rangle }{{\mathbf{K}}_{i}}\left| {{\mathbf{u}}_{j}} \right\rangle \right].\label{eq:57}\end{equation}

Finally, the Hadamard gate produces

\begin{equation}\left| \psi \right\rangle =\frac{1}{2}\frac{1}{\sqrt{\sum\limits_{i=0}^{B-1}{{a}_{i}}}}\left[ \left| 0 \right\rangle \left( \sum\limits_{i=0}^{B-1}{\sqrt{{a}_{i}}\left| i \right\rangle \left( \left| \mathbf{u} \right\rangle +{{\mathbf{K}}_{i}}\left| {{\mathbf{u}}_{j}} \right\rangle \right)} \right)+\left| 1 \right\rangle \left( \sum\limits_{i=0}^{B-1}{\sqrt{{a}_{i}}\left| i \right\rangle \left( \left| \mathbf{u} \right\rangle -{{\mathbf{K}}_{i}}\left| {{\mathbf{u}}_{j}} \right\rangle \right)} \right) \right].\label{eq:58}\end{equation}

Measure the first qubit and denote the probabilities of $|0\rangle$ and $|1\rangle$ by $\zeta_0$ and $\zeta_1$, respectively. Then

\begin{equation}{{\zeta }_{0}}=\frac{1}{2}\left[ 1+\frac{1}{\sum\limits_{i=0}^{B-1}{{a}_{i}}}\operatorname{Re}\left( \sum\limits_{i=0}^{B-1}{{{a}_{i}}\left\langle \mathbf{u} \right|{{\mathbf{K}}_{i}}\left| {{\mathbf{u}}_{j}} \right\rangle } \right) \right],\ \ \ {{\zeta }_{1}}=\frac{1}{2}\left[ 1-\frac{1}{\sum\limits_{i=0}^{B-1}{{a}_{i}}}\operatorname{Re}\left( \sum\limits_{i=0}^{B-1}{{{a}_{i}}\left\langle \mathbf{u} \right|{{\mathbf{K}}_{i}}\left| {{\mathbf{u}}_{j}} \right\rangle } \right) \right].\label{eq:59}\end{equation}

Therefore,

\begin{equation}\operatorname{Re}\left( \sum\limits_{i=0}^{B-1}{{{a}_{i}}\left\langle \mathbf{u} \right|{{\mathbf{K}}_{i}}\left| {{\mathbf{u}}_{j}} \right\rangle } \right)=\left( 2{{\zeta }_{0}}-1 \right)\sum\limits_{i=0}^{B-1}{{a}_{i}}=\left( 1-2{{\zeta }_{1}} \right)\sum\limits_{i=0}^{B-1}{{a}_{i}}.\label{eq:60}\end{equation}

The remaining blocks in Eq.~\eqref{eq:54} are treated similarly. If $B=M$, one circuit configuration estimates $\langle\mathbf{u}|\mathbf{K}|\mathbf{u}_j\rangle$. To estimate $\langle\mathbf{u}|\mathbf{K}|\mathbf{u}\rangle$, replace $\mathbf{V}_j$ in Fig.~\ref{fig:7} with $\mathbf{V}$.
\paragraph{(b)
$a_i$ are general complex numbers}

\begin{figure*}[htbp]
\centering
\includegraphics[width=10.28cm]{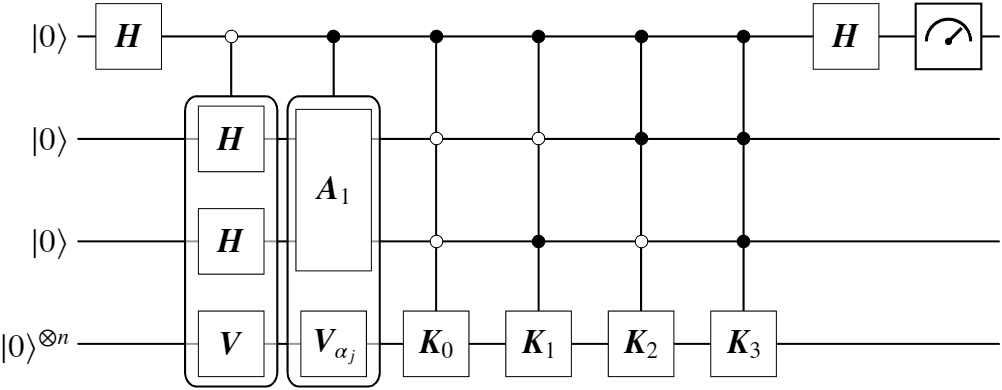}
\caption{Block-Hadamard circuit for the weighted sum in the first block with general complex coefficients ($B=4$)}\label{fig:8}
\end{figure*}

The LCU coefficients $a_i$ in Eq.~\eqref{eq:9} are real for the present problem, but for completeness we also consider complex coefficients. As in (a), amplitude-encode the coefficients $(a_0,\ldots,a_{B-1})$ of the first block:

\begin{equation}\left| {{\hat{\mathbf{a}}}_{1}} \right\rangle =\frac{1}{\sqrt{\sum\limits_{i=0}^{B-1}{{{\left| {{a}_{i}} \right|}^{2}}}}}\sum\limits_{i=0}^{B-1}{{{a}_{i}}\left| i \right\rangle }={{\mathbf{A}}_{1}}{{\left| 0 \right\rangle }^{\otimes b}}.\label{eq:61}\end{equation}
Figure~\ref{fig:8} gives the circuit for general complex coefficients. Again, $\mathbf{V}_j$ prepares the normalized parameter-shifted state.

The input state is $|\psi\rangle=|0\rangle|0\rangle^{\otimes b}|0\rangle^{\otimes n}$. After the first Hadamard gate, it becomes $|\psi\rangle=(|0\rangle+|1\rangle)|0\rangle^{\otimes b}|0\rangle^{\otimes n}/\sqrt{2}$. The next two layers are controlled. On the $|0\rangle$ branch of the first qubit, apply $\mathbf{H}^{\otimes b}\otimes\mathbf{V}(\boldsymbol{\alpha})$ to the target registers. On the $|1\rangle$ branch, apply $\mathbf{A}_1\otimes\mathbf{V}_j$. The state then becomes

\begin{equation}\left| \psi \right\rangle =\frac{1}{\sqrt{2}}\left( \left| 0 \right\rangle {{\mathbf{H}}^{\otimes b}}{{\left| 0 \right\rangle }^{\otimes b}}\left| \mathbf{u} \right\rangle +\left| 1 \right\rangle \left| {{\hat{\mathbf{a}}}_{1}} \right\rangle \left| {{\mathbf{u}}_{j}} \right\rangle \right).\label{eq:62}\end{equation}

After the quantum multiplexors, the state becomes

\begin{equation}\left| \psi \right\rangle =\frac{1}{\sqrt{2}}\left[ \left| 0 \right\rangle \frac{1}{{\left( \sqrt{2} \right)}^{b}}\sum\limits_{i=0}^{B-1}{\left| i \right\rangle }\left| \mathbf{u} \right\rangle +\left| 1 \right\rangle \frac{1}{\sqrt{\sum\limits_{i=0}^{B-1}{{{\left| {{a}_{i}} \right|}^{2}}}}}\sum\limits_{i=0}^{B-1}{{{a}_{i}}\left| i \right\rangle }{{\mathbf{K}}_{i}}\left| {{\mathbf{u}}_{j}} \right\rangle \right].\label{eq:63}\end{equation}

The final Hadamard gate produces

\begin{align}& \left| \psi \right\rangle =\frac{1}{2}\left| 0 \right\rangle \left( \frac{1}{{\left( \sqrt{2} \right)}^{b}}\sum\limits_{i=0}^{B-1}{\left| i \right\rangle }\left| \mathbf{u} \right\rangle +\frac{1}{\sqrt{\sum\limits_{i=0}^{B-1}{{{\left| {{a}_{i}} \right|}^{2}}}}}\sum\limits_{i=0}^{B-1}{{{a}_{i}}\left| i \right\rangle }{{\mathbf{K}}_{i}}\left| {{\mathbf{u}}_{j}} \right\rangle \right) \notag \\ & \ \ \ \ \ \ \ +\frac{1}{2}\left| 1 \right\rangle \left( \frac{1}{{\left( \sqrt{2} \right)}^{b}}\sum\limits_{i=0}^{B-1}{\left| i \right\rangle }\left| \mathbf{u} \right\rangle -\frac{1}{\sqrt{\sum\limits_{i=0}^{B-1}{{{\left| {{a}_{i}} \right|}^{2}}}}}\sum\limits_{i=0}^{B-1}{{{a}_{i}}\left| i \right\rangle }{{\mathbf{K}}_{i}}\left| {{\mathbf{u}}_{j}} \right\rangle \right).\label{eq:64}\end{align}

Measure the first qubit and denote the probabilities of $|0\rangle$ and $|1\rangle$ by $\zeta_0$ and $\zeta_1$. Then

\begin{align}& {{\zeta }_{0}}=\frac{1}{2}\left( 1+\frac{1}{\sqrt{B\sum\limits_{i=0}^{B-1}{{{\left| {{a}_{i}} \right|}^{2}}}}}\operatorname{Re}\left( \sum\limits_{i=0}^{B-1}{{{a}_{i}}\left\langle \mathbf{u} \right|{{\mathbf{K}}_{i}}\left| {{\mathbf{u}}_{j}} \right\rangle } \right) \right) \notag \\ & {{\zeta }_{1}}=\frac{1}{2}\left( 1-\frac{1}{\sqrt{B\sum\limits_{i=0}^{B-1}{{{\left| {{a}_{i}} \right|}^{2}}}}}\operatorname{Re}\left( \sum\limits_{i=0}^{B-1}{{{a}_{i}}\left\langle \mathbf{u} \right|{{\mathbf{K}}_{i}}\left| {{\mathbf{u}}_{j}} \right\rangle } \right) \right).\label{eq:65}\end{align}

Hence,

\begin{equation}\operatorname{Re}\left( \sum\limits_{i=0}^{B-1}{{{a}_{i}}\left\langle \mathbf{u} \right|{{\mathbf{K}}_{i}}\left| {{\mathbf{u}}_{j}} \right\rangle } \right)=\left( 2{{\zeta }_{0}}-1 \right)\sqrt{B\sum\limits_{i=0}^{B-1}{{{\left| {{a}_{i}} \right|}^{2}}}}=\left( 1-2{{\zeta }_{1}} \right)\sqrt{B\sum\limits_{i=0}^{B-1}{{{\left| {{a}_{i}} \right|}^{2}}}}.\label{eq:66}\end{equation}

Other blocks use the same circuit structure. If $B=M$, one circuit configuration estimates $\langle\mathbf{u}|\mathbf{K}|\mathbf{u}_j\rangle$. For $\langle\mathbf{u}|\mathbf{K}|\mathbf{u}\rangle$, replace $\mathbf{V}_j$ in Fig.~\ref{fig:7} with $\mathbf{V}$.

\subsection{Block-Hadamard test for batched derivatives}

Section 4.1 combined several unitary terms within one block, so that the potential energy and its derivative with respect to one parameter can each be estimated with one circuit configuration. With $B=M$, this nominally requires two configurations. A gradient-based optimizer, however, needs derivatives with respect to all $\alpha_j$. Separate circuits for each parameter would still make the configuration count grow with $N_\alpha$. We therefore encode the current state and each state shifted once by $\pi$ in a state-index register. This enables batched estimation of all potential-energy derivatives.

Section 3.2 gave two ways to calculate the potential-energy gradient, in Eqs.~\eqref{eq:49} and \eqref{eq:50}: Eq.~\eqref{eq:49} requires batched estimates of $\langle\mathbf{u}(\boldsymbol{\alpha})|\mathbf{K}|\mathbf{u}_{\alpha_j}(\boldsymbol{\alpha})\rangle$ and $\langle\mathbf{f}|\mathbf{u}_{\alpha_j}(\boldsymbol{\alpha})\rangle$ for $1\le j\le N_\alpha$, whereas Eq.~\eqref{eq:50} requires $\langle\mathbf{u}(\boldsymbol{\alpha}_{j,\Delta\alpha})|\mathbf{K}|\mathbf{u}(\boldsymbol{\alpha}_{j,\Delta\alpha})\rangle$ and $\langle\mathbf{f}|\mathbf{u}(\boldsymbol{\alpha}_{j,\Delta\alpha})\rangle$. To organize these calculations, note that

\begin{equation}\left\{ \begin{aligned} & \left\langle \mathbf{u}\left( \boldsymbol{\alpha } \right) \right|\mathbf{K}\left| {{\mathbf{u}}_{{{\alpha }_{j}}}}\left( \boldsymbol{\alpha } \right) \right\rangle ={{\left\langle 0 \right|}^{\otimes n}}{{\mathbf{V}}^{\mathrm{H}}}\left( \boldsymbol{\alpha } \right)\mathbf{K}{{\mathbf{V}}_{{{\alpha }_{j}}}}\left( \boldsymbol{\alpha } \right){{\left| 0 \right\rangle }^{\otimes n}} \\ & \left\langle \mathbf{u}\left( {{\boldsymbol{\alpha }}_{j,\Delta \alpha }} \right) \right|\mathbf{K}\left| \mathbf{u}\left( {{\boldsymbol{\alpha }}_{j,\Delta \alpha }} \right) \right\rangle ={{\left\langle 0 \right|}^{\otimes n}}{{\mathbf{V}}^{\mathrm{H}}}\left( {{\boldsymbol{\alpha }}_{j,\Delta \alpha }} \right)\mathbf{KV}\left( {{\boldsymbol{\alpha }}_{j,\Delta \alpha }} \right){{\left| 0 \right\rangle }^{\otimes n}} \\ & \left\langle \mathbf{f} | {{\mathbf{u}}_{{{\alpha }_{j}}}}\left( \boldsymbol{\alpha } \right) \right\rangle ={{\left\langle 0 \right|}^{\otimes n}}{{\mathbf{F}}^{\mathrm{H}}}{{\mathbf{V}}_{{{\alpha }_{j}}}}{{\left| 0 \right\rangle }^{\otimes n}},\ \ \ \left\langle \mathbf{f} | \mathbf{u}\left( {{\boldsymbol{\alpha }}_{j,\Delta \alpha }} \right) \right\rangle ={{\left\langle 0 \right|}^{\otimes n}}{{\mathbf{F}}^{\mathrm{H}}}\mathbf{V}\left( {{\boldsymbol{\alpha }}_{j,\Delta \alpha }} \right){{\left| 0 \right\rangle }^{\otimes n}} \end{aligned} \right..\label{eq:67}\end{equation}

By Eq.~\eqref{eq:48}, the derivative state differs from the normalized parameter-shifted state only by the classical factor $1/2$, so that batched evaluation of $\langle\mathbf{u}(\boldsymbol{\alpha})|\mathbf{K}|\mathbf{u}_{\alpha_j}(\boldsymbol{\alpha})\rangle$ reduces to batched evaluation of $\langle0|^{\otimes n}\mathbf{V}^{\mathrm H}\mathbf{K}\mathbf{V}_j|0\rangle^{\otimes n}$. In addition, the finite-difference quadratic term reduces to $\langle0|^{\otimes n}\mathbf{V}_j^{\mathrm H}\mathbf{K}\mathbf{V}_j|0\rangle^{\otimes n}$, and both overlaps with the load state are of the form $\langle0|^{\otimes n}\mathbf{F}^{\mathrm H}\mathbf{V}_j|0\rangle^{\otimes n}$, up to the factor $1/2$ in the analytic derivative. We treat these three cases in turn.

\paragraph{(a)
Batched evaluation of $\langle0|^{\otimes n}\mathbf{V}^{\mathrm H}\mathbf{K}\mathbf{V}_j|0\rangle^{\otimes n}$}

We need these overlaps for $\mathbf{V}_0,\mathbf{V}_1,\ldots,\mathbf{V}_{N_\alpha-1}$. Decompose the stiffness matrix as $\mathbf{K}=\sum_{i=0}^{M-1}a_i\mathbf{K}_i$, so that the batched calculation becomes

\begin{align}& {{\left\langle 0 \right|}^{\otimes n}}{{\mathbf{V}}^{\mathrm{H}}}\mathbf{K}{{\mathbf{V}}_{j}}{{\left| 0 \right\rangle }^{\otimes n}}=\sum\limits_{i=0}^{M-1}{{{a}_{i}}{{\left\langle 0 \right|}^{\otimes n}}{{\mathbf{V}}^{\mathrm{H}}}{{\mathbf{K}}_{i}}{{\mathbf{V}}_{j}}{{\left| 0 \right\rangle }^{\otimes n}}} \notag \\ & \ \ \ \ \ \ \ \ \ \ \ \ \ \ \ \ \ \ \ \ \ \ \ \ \ \ =\sum\limits_{i=0}^{B-1}{{{a}_{i}}{{\left\langle 0 \right|}^{\otimes n}}{{\mathbf{V}}^{\mathrm{H}}}{{\mathbf{K}}_{i}}{{\mathbf{V}}_{j}}{{\left| 0 \right\rangle }^{\otimes n}}}+\sum\limits_{i=B}^{2B-1}{{{a}_{i}}{{\left\langle 0 \right|}^{\otimes n}}{{\mathbf{V}}^{\mathrm{H}}}{{\mathbf{K}}_{i}}{{\mathbf{V}}_{j}}{{\left| 0 \right\rangle }^{\otimes n}}} \notag \\ & \ \ \ \ \ \ \ \ \ \ \ \ \ \ \ \ \ \ \ \ \ \ \ \ \ \ +\cdots +\sum\limits_{i=M-B}^{M-1}{{{a}_{i}}{{\left\langle 0 \right|}^{\otimes n}}{{\mathbf{V}}^{\mathrm{H}}}{{\mathbf{K}}_{i}}{{\mathbf{V}}_{j}}{{\left| 0 \right\rangle }^{\otimes n}}}.\label{eq:68}\end{align}

Partition $j\in\{0,1,\ldots,N_\alpha-1\}$ into $N_\alpha/D$ groups of $D=2^d$ indices. Without loss of generality, consider the first group:

\begin{equation}\sum\limits_{i=0}^{B-1}{{{a}_{i}}{{\left\langle 0 \right|}^{\otimes n}}{{\mathbf{V}}^{\mathrm{H}}}{{\mathbf{K}}_{i}}{{\mathbf{V}}_{0}}{{\left| 0 \right\rangle }^{\otimes n}}},\ \ \sum\limits_{i=0}^{B-1}{{{a}_{i}}{{\left\langle 0 \right|}^{\otimes n}}{{\mathbf{V}}^{\mathrm{H}}}{{\mathbf{K}}_{i}}{{\mathbf{V}}_{1}}{{\left| 0 \right\rangle }^{\otimes n}}},\ \ \cdots ,\ \ \sum\limits_{i=0}^{B-1}{{{a}_{i}}{{\left\langle 0 \right|}^{\otimes n}}{{\mathbf{V}}^{\mathrm{H}}}{{\mathbf{K}}_{i}}{{\mathbf{V}}_{D-1}}{{\left| 0 \right\rangle }^{\otimes n}}}.\label{eq:69}\end{equation}

\begin{figure*}[htbp]
\centering
\includegraphics[width=14.00cm]{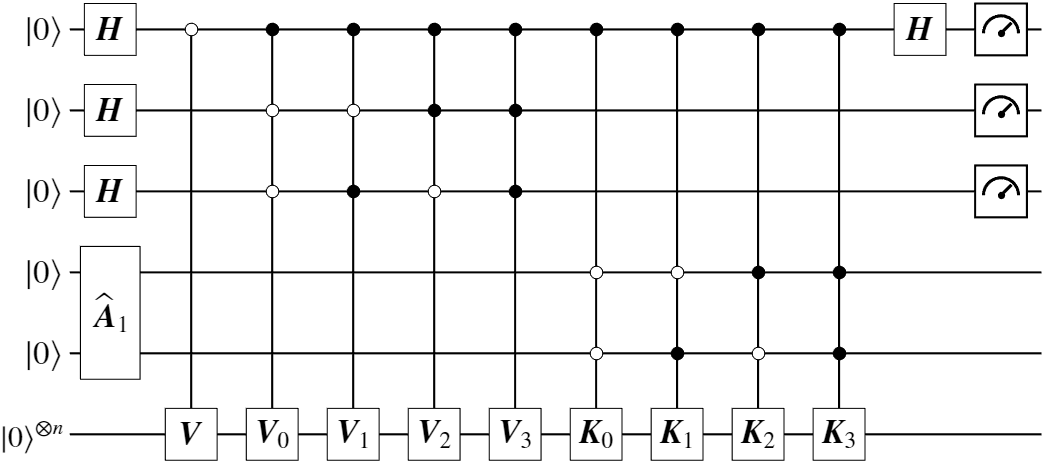}
\caption{Cross-state block-Hadamard circuit for batched estimation of $\operatorname{Re}\langle\mathbf{u}(\boldsymbol{\alpha})|\mathbf{K}|\mathbf{u}_s\rangle$ ($D_K=B=4$)}\label{fig:10}
\end{figure*}

The terms in Eq.~\eqref{eq:69} can be estimated with the circuit in Fig.~\ref{fig:10}, illustrated for $D=B=4$. To simplify the quantum multiplexor, we assume nonnegative real $a_i$. For negative or complex coefficients, their phases can be absorbed into $\mathbf{K}_i$ as in Section 4.1(b).

The input is $|\psi\rangle=|0\rangle|0\rangle^{\otimes d}|0\rangle^{\otimes b}|0\rangle^{\otimes n}$. After the circuit layers, the final state is

\begin{equation}\left| \psi \right\rangle =\frac{1}{{\left( \sqrt{2} \right)}^{d+2}\sqrt{\sum\limits_{i=0}^{B-1}{{{a}_{i}}}}}\left( \left| 0 \right\rangle \sum\limits_{j=0}^{D-1}{\left| j \right\rangle \sum\limits_{i=0}^{B-1}{\sqrt{{{a}_{i}}}\left| i \right\rangle \left( \left| \mathbf{u} \right\rangle +{{\mathbf{K}}_{i}}{{\mathbf{V}}_{j}}{{\left| 0 \right\rangle }^{\otimes n}} \right)}}+\left| 1 \right\rangle \sum\limits_{j=0}^{D-1}{\left| j \right\rangle \sum\limits_{i=0}^{B-1}{\sqrt{{{a}_{i}}}\left| i \right\rangle \left( \left| \mathbf{u} \right\rangle -{{\mathbf{K}}_{i}}{{\mathbf{V}}_{j}}{{\left| 0 \right\rangle }^{\otimes n}} \right)}} \right).\label{eq:70}\end{equation}

Measure the first ancilla and the state-index register. Denote the joint probabilities of $|0\rangle|j\rangle$ and $|1\rangle|j\rangle$ by $\zeta_{0j}$ and $\zeta_{1j}$, respectively. They are

\begin{equation}{{\zeta }_{0j}}=\frac{1}{2D}\left[ 1+\frac{\operatorname{Re}\left( \sum\limits_{i=0}^{B-1}{{{a}_{i}}{{\left\langle 0 \right|}^{\otimes n}}{{\mathbf{V}}^{\mathrm{H}}}{{\mathbf{K}}_{i}}{{\mathbf{V}}_{j}}{{\left| 0 \right\rangle }^{\otimes n}}} \right)}{\sum\limits_{i=0}^{B-1}{{{a}_{i}}}} \right],\ \ \ {{\zeta }_{1j}}=\frac{1}{2D}\left[ 1-\frac{\operatorname{Re}\left( \sum\limits_{i=0}^{B-1}{{{a}_{i}}{{\left\langle 0 \right|}^{\otimes n}}{{\mathbf{V}}^{\mathrm{H}}}{{\mathbf{K}}_{i}}{{\mathbf{V}}_{j}}{{\left| 0 \right\rangle }^{\otimes n}}} \right)}{\sum\limits_{i=0}^{B-1}{{{a}_{i}}}} \right].\label{eq:71}\end{equation}

Therefore,

\begin{align}& \operatorname{Re}\left( \sum\limits_{i=0}^{B-1}{{{a}_{i}}{{\left\langle 0 \right|}^{\otimes n}}{{\mathbf{V}}^{\mathrm{H}}}{{\mathbf{K}}_{i}}{{\mathbf{V}}_{j}}{{\left| 0 \right\rangle }^{\otimes n}}} \right)=\left( 1-2D{{\zeta }_{1j}} \right)\sum\limits_{i=0}^{B-1}{{{a}_{i}}}=\left( 2D{{\zeta }_{0j}}-1 \right)\sum\limits_{i=0}^{B-1}{{{a}_{i}}} \notag \\ & \ \ \ \ \ \ \ \ \ \ \ \ \ \ \ \ \ \ \ \ \ \ \ \ \ \ \ \ \ \ \ \ \ \ \ \ \ \ \ \ \ =D\left( {{\zeta }_{0j}}-{{\zeta }_{1j}} \right)\sum\limits_{i=0}^{B-1}{{{a}_{i}}},\ \ \ \ 0\le j\le D-1.\label{eq:72}\end{align}

The same circuit structure estimates all $\langle0|^{\otimes n}\mathbf{V}^{\mathrm H}\mathbf{K}\mathbf{V}_j|0\rangle^{\otimes n}$. If $B=M$ and $D=N_\alpha$, one circuit configuration suffices for all of them.

\paragraph{(b)
Batched evaluation of $\langle0|^{\otimes n}\mathbf{V}_j^{\mathrm H}\mathbf{K}\mathbf{V}_j|0\rangle^{\otimes n}$}

For the required quadratic terms, decompose $\mathbf{K}$ into its LCU terms. Then

\begin{align}& {{\left\langle 0 \right|}^{\otimes n}}\mathbf{V}_{j}^{\mathrm{H}}\mathbf{K}{{\mathbf{V}}_{j}}{{\left| 0 \right\rangle }^{\otimes n}}=\sum\limits_{i=0}^{M-1}{{{a}_{i}}{{\left\langle 0 \right|}^{\otimes n}}\mathbf{V}_{j}^{\mathrm{H}}{{\mathbf{K}}_{i}}{{\mathbf{V}}_{j}}{{\left| 0 \right\rangle }^{\otimes n}}} \notag \\ & \ \ \ \ \ \ \ \ \ \ \ \ \ \ \ \ \ \ \ \ \ \ \ \ \ \ =\sum\limits_{i=0}^{B-1}{{{a}_{i}}{{\left\langle 0 \right|}^{\otimes n}}\mathbf{V}_{j}^{\mathrm{H}}{{\mathbf{K}}_{i}}{{\mathbf{V}}_{j}}{{\left| 0 \right\rangle }^{\otimes n}}}+\sum\limits_{i=B}^{2B-1}{{{a}_{i}}{{\left\langle 0 \right|}^{\otimes n}}\mathbf{V}_{j}^{\mathrm{H}}{{\mathbf{K}}_{i}}{{\mathbf{V}}_{j}}{{\left| 0 \right\rangle }^{\otimes n}}} \notag \\ & \ \ \ \ \ \ \ \ \ \ \ \ \ \ \ \ \ \ \ \ \ \ \ \ \ \ +\cdots +\sum\limits_{i=M-B}^{M-1}{{{a}_{i}}{{\left\langle 0 \right|}^{\otimes n}}\mathbf{V}_{j}^{\mathrm{H}}{{\mathbf{K}}_{i}}{{\mathbf{V}}_{j}}{{\left| 0 \right\rangle }^{\otimes n}}}.\label{eq:73}\end{align}

Again, partition $j\in\{0,1,\ldots,N_\alpha-1\}$ into $N_\alpha/D$ groups of $D=2^d$ indices and consider the first group. A circuit with the structure of Fig.~\ref{fig:10} evaluates its terms. Replace state preparation $\mathbf{V}(\boldsymbol{\alpha})$ in the first branch by $\mathbf{V}_j$. The two branches then prepare $\mathbf{V}_j|0\rangle^{\otimes n}$ and $\mathbf{K}_i\mathbf{V}_j|0\rangle^{\otimes n}$. The final state is

\begin{align}& \left| \psi \right\rangle =\frac{1}{{\left( \sqrt{2} \right)}^{d+2}\sqrt{\sum\limits_{i=0}^{B-1}{{{a}_{i}}}}}\left[ \left| 0 \right\rangle \sum\limits_{j=0}^{D-1}{\left| j \right\rangle \sum\limits_{i=0}^{B-1}{\sqrt{{{a}_{i}}}\left| i \right\rangle \left( {{\mathbf{V}}_{j}}{{\left| 0 \right\rangle }^{\otimes n}}+{{\mathbf{K}}_{i}}{{\mathbf{V}}_{j}}{{\left| 0 \right\rangle }^{\otimes n}} \right)}} \right. \notag \\ & \ \ \ \ \ \ \ \left. +\left| 1 \right\rangle \sum\limits_{j=0}^{D-1}{\left| j \right\rangle \sum\limits_{i=0}^{B-1}{\sqrt{{{a}}_{i}}\left| i \right\rangle \left( {{\mathbf{V}}_{j}}{{\left| 0 \right\rangle }^{\otimes n}}-{{\mathbf{K}}_{i}}{{\mathbf{V}}_{j}}{{\left| 0 \right\rangle }^{\otimes n}} \right)}} \right].\label{eq:74}\end{align}

The joint probabilities $\zeta_{0j}$ and $\zeta_{1j}$ for $|0\rangle|j\rangle$ and $|1\rangle|j\rangle$ have the form of Eq.~\eqref{eq:71}, with $\mathbf{V}^{\mathrm H}$ replaced by $\mathbf{V}_j^{\mathrm H}$. Accordingly, Eq.~\eqref{eq:72} becomes $\operatorname{Re}\sum_{i=0}^{B-1}a_i\langle0|^{\otimes n}\mathbf{V}_j^{\mathrm H}\mathbf{K}_i\mathbf{V}_j|0\rangle^{\otimes n}=D(\zeta_{0j}-\zeta_{1j})\sum_{i=0}^{B-1}a_i$, for $0\le j\le D-1$. If $B=M$ and $D=N_\alpha$, one circuit configuration suffices for all such quadratic terms.

\paragraph{(c)
Batched evaluation of $\langle0|^{\otimes n}\mathbf{F}^{\mathrm H}\mathbf{V}_j|0\rangle^{\otimes n}$}

We need an overlap of this form for $\mathbf{V}_0,\mathbf{V}_1,\ldots,\mathbf{V}_{N_\alpha-1}$. No stiffness matrix is involved, so a unitary-term index register is unnecessary. Partition $j\in\{0,1,\ldots,N_\alpha-1\}$ into $N_\alpha/D$ groups of $D=2^d$ indices. For the first group,

\begin{equation}{{\left\langle 0 \right|}^{\otimes n}}{{\mathbf{F}}^{\mathrm{H}}}{{\mathbf{V}}_{0}}{{\left| 0 \right\rangle }^{\otimes n}},\ \ {{\left\langle 0 \right|}^{\otimes n}}{{\mathbf{F}}^{\mathrm{H}}}{{\mathbf{V}}_{1}}{{\left| 0 \right\rangle }^{\otimes n}},\ \ \cdots ,\ \ {{\left\langle 0 \right|}^{\otimes n}}{{\mathbf{F}}^{\mathrm{H}}}{{\mathbf{V}}_{D-1}}{{\left| 0 \right\rangle }^{\otimes n}}.\label{eq:75}\end{equation}

\begin{figure*}[htbp]
\centering
\includegraphics[width=10.62cm]{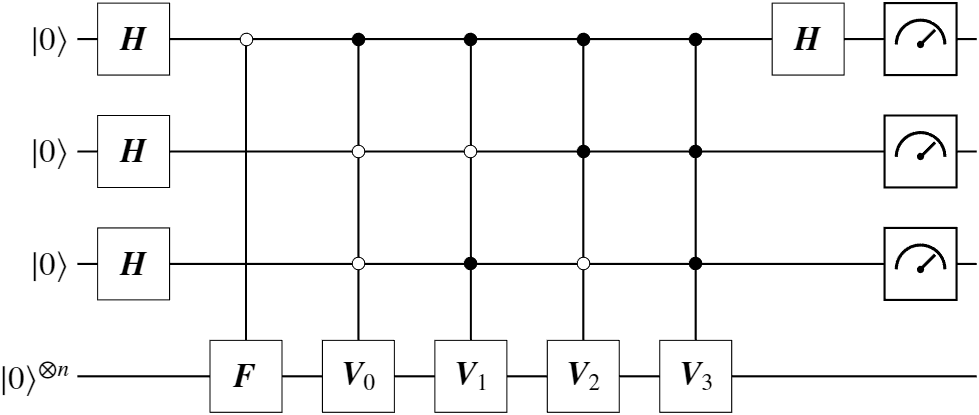}
\caption{Block-Hadamard circuit for batched estimation of $\operatorname{Re}\langle0|^{\otimes n}\mathbf{F}^{\mathrm H}\mathbf{V}_s|0\rangle^{\otimes n}$ ($D_f=4$)}\label{fig:11}
\end{figure*}

The terms in Eq.~\eqref{eq:75} can be estimated by the circuit in Fig.~\ref{fig:11}, shown for $D=4$.

The input is $|\psi\rangle=|0\rangle|0\rangle^{\otimes d}|0\rangle^{\otimes n}$. After the circuit layers, the final state is

\begin{equation}\left| \psi \right\rangle =\frac{1}{{\left( \sqrt{2} \right)}^{d+2}}\left( \left| 0 \right\rangle \sum\limits_{j=0}^{D-1}{\left| j \right\rangle }\left( \mathbf{F}+{{\mathbf{V}}_{j}} \right){{\left| 0 \right\rangle }^{\otimes n}}+\left| 1 \right\rangle \sum\limits_{j=0}^{D-1}{\left| j \right\rangle }\left( \mathbf{F}-{{\mathbf{V}}_{j}} \right){{\left| 0 \right\rangle }^{\otimes n}} \right).\label{eq:76}\end{equation}

Measure the first ancilla and the state-index register. Denote the joint probabilities of $|0\rangle|j\rangle$ and $|1\rangle|j\rangle$ by $\zeta_{0j}$ and $\zeta_{1j}$, respectively. They are

\begin{equation}{{\zeta }_{0j}}=\frac{1}{2D}\left[ 1+\operatorname{Re}\left( {{\left\langle 0 \right|}^{\otimes n}}{{\mathbf{F}}^{\mathrm{H}}}{{\mathbf{V}}_{j}}{{\left| 0 \right\rangle }^{\otimes n}} \right) \right],\ \ \ {{\zeta }_{1j}}=\frac{1}{2D}\left[ 1-\operatorname{Re}\left( {{\left\langle 0 \right|}^{\otimes n}}{{\mathbf{F}}^{\mathrm{H}}}{{\mathbf{V}}_{j}}{{\left| 0 \right\rangle }^{\otimes n}} \right) \right].\label{eq:77}\end{equation}

Hence,

\begin{align}& \operatorname{Re}\left( {{\left\langle 0 \right|}^{\otimes n}}{{\mathbf{F}}^{\mathrm{H}}}{{\mathbf{V}}_{j}}{{\left| 0 \right\rangle }^{\otimes n}} \right)=1-2D{{\zeta }_{1j}}=2D{{\zeta }_{0j}}-1 \notag \\ & \ \ \ \ \ \ \ \ \ \ \ \ \ \ \ \ \ \ \ \ \ \ \ \ \ \ \ \ \ \ =D\left( {{\zeta }_{0j}}-{{\zeta }_{1j}} \right),\ \ \ \ 0\le j\le D-1.\label{eq:78}\end{align}

The same circuit structure estimates all $\langle0|^{\otimes n}\mathbf{F}^{\mathrm H}\mathbf{V}_j|0\rangle^{\otimes n}$. If $D=N_\alpha$, one circuit configuration suffices, so that the total potential energy and all its parameter derivatives can be estimated in batches with a constant number of circuit configurations. Define

\begin{equation}A_s^{(K)}=\operatorname{Re}\left( {{\left\langle 0 \right|}^{\otimes n}}{{\mathbf{V}}^{\mathrm{H}}}\mathbf{K}{{\mathbf{V}}_{s}}{{\left| 0 \right\rangle }^{\otimes n}} \right),\ \ \ L_s^{(f)}=\operatorname{Re}\left( {{\left\langle 0 \right|}^{\otimes n}}{{\mathbf{F}}^{\mathrm{H}}}{{\mathbf{V}}_{s}}{{\left| 0 \right\rangle }^{\otimes n}} \right),\ \ \ 0\le s\le {{N}_{\alpha }}.\label{eq:79}\end{equation}
Here, $A_s^{(K)}$ is recovered from the batched result in Eq.~\eqref{eq:72}. For the finite-difference method in Eq.~\eqref{eq:50}, use the corresponding diagonal term in Eq.~\eqref{eq:73}. The quantity $L_s^{(f)}$ is recovered from Eq.~\eqref{eq:78}, and the potential-energy gradient is then reconstructed as

\begin{equation}\frac{\partial\Pi}{\partial c}=cA_0^{(K)}-\|\mathbf{f}\|_2L_0^{(f)},\qquad \frac{\partial\Pi}{\partial\alpha_j}=\frac{1}{2}\left[c^2A_j^{(K)}-c\|\mathbf{f}\|_2L_j^{(f)}\right],\quad 1\le j\le N_{\alpha}.\label{eq:80}\end{equation}

This reduction concerns the number of distinct circuit configurations: each configuration still requires repeated sampling, and the joint probability for an individual state index decreases as $1/D$. The shots required for a specified statistical precision therefore grow with the index-space size, variance, and target confidence level.

\textbf{Remark 3}
The batched method relies on quantum multiplexors implementing $\bigoplus_{j=0}^{D-1}\mathbf{V}_j$ and $\bigoplus_{i=0}^{B-1}\mathbf{K}_i$. Their realization is nontrivial and generally uses controlled gates and uniformly controlled rotation decompositions \cite{Mottonen2005}. Noise in current hardware may limit their implementation. Practical deployment of the algorithm thus depends on hardware progress. Nevertheless, the LCU decomposition in Section 3.1 makes the $\mathbf{K}_i$ multiplexor primarily a combination of cyclic permutations, $R_y$ gates, and diagonal unitaries. If the ansatz in Fig.~\ref{fig:1} is used, the $\mathbf{V}_j$ multiplexor also mainly comprises $R_y$ gates. Remark 2 and Appendix 2 give the assumptions under which these gates can be implemented in $O(\mathrm{poly}(\log N))$ time.

\textbf{Remark 4}
The block-Hadamard test presented here batches overlaps of the form $\langle\mathbf{u}|\mathbf{K}_i|\mathbf{u}_j\rangle$, and the same idea also applies to $\langle\mathbf{u}|\mathbf{K}_i^{\mathrm H}\mathbf{K}_j|\mathbf{u}\rangle$ in conventional loss functions. Appendix 3 gives the corresponding quantum circuit.
\section{Numerical examples}

This section presents three proof-of-concept examples: a two-dimensional truss, a plane-stress plate, and a three-dimensional steady-state heat-conduction problem. The first two test static structural problems, and the third tests a three-dimensional system with one degree of freedom per node. Together, they assess the voxel-grid LCU decomposition and minimum-potential-energy formulation from Section 3. All results were obtained with classical numerical programs in MATLAB.

Let $\mathbf{u}_{\mathrm{ref}}$ denote the displacement or nodal-temperature vector from classical finite-element analysis, and let $\mathbf{u}$ denote the corresponding field reconstructed from the ansatz. The relative error is
\begin{equation}
\varepsilon=\frac{\|\mathbf{u}-\mathbf{u}_{\mathrm{ref}}\|_1}{\|\mathbf{u}_{\mathrm{ref}}\|_1}.
\label{eq:81}
\end{equation}

Voxel-BVQLS uses the minimum-potential-energy objective in Eq.~\eqref{eq:44} and the voxel-grid LCU decomposition. For linear static problems, this objective is the total potential energy. The comparison methods VQLS-G, VQLS-L, and VQLS-R use the global, local, and residual losses from Section 2.3.2, respectively. Pauli-VQLS denotes the comparison method based on a Pauli decomposition.

\textbf{Remark 5}
The proposed block-Hadamard circuits estimate the total potential energy and its derivatives in batches. Under ideal noiseless conditions, different circuits estimating the same physical quantity yield the same result. The present examples therefore do not quantify the computational-efficiency gain of block-Hadamard tests over conventional Hadamard tests. We provide only a qualitative assessment. As shown in Section 4, block-Hadamard tests reduce the number of circuit configurations needed for the total potential energy and its derivatives. If the block size equals the number of unitary terms, two types of circuit configuration are sufficient in principle.

\textbf{Remark 6}
The advantage of the proposed voxel-grid LCU decomposition is measured by its number of unitary terms. In each example, we decompose the stiffness matrix both by this method and by a Pauli decomposition, then compare the numbers of nonzero unitary terms.

\textbf{Remark 7}
The minimum-potential-energy objective in Eq.~\eqref{eq:44} is strictly convex over the full physical-field space, but its composition with a parameterized ansatz is generally nonconvex. The following comparisons describe numerical behavior for the specified ansatz, initial values, and stopping criteria and do not provide a general convergence guarantee for any objective or optimizer.

\subsection{Truss problem}

\begin{figure}[htbp]
\centering
\includegraphics[width=6.65cm]{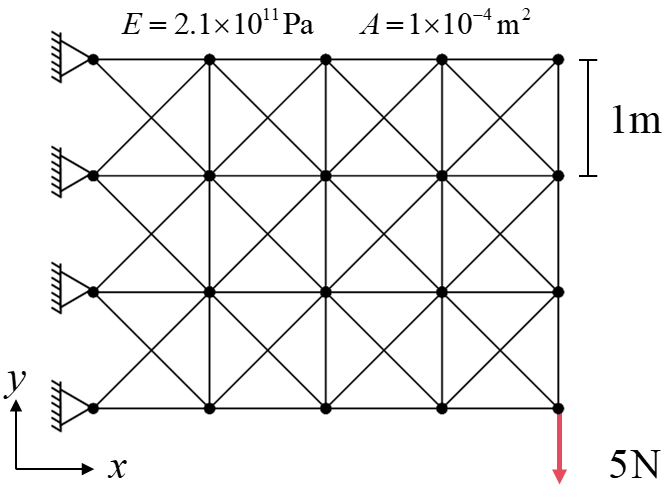}
\caption{Geometry, boundary conditions, and applied load of the two-dimensional truss}\label{fig:12}
\end{figure}

\begin{figure*}[htbp]
\centering
\includegraphics[width=14.60cm]{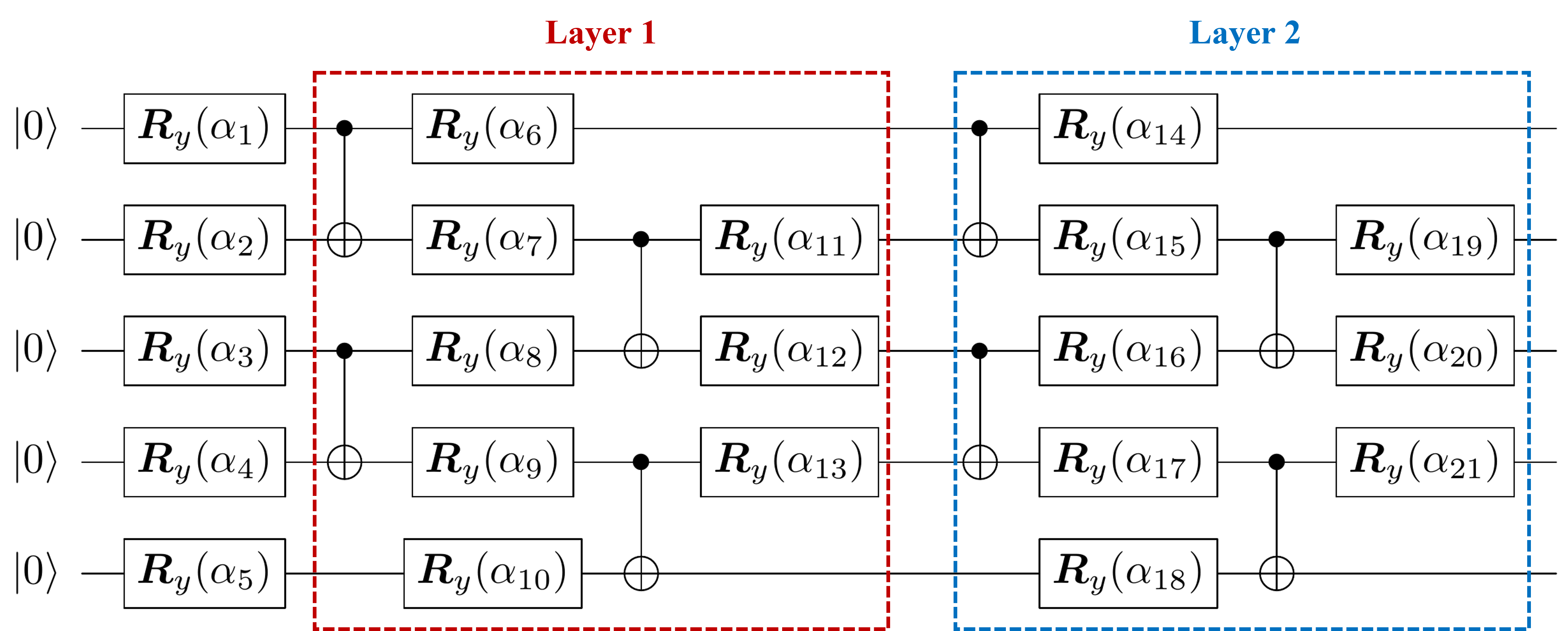}
\caption{Ansatz circuit for the two-dimensional truss example}\label{fig:13}
\end{figure*}

\begin{figure*}[htbp]
\centering
\includegraphics[width=14.66cm]{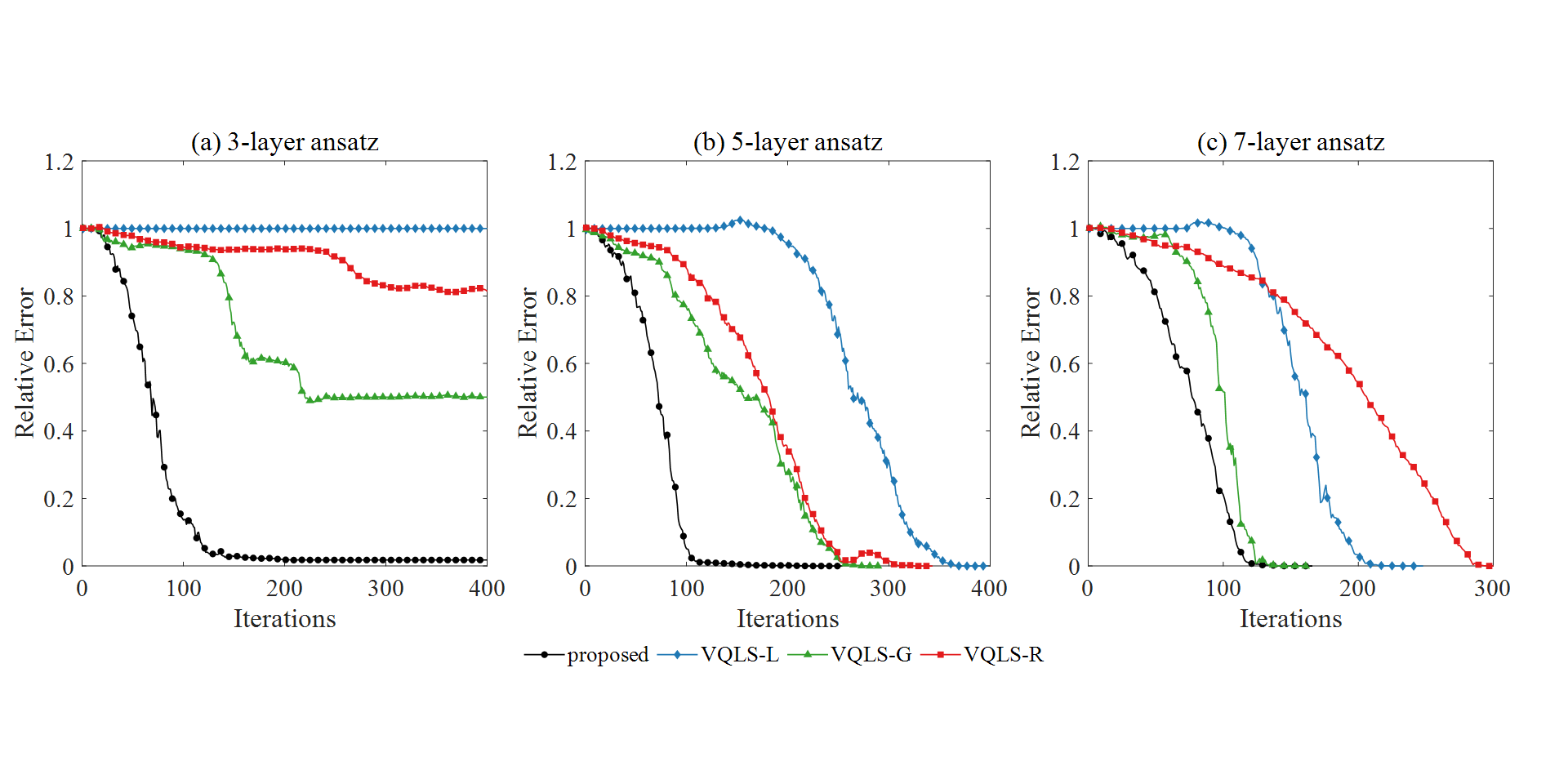}
\caption{Displacement $L_1$ relative-error curves for four methods with 3-, 5-, and 7-layer ansatz circuits in the two-dimensional truss example}\label{fig:14}
\end{figure*}

The first example is the two-dimensional truss in Fig.~\ref{fig:12}. It has 20 nodes, two degrees of freedom per node, and a nodal spacing of $1\,\mathrm{m}$. The bars have cross-sectional area $A=10^{-4}\,\mathrm{m}^{2}$ and Young's modulus $E=2.1\times10^{11}\,\mathrm{Pa}$. The nodes on the left edge are fixed. A concentrated force of $5\,\mathrm{N}$ acts downward at the lower-right node. After imposing the constraints, $N=32$ degrees of freedom remain. This is a two-dimensional problem with two degrees of freedom per node. Table~\ref{tab:3} shows 51 nonzero unitary terms from the Pauli decomposition and 18 from the proposed voxel-grid LCU decomposition, a reduction of about 65\%. We solve the problem with Voxel-BVQLS, VQLS-G, VQLS-L, and VQLS-R. All four methods use the ansatz in Fig.~\ref{fig:13}, with every initial parameter set to 1 and with 3, 5, or 7 layers. Sequential quadratic programming (SQP) is used with at most 500 iterations. The allowed difference between successive parameter iterates (StepTolerance) is $10^{-10}$. Figure~\ref{fig:14} plots relative error against iteration count for the four methods and three ansatz depths.

\begin{table}[htbp]
\centering
\caption{Nonzero unitary terms in Pauli and voxel-grid decompositions of the truss stiffness matrix}\label{tab:3}
\begin{tabular}{p{0.36\linewidth}p{0.19\linewidth}p{0.19\linewidth}}
\toprule
Decomposition & Pauli LCU & Voxel-grid LCU \\
\midrule
Number of nonzero unitary terms & 51 & 18 \\
\bottomrule
\end{tabular}
\end{table}

\FloatBarrier

Figure~\ref{fig:14} shows that, with the same initial values and optimization settings, Voxel-BVQLS reduces the relative error rapidly and stabilizes after about 100--150 iterations at all three ansatz depths. With five layers, it reaches low error after about 110 iterations, whereas VQLS-G, VQLS-R, and VQLS-L need about 250, 320, and 370 iterations, respectively, to approach zero error. With three layers, none of those three comparison methods finds a low-error solution within the plotted range. With seven layers, VQLS-G also reduces the error relatively quickly, while VQLS-L and VQLS-R require more iterations. Under the ansatz, initialization, and SQP settings of this example, these results indicate that the minimum-potential-energy objective improves optimization and alleviates the high-error stagnation associated here with barren plateaus at shallow depth.

\begin{figure}[H]
\centering
\includegraphics[width=13.50cm]{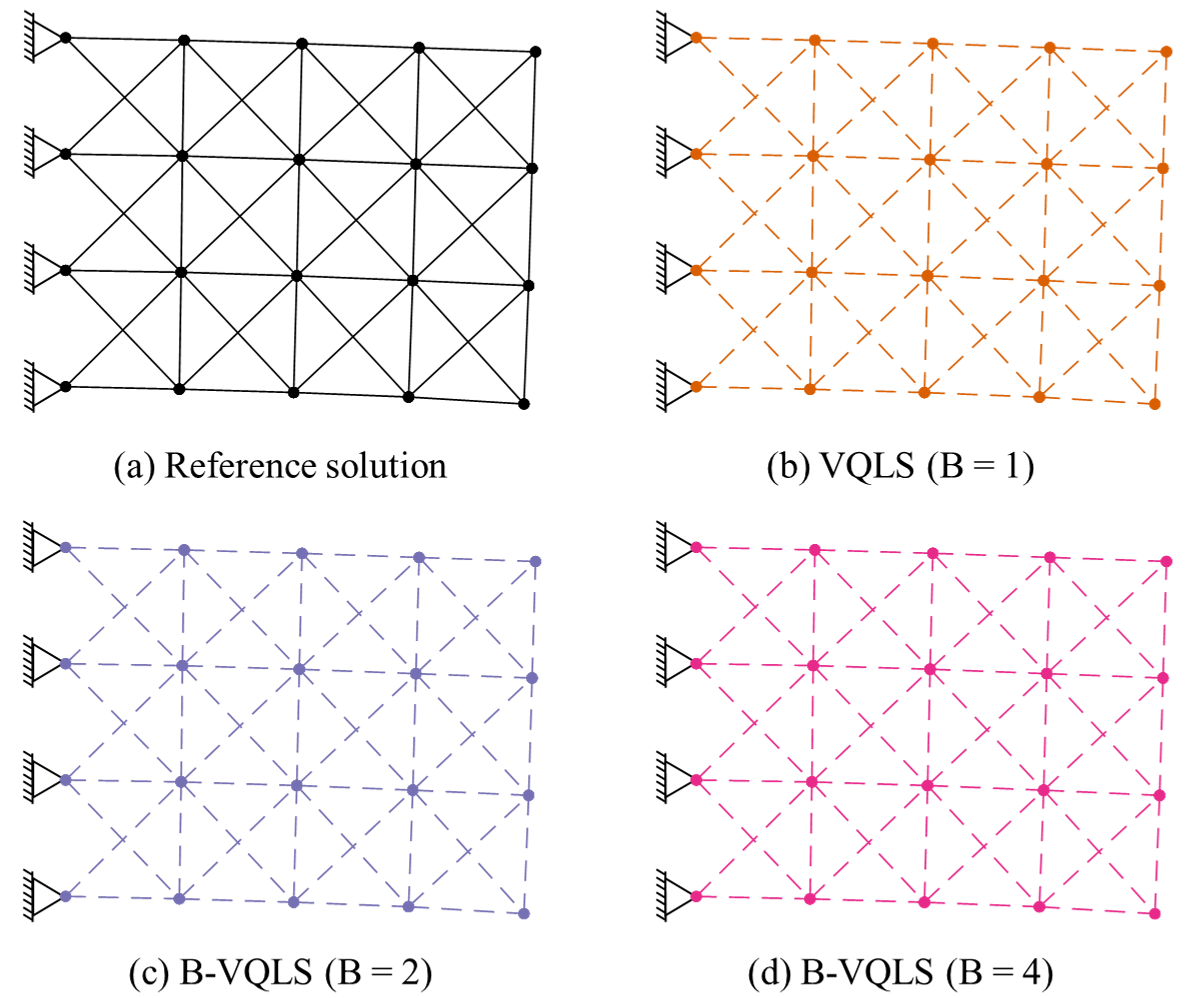}
\caption{Reference and computed deformed truss configurations for block sizes $B=1$, $2$, and $4$}\label{fig:truss-displacement}
\end{figure}

\FloatBarrier

Figure~\ref{fig:truss-displacement} compares the truss-node positions from the classical finite element reference, the unblocked scheme ($B=1$), and two block-Hadamard schemes ($B=2,4$). All three schemes reproduce the overall deformation under load. At the plotted scale, the loaded and interior nodes are close to the reference positions. The $B=1$, $B=2$, and $B=4$ results coincide with the finite element reference in the figure. Thus, changing the unitary-term block size does not visibly degrade nodal-displacement accuracy in this noiseless example. Blocking changes how the overlaps are organized and estimated. With exact summation, the stiffness matrix and variational objective are unchanged. The block-Hadamard test therefore combines several term-wise overlap tests into fewer circuit configurations while retaining the plotted nodal-displacement accuracy in this example.
\FloatBarrier

\subsection{Plane-stress problem}

\begin{figure*}[htbp]
\centering
\includegraphics[width=10.00cm]{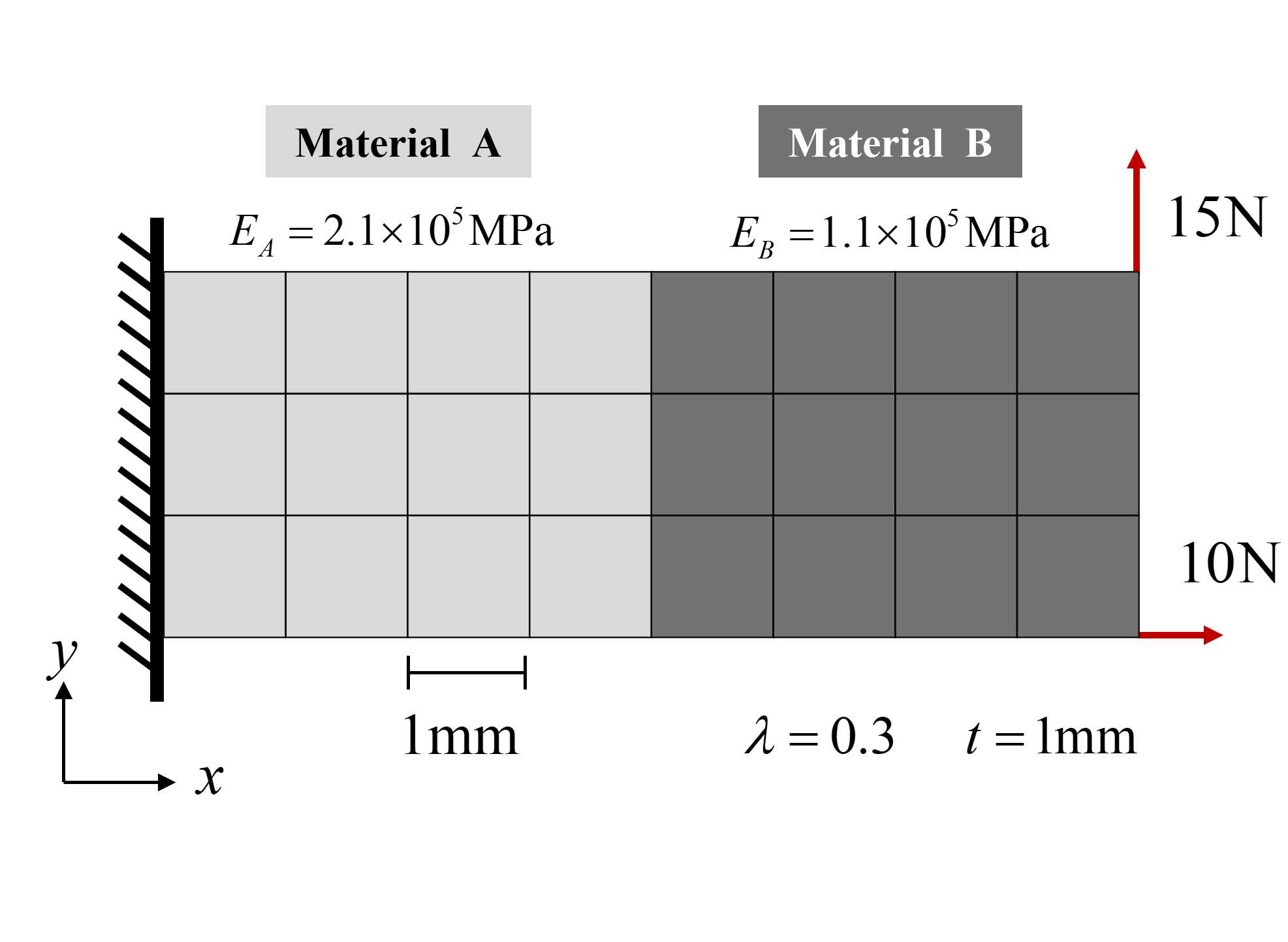}
\caption{Geometry, material layout, boundary conditions, and applied loads of the plane-stress model}\label{fig:15}
\end{figure*}

\begin{figure*}[htbp]
\centering
\includegraphics[width=14.00cm]{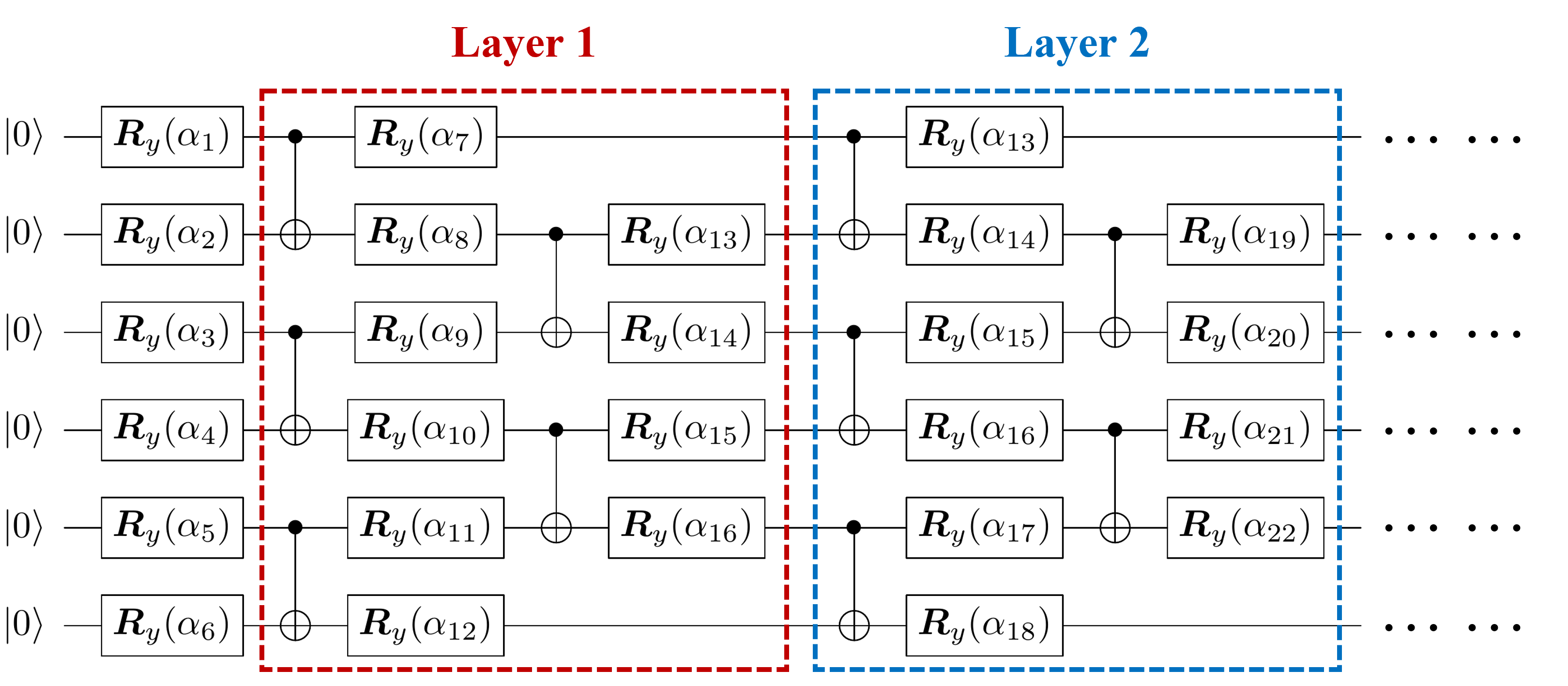}
\caption{Ansatz circuit for the plane-stress example}\label{fig:16}
\end{figure*}

\begin{figure*}[htbp]
\centering
\includegraphics[width=16.00cm]{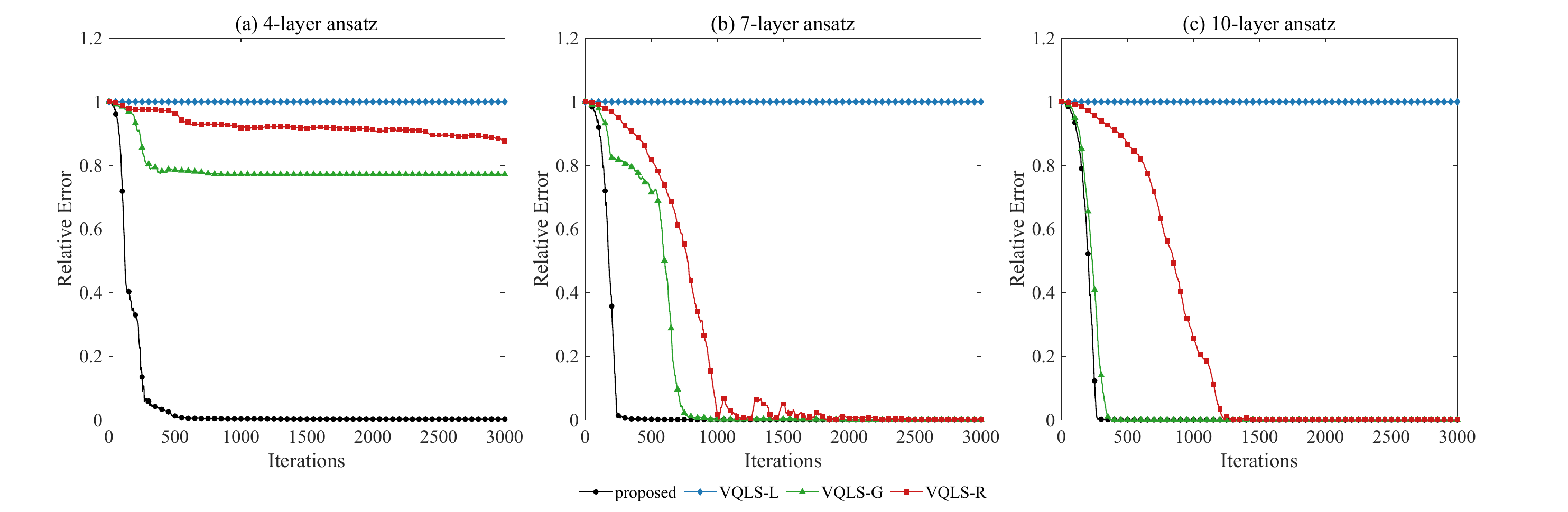}
\caption{Displacement $L_1$ relative-error curves for four methods with 4-, 7-, and 10-layer ansatz circuits in the plane-stress example}\label{fig:17}
\end{figure*}

\begin{figure*}[htbp]
\centering
\includegraphics[width=14.66cm]{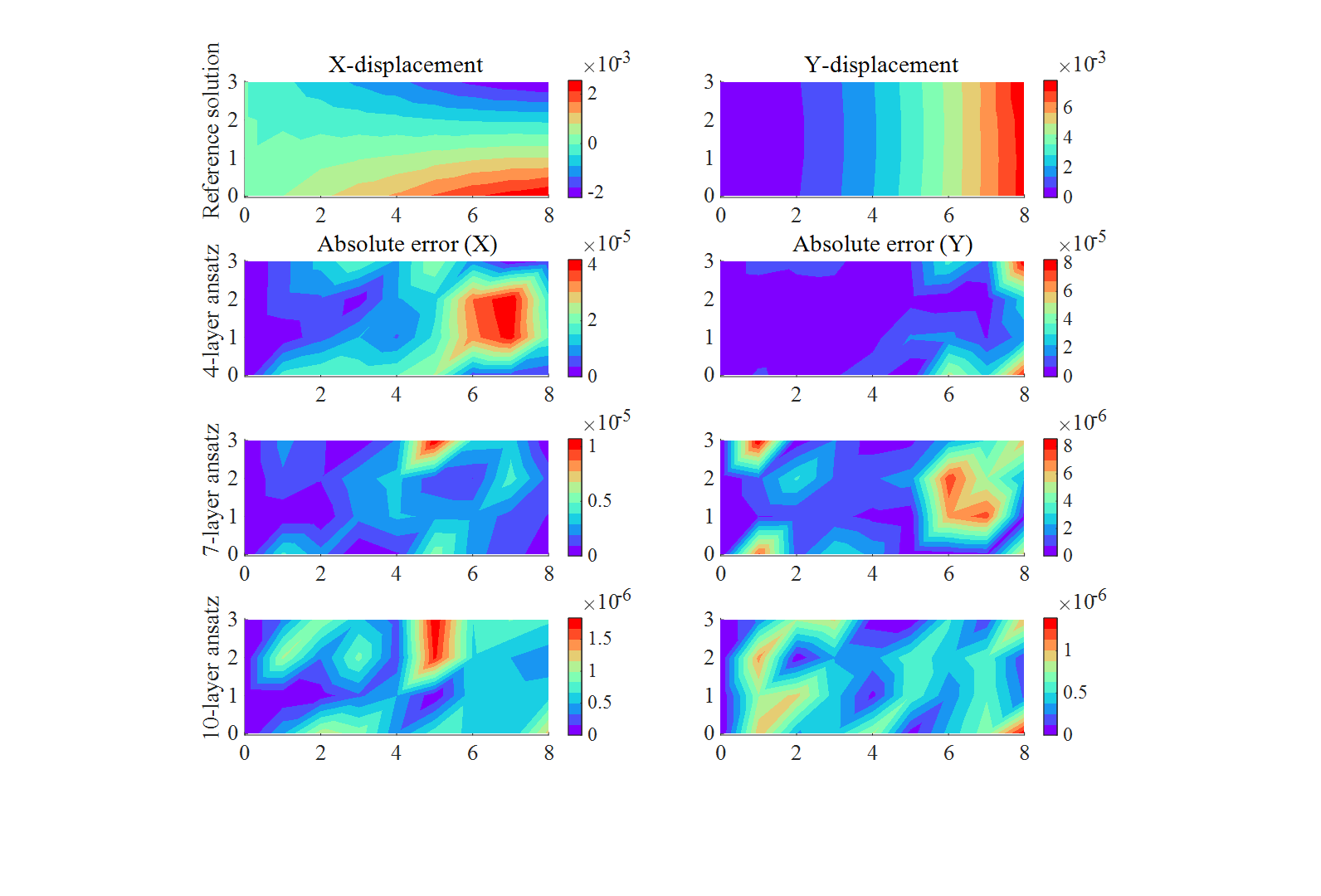}
\caption{Reference $x$- and $y$-displacement fields and absolute errors for 4-, 7-, and 10-layer ansatz circuits in the plane-stress example}\label{fig:18}
\end{figure*}

The second example is the plane-stress plate in Fig.~\ref{fig:15}. It has 36 nodes with two degrees of freedom per node. Its length, width, and thickness are $7\,\mathrm{mm}$, $3\,\mathrm{mm}$, and $1\,\mathrm{mm}$, respectively. Materials A and B have Young's moduli of $2.1\times10^5\,\mathrm{MPa}$ and $1.1\times10^5\,\mathrm{MPa}$, respectively, and both have a Poisson's ratio of $0.3$. The left-edge nodes are fixed. A $10\,\mathrm{N}$ force acts in the positive $x$ direction at the lower-right node, and a $15\,\mathrm{N}$ force acts in the positive $y$ direction at the upper-right node. After imposing the constraints, $N=64$ degrees of freedom remain. This is a two-dimensional problem with two degrees of freedom per node. Table~\ref{tab:5} lists 224 nonzero unitary terms from the Pauli decomposition and 18 from the proposed voxel-grid LCU decomposition, a reduction of about 92\%. The four methods use the ansatz in Fig.~\ref{fig:16}, with every initial parameter set to 1 and depths of 4, 7, and 10 layers. SQP is run for at most 3000 iterations. To make every curve in Fig.~\ref{fig:17} cover the same iteration range, the early-stopping tolerances for function value, step size, and first-order optimality are all set to zero. Figure~\ref{fig:17} compares relative-error curves across methods and depths. Figure~\ref{fig:18} shows spatial displacement-field errors for Voxel-BVQLS at the same depths.

\begin{table}[htbp]
\centering
\caption{Nonzero unitary terms in Pauli and voxel-grid decompositions of the plane-stress stiffness matrix}\label{tab:5}
\begin{tabular}{p{0.36\linewidth}p{0.19\linewidth}p{0.19\linewidth}}
\toprule
Decomposition & Pauli LCU & Voxel-grid LCU \\
\midrule
Number of nonzero unitary terms & 224 & 18 \\
\bottomrule
\end{tabular}
\end{table}

\FloatBarrier

Figure~\ref{fig:17} shows that, with the same initial values and a 3000-iteration budget, Voxel-BVQLS reaches a low displacement-error range earlier at all three depths. With four layers, its final error is about $2.15\times10^{-3}$. None of the other methods reaches $10^{-2}$ in the plotted range. With seven layers, VQLS-G and VQLS-R reach that threshold after about 780 and 1000 iterations, respectively, later than Voxel-BVQLS. The VQLS-L error remains near 1. With ten layers, VQLS-G reaches $10^{-2}$ after about 350 iterations, whereas VQLS-R needs about 1220. Under the ansatz, initialization, and SQP settings used here, the minimum-potential-energy objective reaches a 1\% displacement relative error earlier at each depth and alleviates the high-error stagnation seen at shallow depth.

The upper row of Fig.~\ref{fig:18} shows the classical finite element reference displacements in the $x$ and $y$ directions, with values on the order of $10^{-3}$. The other rows show absolute errors for Voxel-BVQLS with 4, 7, and 10 layers. At four layers, the upper color-scale limits for the $x$- and $y$-direction errors are about $4\times10^{-5}$ and $8\times10^{-5}$. The larger errors occur mainly on the right side of the plate. As depth increases, both maximum color-scale values decline. At ten layers, they are about $1.5\times10^{-6}$ and $10^{-6}$. Under the tested depths and optimization settings, the method reproduces the overall reference displacement distribution, and deeper ansatz circuits reduce local error. These are field approximations for the tested problem, not a general accuracy guarantee.

\FloatBarrier

\subsection{Three-dimensional heat-conduction problem}

\begin{figure*}[p]
\centering
\includegraphics[width=10.5cm]{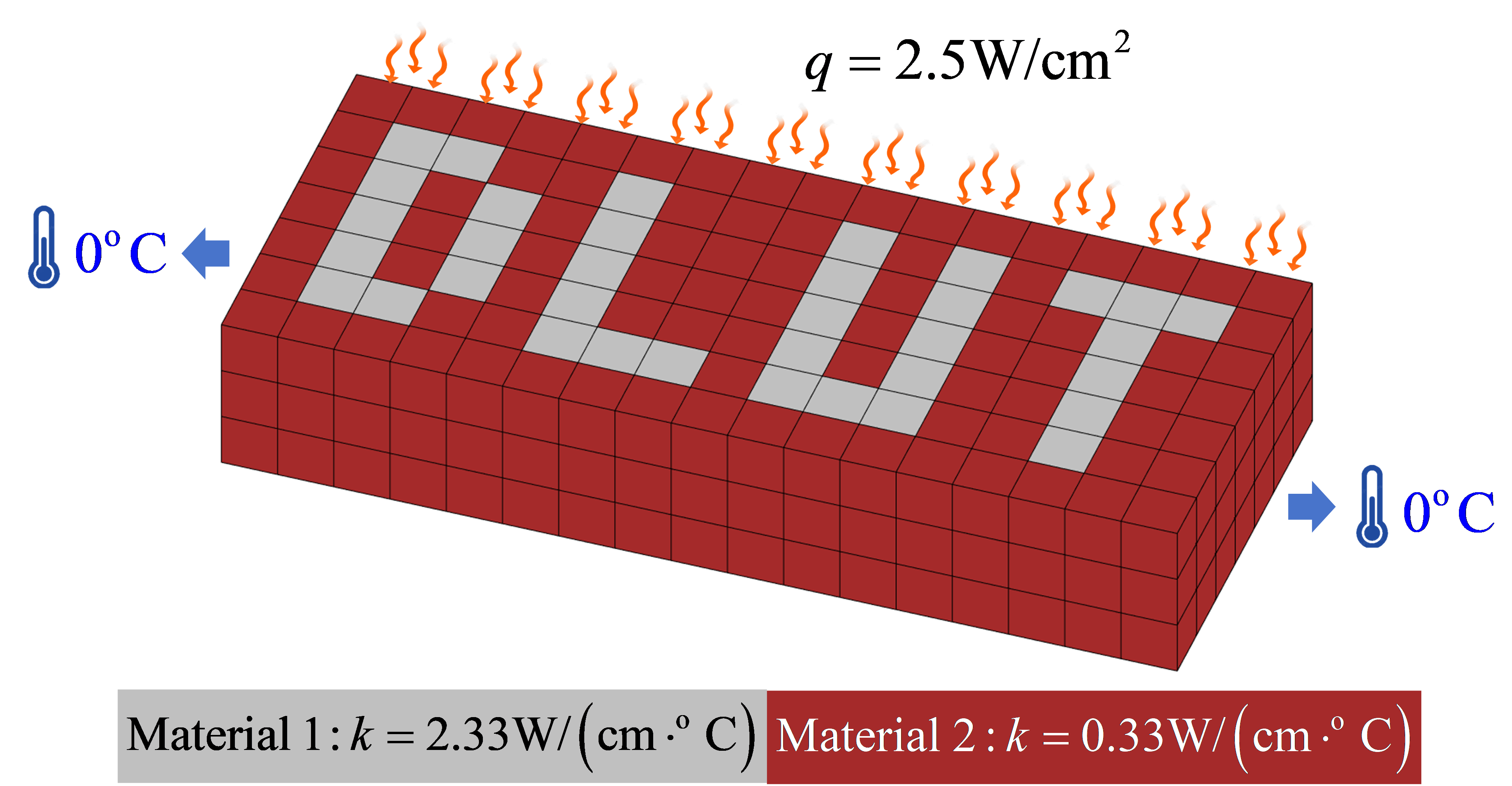}
\caption{Voxel geometry, material distribution, applied heat flux, and temperature boundary conditions of the three-dimensional heat-conduction model}\label{fig:19}
\vspace{0.2cm}
\includegraphics[width=12.0cm]{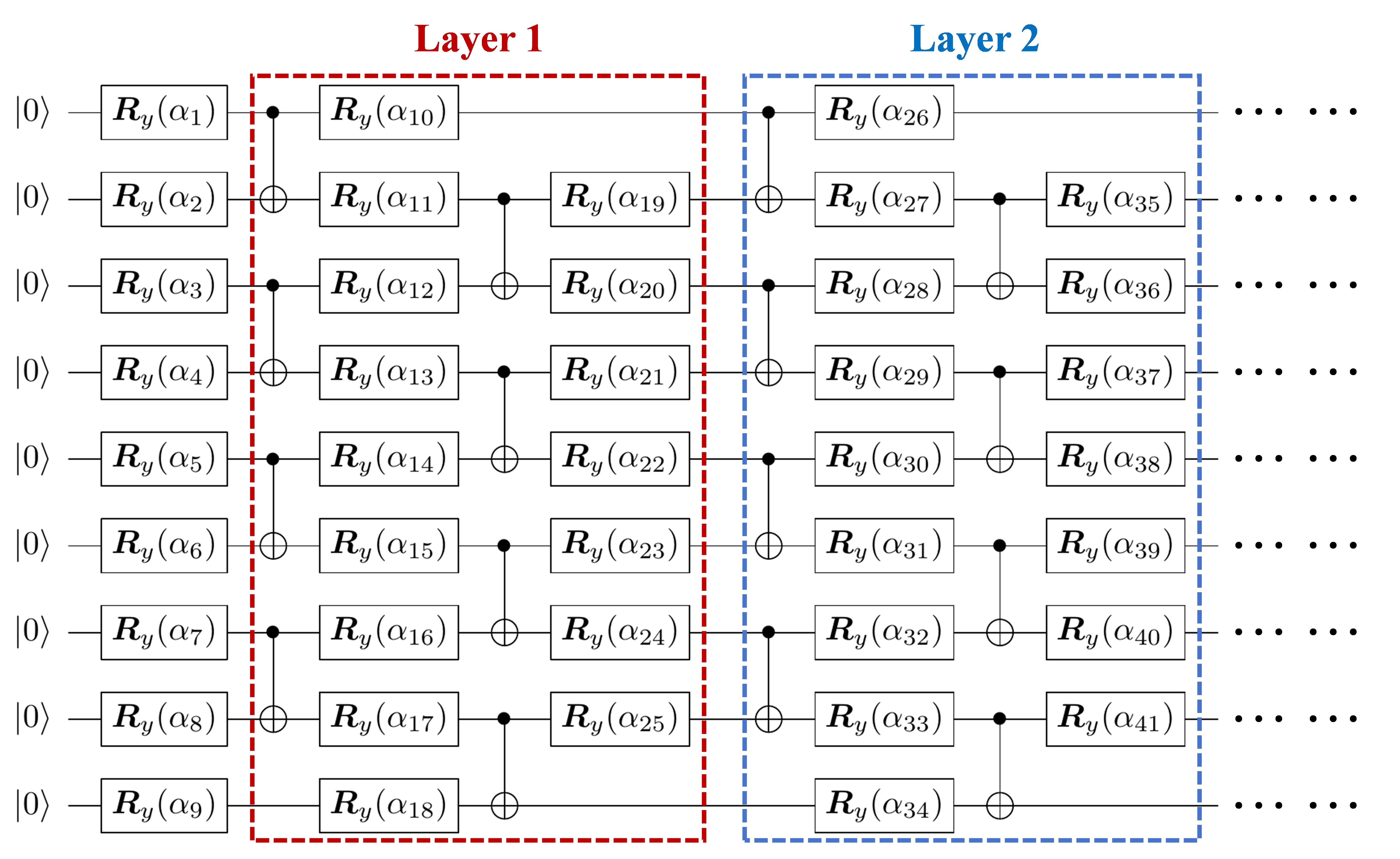}
\caption{Ansatz circuit for the three-dimensional heat-conduction example}\label{fig:20}
\vspace{0.2cm}
\includegraphics[width=12.5cm]{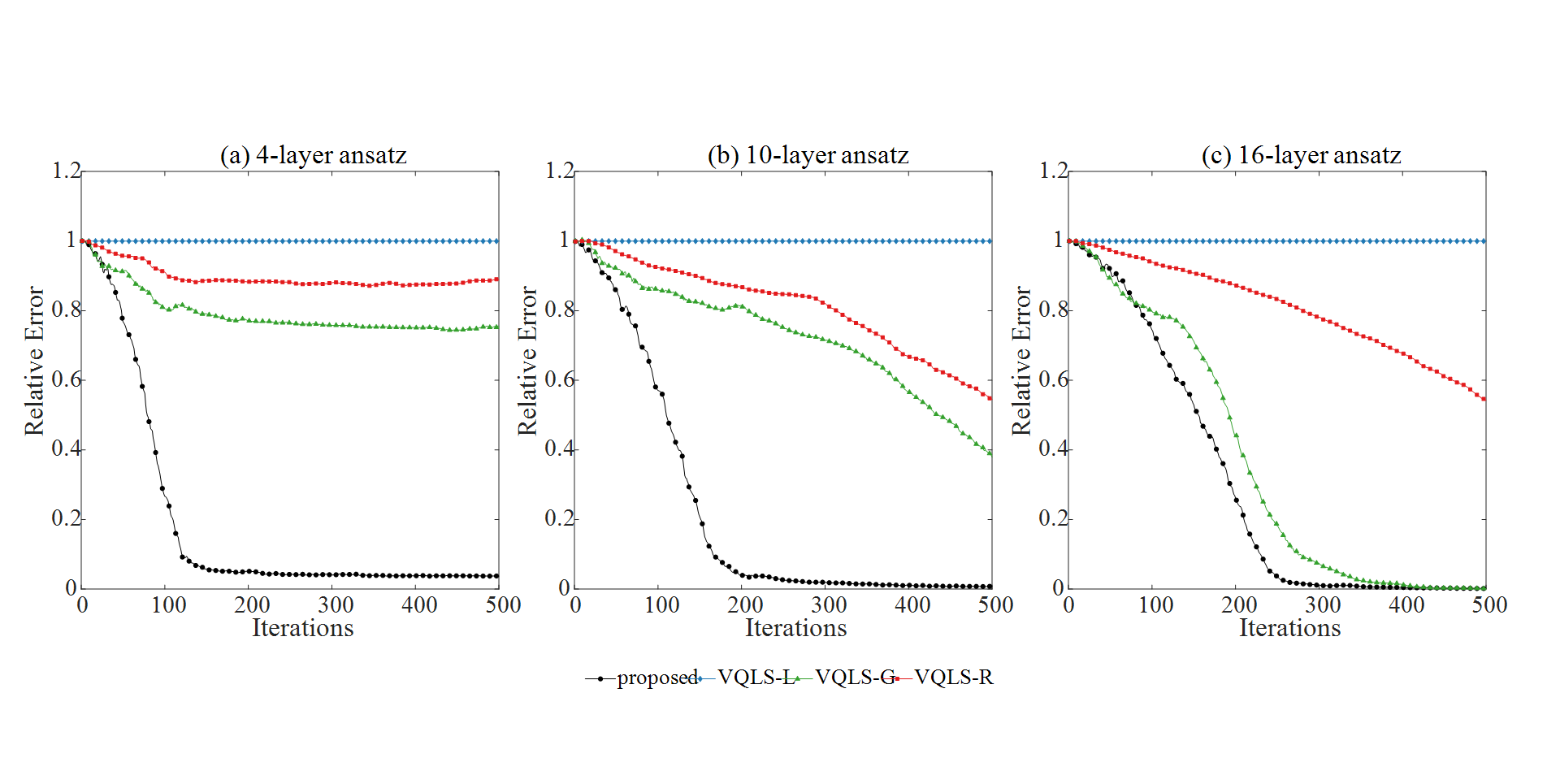}
\caption{Temperature $L_1$ relative-error curves for four methods with 4-, 10-, and 16-layer ansatz circuits in the three-dimensional heat-conduction example}\label{fig:21}
\end{figure*}

The third example is the three-dimensional steady-state heat-conduction model in Fig.~\ref{fig:19}. It contains $17\times7\times3$ eight-node voxel elements and $18\times8\times4=576$ nodes, each with one temperature degree of freedom. The two materials have thermal conductivities $k_1=2.33\,\mathrm{W}\,\mathrm{cm}^{-1}\,{}^{\circ}\mathrm{C}^{-1}$ and $k_2=0.33\,\mathrm{W}\,\mathrm{cm}^{-1}\,{}^{\circ}\mathrm{C}^{-1}$. A uniform heat flux of $q=2.5\,\mathrm{W}\,\mathrm{cm}^{-2}$ is applied to the top surface. The left and right end faces are held at $0\,^{\circ}\mathrm{C}$. After applying these boundary conditions, $N=512$ temperature degrees of freedom remain. This is a three-dimensional problem with one degree of freedom per node. Table~\ref{tab:6} gives 228 nonzero unitary terms for the Pauli decomposition and 18 for the voxel-grid LCU decomposition, a reduction of about 92\%. All four methods use the ansatz in Fig.~\ref{fig:20}, with initial parameters equal to 1 and depths of 4, 10, and 16 layers. SQP is run for at most 500 iterations with StepTolerance $10^{-10}$.

Figure~\ref{fig:21} shows that the Voxel-BVQLS relative error starts to decrease markedly earlier at all three depths. With four layers, it falls below $10^{-1}$ after about 120 iterations and stabilizes near $4\times10^{-2}$. With ten and sixteen layers, it reaches a lower level after about 200 and 250 iterations, respectively, and continues to decrease. By comparison, VQLS-L stays near an error of 1 at all depths, and VQLS-R finds no low-error solution within 500 iterations. VQLS-G retains a noticeable residual error at four and ten layers. At sixteen layers, it also reduces the error close to zero, but requires more iterations than Voxel-BVQLS. For this three-dimensional example and the stated initialization and SQP settings, the minimum-potential-energy objective alleviates high-error stagnation. Greater ansatz depth also improves the solution accuracy.

\begin{figure*}[htbp]
\centering
\includegraphics[width=14.66cm]{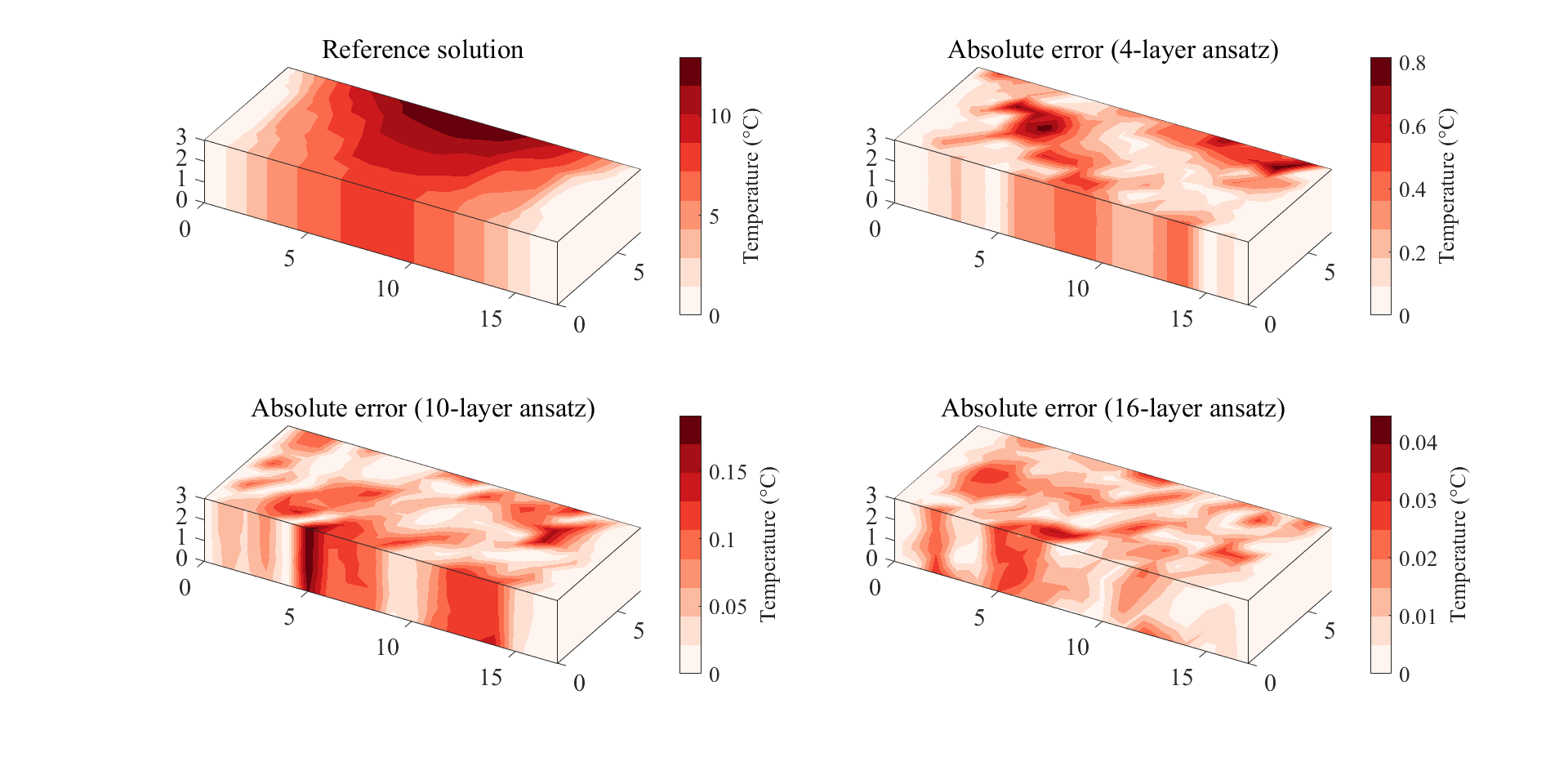}
\caption{Reference temperature field and absolute errors at different ansatz depths for the three-dimensional heat-conduction problem}\label{fig:22}
\end{figure*}

The classical finite element reference in Fig.~\ref{fig:22} shows temperature rising from the two $0\,^{\circ}\mathrm{C}$ end faces toward the interior, with a hot region near the heated surface and the center. At four layers, the absolute temperature error of Voxel-BVQLS has an upper color-scale limit of about $0.8\,^{\circ}\mathrm{C}$. This falls to about $0.15\,^{\circ}\mathrm{C}$ at ten layers and $0.04\,^{\circ}\mathrm{C}$ at sixteen layers. The spatial decline in error with depth is consistent with the relative-error trends in Fig.~\ref{fig:21}. Thus, the solutions approach the reference both in overall vector norm and in the temperature-field pattern.

\begin{table}[htbp]
\centering
\caption{Nonzero unitary terms in Pauli and voxel-grid decompositions of the heat-conduction stiffness matrix}\label{tab:6}
\begin{tabular}{p{0.36\linewidth}p{0.19\linewidth}p{0.19\linewidth}}
\toprule
Decomposition & Pauli LCU & Voxel-grid LCU \\
\midrule
Number of nonzero unitary terms & 228 & 18 \\
\bottomrule
\end{tabular}
\end{table}

These three noiseless examples show fewer nonzero unitary terms from the voxel-grid LCU decomposition than from the Pauli decomposition. Optimization of the minimum-potential-energy objective also yields approximations to the displacement and temperature fields.

\FloatBarrier

\section{Conclusions}

To address large sparse linear systems in static solid mechanics, we developed a voxel-based block variational quantum linear solver (Voxel-BVQLS) on regular voxel grids. It combines structured matrix decomposition, the principle of minimum potential energy, and batched quantum tests in a hybrid quantum--classical finite element workflow. First, for block-banded voxel-grid stiffness matrices, we use cyclic permutation matrices and block-diagonal matrices to construct an LCU decomposition whose upper bound on the number of unitary terms does not grow with the number of mesh nodes. Second, using the linear-static total potential energy as the variational objective mitigates barren plateaus in the studied settings. Third, we propose a block-Hadamard test that directly estimates weighted sums of multiple inner products, reducing the number of circuit configurations required per iteration.

Three numerical examples assessed the method. In the two-dimensional truss, plane-stress, and three-dimensional steady-state heat-conduction problems, the voxel-grid LCU decomposition has 18 nonzero unitary terms in each case, compared with 51, 224, and 228 Pauli terms, respectively. The computed displacement and temperature fields also approximate the reference solutions with the errors reported above. Within these examples, the method reduces the numbers of unitary terms and circuit configurations and attains the reported field errors in fewer optimization iterations than the comparison VQLS methods.

Several limitations remain. All calculations were performed on a classical computer. Finite-shot sampling, quantum gate noise, and measurement noise were not included. The block-Hadamard test relies on quantum multiplexors, whose gate-level cost and compatibility with current hardware require separate assessment. Future work will add gate-level resource estimates and noisy simulation, and develop improved optimizers for engineering applications.

\section*{Appendix 1. Gate representation of the singular value decomposition of a real $2\times2$ matrix (proof of Lemma 1)}

\textbf{Lemma 1.} The singular value decomposition of any real $2\times2$ matrix $\mathbf{A}=\left[ \begin{matrix} a & b \\ c & d \\ \end{matrix} \right]$ can be written as
\begin{equation}
\mathbf{A}=\mathbf{U}\boldsymbol{\Sigma}{{\mathbf{V}}^{\mathrm{T}}},\qquad
\mathbf{U}={{\mathbf{R}}_{y}}\left( 2\gamma  \right),\qquad
\mathbf{V}={{\mathbf{R}}_{y}}\left( 2\theta  \right),\qquad
\boldsymbol{\Sigma}=\operatorname{diag}\left( {{\sigma }_{1}},\ {{\sigma }_{2}} \right),
\label{eq:82}
\end{equation}
where $\sigma_1\ge\sigma_2\ge0$, and $\mathbf{R}_y(\varphi)$ is defined in Table~\ref{tab:1}:
$\mathbf{R}_y(\varphi)=\left[ \begin{matrix} \cos \left( \varphi /2 \right) & -\sin \left( \varphi /2 \right) \\ \sin \left( \varphi /2 \right) & \cos \left( \varphi /2 \right) \\ \end{matrix} \right]$.
The rotation angles $2\theta$ and $2\gamma$ are given by Eqs.~\eqref{eq:93} and \eqref{eq:94}. If $\det\mathbf{A}\ge0$, both $\mathbf{U}$ and $\mathbf{V}$ can have the pure-rotation form of Eq.~\eqref{eq:82}. If $\det\mathbf{A}<0$, exactly one of the two factors requires an additional $\mathbf{Z}$ gate as in Eq.~\eqref{eq:95}. The proof follows.

The key step is an orthogonal diagonalization of the real symmetric positive-semidefinite matrix $\mathbf{A}^{\mathrm T}\mathbf{A}$. First, expand it as
\begin{equation}
\mathbf{A}^{\mathrm{T}}\mathbf{A}=\left[ \begin{matrix} a & c \\ b & d \\ \end{matrix} \right]\left[ \begin{matrix} a & b \\ c & d \\ \end{matrix} \right]=\left[ \begin{matrix} {{a}^{2}}+{{c}^{2}} & ab+cd \\ ab+cd & {{b}^{2}}+{{d}^{2}} \\ \end{matrix} \right].
\label{eq:83}
\end{equation}

Let $T=\operatorname{tr}(\mathbf{A}^{\mathrm T}\mathbf{A})=a^2+b^2+c^2+d^2$ and $D=\det(\mathbf{A}^{\mathrm T}\mathbf{A})=(\det\mathbf{A})^2=(ad-bc)^2$. The discriminant of the characteristic equation is
\begin{equation}
\Delta =\sqrt{{{T}^{2}}-4D}=\sqrt{{{\left( {{a}^{2}}+{{c}^{2}}-{{b}^{2}}-{{d}^{2}} \right)}^{2}}+4{{\left( ab+cd \right)}^{2}}}\ge 0.
\label{eq:84}
\end{equation}

This quantity is nonnegative, ensuring that the eigenvalues of $\mathbf{A}^{\mathrm T}\mathbf{A}$ are real and nonnegative. Solving $\lambda^2-T\lambda+D=0$ gives
\begin{equation}
{{\lambda }_{1}}=\frac{T+\Delta }{2},\qquad {{\lambda }_{2}}=\frac{T-\Delta }{2}.
\label{eq:85}
\end{equation}

The singular values are the nonnegative square roots of the eigenvalues. In descending order, they form
\begin{equation}
\boldsymbol{\Sigma}=\left[ \begin{matrix} {{\sigma }_{1}} & {} \\ {} & {{\sigma }_{2}} \\ \end{matrix} \right],\qquad
{{\sigma }_{1}}=\sqrt{{{\lambda }_{1}}}=\sqrt{\frac{T+\Delta }{2}},\qquad
{{\sigma }_{2}}=\sqrt{{{\lambda }_{2}}}=\sqrt{\frac{T-\Delta }{2}}.
\label{eq:86}
\end{equation}

The columns of the matrix of right singular vectors $\mathbf{V}$ are orthonormal eigenvectors of $\mathbf{A}^{\mathrm T}\mathbf{A}$ satisfying
\begin{equation}
\left( \mathbf{A}^{\mathrm{T}}\mathbf{A}-{{\lambda }_{i}}{{\mathbf{I}}_{2}} \right){{\mathbf{v}}_{i}}=\mathbf{0},\qquad i=1,\ 2.
\label{eq:87}
\end{equation}

For the largest eigenvalue $\lambda_1$, substitute Eq.~\eqref{eq:83} into Eq.~\eqref{eq:87}. A normalized eigenvector satisfying the result is
\begin{equation}
{{\mathbf{v}}_{1}}=\frac{1}{{{N}_{1}}}\left[ \begin{matrix} ab+cd \\ {{\lambda }_{1}}-{{a}^{2}}-{{c}^{2}} \\ \end{matrix} \right],\qquad
{{N}_{1}}=\sqrt{{{\left( ab+cd \right)}^{2}}+{{\left( {{\lambda }_{1}}-{{a}^{2}}-{{c}^{2}} \right)}^{2}}}.
\label{eq:88}
\end{equation}

By orthogonality of eigenvectors of a real symmetric matrix, an eigenvector for $\lambda_2$ can be chosen as $\mathbf{v}_2=(-v_{12},v_{11})^{\mathrm T}$, where $v_{1i}$ is component $i$ of $\mathbf{v}_1$. Thus, the matrix of right singular vectors is
\begin{equation}
\mathbf{V}=\left[ \begin{matrix} {{\mathbf{v}}_{1}} & {{\mathbf{v}}_{2}} \\ \end{matrix} \right]=\frac{1}{{{N}_{1}}}\left[ \begin{matrix} ab+cd & -\left( {{\lambda }_{1}}-{{a}^{2}}-{{c}^{2}} \right) \\ {{\lambda }_{1}}-{{a}^{2}}-{{c}^{2}} & ab+cd \\ \end{matrix} \right].
\label{eq:89}
\end{equation}

The matrix of left singular vectors follows from $\mathbf{AV}=\mathbf{U}\boldsymbol{\Sigma}$: if $\sigma_2>0$, then $\mathbf{A}$ has full rank, $\mathbf{U}=\mathbf{AV}\boldsymbol{\Sigma}^{-1}$, and hence $\mathbf{u}_i=\mathbf{A}\mathbf{v}_i/\sigma_i$ for $i=1,2$, giving
\begin{equation}
\mathbf{U}=\left[ \begin{matrix} {{\mathbf{u}}_{1}} & {{\mathbf{u}}_{2}} \\ \end{matrix} \right]=\frac{1}{{{\sigma }_{1}}{{\sigma }_{2}}}\left[ \begin{matrix} \left( a{{v}_{11}}+b{{v}_{12}} \right){{\sigma }_{2}} & \left( a{{v}_{21}}+b{{v}_{22}} \right){{\sigma }_{1}} \\ \left( c{{v}_{11}}+d{{v}_{12}} \right){{\sigma }_{2}} & \left( c{{v}_{21}}+d{{v}_{22}} \right){{\sigma }_{1}} \\ \end{matrix} \right].
\label{eq:90}
\end{equation}

If $\sigma_2=0$, $\mathbf{u}_1=\mathbf{A}\mathbf{v}_1/\sigma_1$ still determines the first vector. Choose $\mathbf{u}_2$ as a unit vector orthogonal to $\mathbf{u}_1$.

We now express the orthogonal factors $\mathbf{U}$ and $\mathbf{V}$ as rotation gates directly implementable in a quantum circuit. Since $\mathbf{A}^{\mathrm T}\mathbf{A}=\mathbf{V}\operatorname{diag}(\lambda_1,\lambda_2)\mathbf{V}^{\mathrm T}$, choose $\mathbf{V}$ as the rotation matrix
\begin{equation}
\mathbf{V}={{\mathbf{R}}_{y}}\left( 2\theta  \right)=\left[ \begin{matrix} \cos \theta  & -\sin \theta  \\ \sin \theta  & \cos \theta  \\ \end{matrix} \right].
\label{eq:91}
\end{equation}

Expand $\mathbf{V}\operatorname{diag}(\lambda_1,\lambda_2)\mathbf{V}^{\mathrm T}$ and equate its entries to those in Eq.~\eqref{eq:83}. This yields
\begin{equation}
ab+cd=\frac{{{\lambda }_{1}}-{{\lambda }_{2}}}{2}\sin 2\theta ,\qquad
\left( {{a}^{2}}+{{c}^{2}} \right)-\left( {{b}^{2}}+{{d}^{2}} \right)=\left( {{\lambda }_{1}}-{{\lambda }_{2}} \right)\cos 2\theta.
\label{eq:92}
\end{equation}

Dividing the two equations eliminates $\lambda_1-\lambda_2$ and gives an explicit constraint on the rotation angle:
$\tan 2\theta =\frac{2\left( ab+cd \right)}{\left( {{a}^{2}}+{{c}^{2}} \right)-\left( {{b}^{2}}+{{d}^{2}} \right)}$.
To retain quadrant information when $a^2+c^2=b^2+d^2$, compute the angle with the two-argument arctangent:
\begin{equation}
2\theta =\operatorname{atan2}\!\left( 2\left( ab+cd \right),\ {{a}^{2}}+{{c}^{2}}-{{b}^{2}}-{{d}^{2}} \right).
\label{eq:93}
\end{equation}

Similarly, substitute Eq.~\eqref{eq:89} into $\mathbf{U}=\mathbf{AV}\boldsymbol{\Sigma}^{-1}$. Setting $\mathbf{U}=\mathbf{R}_y(2\gamma)$ and comparing matrix entries gives the left rotation angle:
\begin{equation}
\gamma =\operatorname{atan2}\!\left( c\cos \theta +d\sin \theta ,\ a\cos \theta +b\sin \theta  \right).
\label{eq:94}
\end{equation}

In the degenerate case $\lambda_1=\lambda_2>0$, we have $\mathbf{A}^{\mathrm T}\mathbf{A}=\lambda_1\mathbf{I}_2$ and $\tan2\theta$ is undefined, so one may choose $\theta=0$, $\mathbf{V}=\mathbf{I}_2=\mathbf{R}_y(0)$, and $\mathbf{U}=\mathbf{A}/\sigma_1$. If $\mathbf{A}=\mathbf{0}$, choose $\mathbf{U}=\mathbf{V}=\mathbf{I}_2$. The rotation angles are not unique in these cases, but the decomposition remains valid.

Finally, consider the determinant signs. Every real orthogonal $2\times2$ matrix $\mathbf{Q}$ can be written as
\begin{equation}
\mathbf{Q}={{\mathbf{R}}_{y}}\left( 2\eta  \right){{\mathbf{Z}}^{s}},\qquad s=\frac{1-\det \left( \mathbf{Q} \right)}{2}\in \left\{ 0,\ 1 \right\},
\label{eq:95}
\end{equation}
where $\mathbf{Z}=\operatorname{diag}(1,-1)$. If $\det\mathbf{A}\ge0$ and $\mathbf{A}$ has full rank, one can consistently choose $\det\mathbf{U}=\det\mathbf{V}=1$, so both factors in Eq.~\eqref{eq:82} are pure rotations. If $\det\mathbf{A}<0$, exactly one factor contains a reflection and requires a $\mathbf{Z}$ gate as in Eq.~\eqref{eq:95}. Thus, regardless of the determinant sign, every $2\times2$ orthogonal block $\mathbf{U}_n$ or $\mathbf{V}_n$ in Eqs.~\eqref{eq:34}--\eqref{eq:38} can be implemented with a single-qubit $R_y$ rotation and at most one $\mathbf{Z}$ gate. This completes the proof.

\section*{Appendix 2. Quantum circuit for a diagonal phase unitary}

Let $N=2^n$ and consider the diagonal unitary
\begin{equation}
{{\mathbf{D}}_{\theta }}=\bigoplus_{i=0}^{N-1}{{{e}^{\mathrm{i}{{\theta }_{i}}}}},\qquad
{{\mathbf{D}}_{\theta }}\left| \mathbf{y} \right\rangle =\sum_{i=0}^{N-1}{{{e}^{\mathrm{i}{{\theta }_{i}}}}{{y}_{i}}\left| i \right\rangle },
\label{eq:96}
\end{equation}
where $|\mathbf{y}\rangle=\sum_i y_i|i\rangle$. By the definition of a quantum multiplexor, Eq.~\eqref{eq:96} is a uniformly controlled phase gate on $n$ qubits. Suppose the $N$ phases take only $N_s$ distinct values, denoted by $\{\hat{\theta}_1,\ldots,\hat{\theta}_{N_s}\}$. Let $g(i)\in\{1,\ldots,N_s\}$ identify the class of $\theta_i$, so that $\theta_i=\hat{\theta}_{g(i)}$. Equation~\eqref{eq:96} can then be implemented with $N_s$ ancilla qubits. First, encode the class in the ancilla register using one-hot encoding. Next, apply rotations to the ancillas in parallel. Finally, uncompute the register.

A reversible lookup oracle $O_g$ writes the class information. It is defined by
\begin{equation}
{{O}_{g}}:\ \left| i \right\rangle {{\left| 0 \right\rangle }^{\otimes {{N}_{s}}}}\longmapsto \left| i \right\rangle \left| {{m}_{g\left( i \right)}} \right\rangle.
\label{eq:97}
\end{equation}
The state $|m_j\rangle$ is a one-hot state on $N_s$ qubits: qubit $j$ is $|1\rangle$ and all others are $|0\rangle$. Each $|m_j\rangle$ corresponds to an angle $\hat{\theta}_j$.

Let the input at the left of the circuit be $|\psi\rangle=(\sum_i y_i|i\rangle)|0\rangle^{\otimes N_s}$. After $O_g$, it becomes
\begin{equation}
\left| \psi \right\rangle =\sum_{j=1}^{{{N}_{s}}}{\sum_{g\left( i \right)=j}{{{y}_{i}}\left| i \right\rangle \left| {{m}_{j}} \right\rangle }}.
\label{eq:98}
\end{equation}

Apply $R_z(s_j)$ in parallel to the $N_s$ ancilla qubits, namely $\bigotimes_{j=1}^{N_s}\mathbf{R}_z(s_j)$. Since $\mathbf{R}_z(s)=\operatorname{diag}(e^{-\mathrm{i}s/2},e^{\mathrm{i}s/2})$ and only qubit $j$ is $|1\rangle$ in $|m_j\rangle$, the phase accumulated by the ancilla register is
\begin{equation}
\exp \left[ \mathrm{i}\left( {{s}_{j}}-\frac{1}{2}\sum_{q=1}^{{{N}_{s}}}{{{s}_{q}}} \right) \right].
\label{eq:99}
\end{equation}

For $N_s\ne2$, choose
\begin{equation}
{{s}_{j}}={{\hat{\theta }}_{j}}+\frac{\Theta }{2-{{N}_{s}}},\qquad \Theta =\sum_{q=1}^{{{N}_{s}}}{{{{\hat{\theta }}}_{q}}}.
\label{eq:100}
\end{equation}

Equation~\eqref{eq:99} then equals $e^{\mathrm{i}\hat{\theta}_j}$. For $N_s=2$, after ignoring the global phase $e^{\mathrm{i}(\hat{\theta}_1+\hat{\theta}_2)/2}$, choose $s_j=\hat{\theta}_j-(\hat{\theta}_1+\hat{\theta}_2)/2$. Finally, apply $O_g^\dagger$ to uncompute the ancillas, obtaining
\begin{equation}
O_{g}^{\dagger }\left( \sum_{j=1}^{{{N}_{s}}}{\sum_{g\left( i \right)=j}{{{y}_{i}}{{e}^{\mathrm{i}{{{\hat{\theta }}}_{j}}}}\left| i \right\rangle \left| {{m}_{j}} \right\rangle }} \right)=\left( {{\mathbf{D}}_{\theta }}\left| \mathbf{y} \right\rangle  \right){{\left| 0 \right\rangle }^{\otimes {{N}_{s}}}}.
\label{eq:101}
\end{equation}

\begin{figure}[htbp]
\centering
\includegraphics[width=7.71cm]{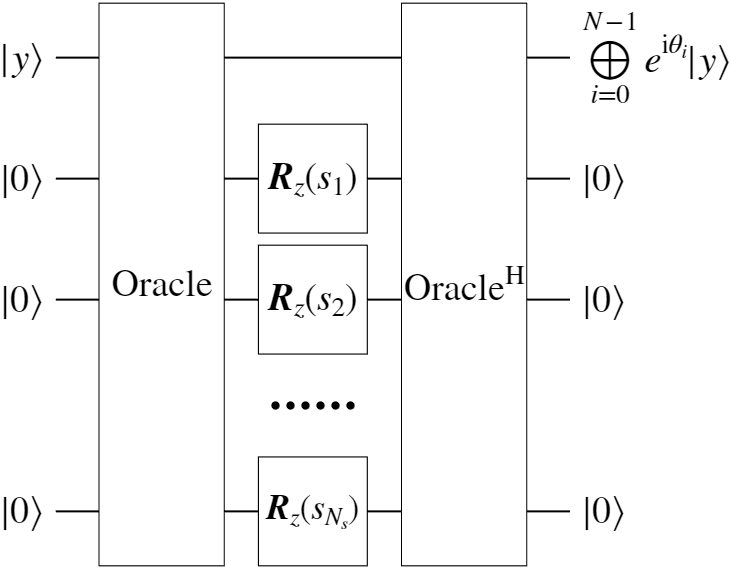}
\caption{Circuit for a diagonal phase unitary using a one-hot lookup oracle}\label{fig:23}
\end{figure}

The corresponding circuit is shown in Fig.~\ref{fig:23}.

Figure~\ref{fig:23} trades space for time. The parallel $R_z$ layer has depth $O(1)$ but requires $N_s$ ancilla qubits and $O(N_s)$ rotation gates. In the worst case, all $N$ phases differ, so $N_s=N$. The oracle $O_g$ can then use bucket-brigade quantum random-access memory, with $O(\log N)$ addressing depth \cite{Giovannetti2008}. Storage, control, and fault-tolerant implementation costs remain linear in $N$. Without structured $g(i)$ or available quantum random-access memory, generic oracle synthesis also generally requires resources growing with $N$. Thus, we assert only that a unitary implementation of Eq.~\eqref{eq:96} exists. Its actual complexity must be reassessed for a specific data-access model and hardware connectivity.

\section*{Appendix 3. Block-Hadamard test for the global loss}

Let $\mathbf{K}=\sum_{i=0}^{M-1}a_i\mathbf{K}_i$. The quadratic form in the global loss is
\begin{equation}
\left\langle \mathbf{u} \right|{{\mathbf{K}}^{\mathrm{H}}}\mathbf{K}\left| \mathbf{u} \right\rangle =\sum_{i=0}^{M-1}{\sum_{j=0}^{M-1}{a_{i}^{*}{{a}_{j}}\left\langle \mathbf{u} \right|\mathbf{K}_{i}^{\mathrm{H}}{{\mathbf{K}}_{j}}\left| \mathbf{u} \right\rangle }}.
\label{eq:102}
\end{equation}

The loss requires the weighted sum of overlaps, not each individual value $\langle\mathbf{u}|\mathbf{K}_i^{\mathrm H}\mathbf{K}_j|\mathbf{u}\rangle$. A separate Hadamard circuit for each term in Eq.~\eqref{eq:102} requires $O(M^2)$ circuit calls. We therefore partition the $M$ unitary terms into $M/B$ blocks of $B=2^b$ terms, padding the term list with zero coefficients if $B$ does not divide $M$. Define the weighted sum for block pair $(k,l)$ as
\begin{equation}
{{c}_{k,l}}=\sum_{i=Bk}^{B\left( k+1 \right)-1}{\sum_{j=Bl}^{B\left( l+1 \right)-1}{a_{i}^{*}{{a}_{j}}\left\langle \mathbf{u} \right|\mathbf{K}_{i}^{\mathrm{H}}{{\mathbf{K}}_{j}}\left| \mathbf{u} \right\rangle }},\qquad
\left\langle \mathbf{u} \right|{{\mathbf{K}}^{\mathrm{H}}}\mathbf{K}\left| \mathbf{u} \right\rangle =\sum_{k=0}^{M/B-1}{\sum_{l=0}^{M/B-1}{{{c}_{k,l}}}}.
\label{eq:103}
\end{equation}

\begin{figure*}[htbp]
\centering
\includegraphics[width=11.75cm]{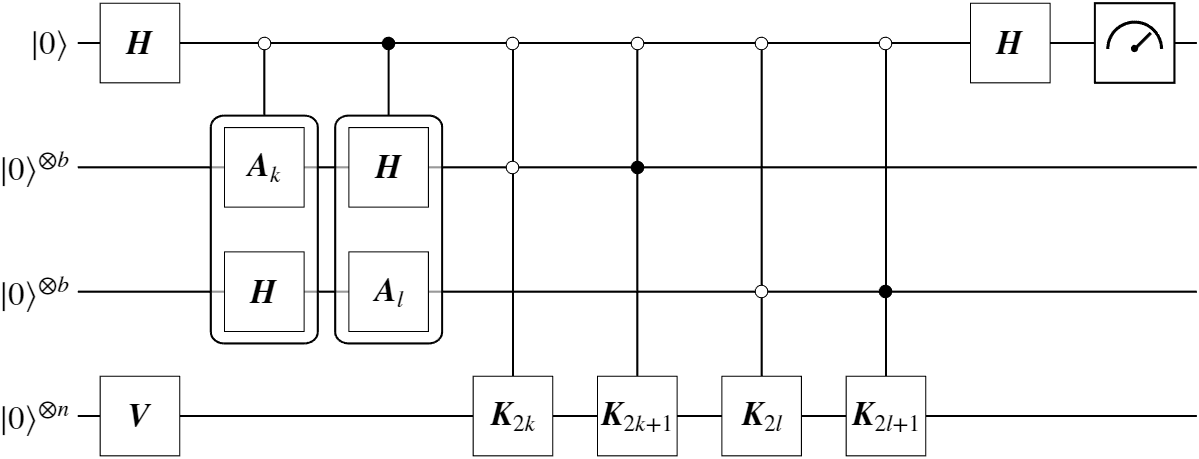}
\caption{Block-Hadamard circuit for estimating $c_{k,l}$ ($B=2$)}\label{fig:24}
\end{figure*}

The quantity $c_{k,l}$ is a weighted sum of $B^2$ overlaps and serves as a basic computational block for Eq.~\eqref{eq:102}. If one circuit directly estimates $c_{k,l}$, the number of circuit configurations per loss evaluation falls from $O(M^2)$ to $O((M/B)^2)$. Define the normalized coefficient state of block $k$ by
\begin{equation}
\left| {{{\hat{\mathbf{a}}}}_{k}} \right\rangle =\frac{1}{{{A}_{k}}}\sum_{i=Bk}^{B\left( k+1 \right)-1}{{{a}_{i}}\left| i-Bk \right\rangle }={{\mathbf{A}}_{k}}{{\left| 0 \right\rangle }^{\otimes b}},\qquad
{{A}_{k}}=\sqrt{\sum_{i=Bk}^{B\left( k+1 \right)-1}{{{\left| {{a}_{i}} \right|}^{2}}}},
\label{eq:104}
\end{equation}
where $\mathbf{A}_k$ is a $B\times B$ unitary. Figure~\ref{fig:24} shows the circuit estimating $c_{k,l}$ for $B=2$, which places coefficient-state preparation and the corresponding SELECT operations for blocks $k$ and $l$ in the two branches of a Hadamard test.

The input at the left of the circuit is $|\psi\rangle=|0\rangle|0\rangle^{\otimes b}|0\rangle^{\otimes b}|0\rangle^{\otimes n}$. After the circuit layers, the final state is
\begin{equation}
\begin{aligned}
\left| \psi \right\rangle ={}&\frac{1}{{{\left( \sqrt{2} \right)}^{b+2}}}\left| 0 \right\rangle \sum_{i=0}^{B-1}{\sum_{j=0}^{B-1}{\left| i \right\rangle \left| j \right\rangle }}\left( \frac{{{a}_{i+Bk}}}{{{A}_{k}}}{{\mathbf{K}}_{i+Bk}}\left| \mathbf{u} \right\rangle +\frac{{{a}_{j+Bl}}}{{{A}_{l}}}{{\mathbf{K}}_{j+Bl}}\left| \mathbf{u} \right\rangle  \right) \\
&+\frac{1}{{{\left( \sqrt{2} \right)}^{b+2}}}\left| 1 \right\rangle \sum_{i=0}^{B-1}{\sum_{j=0}^{B-1}{\left| i \right\rangle \left| j \right\rangle }}\left( \frac{{{a}_{i+Bk}}}{{{A}_{k}}}{{\mathbf{K}}_{i+Bk}}\left| \mathbf{u} \right\rangle -\frac{{{a}_{j+Bl}}}{{{A}_{l}}}{{\mathbf{K}}_{j+Bl}}\left| \mathbf{u} \right\rangle  \right).
\end{aligned}
\label{eq:105}
\end{equation}

Measure the first ancilla and denote the probabilities of $|0\rangle$ and $|1\rangle$ by $\zeta_0$ and $\zeta_1$, respectively. They are
\begin{equation}
{{\zeta }_{0}}=\frac{1}{2}\left[ 1+\frac{\operatorname{Re}\left( {{c}_{k,l}} \right)}{{{A}_{k}}{{A}_{l}}B} \right],\qquad
{{\zeta }_{1}}=\frac{1}{2}\left[ 1-\frac{\operatorname{Re}\left( {{c}_{k,l}} \right)}{{{A}_{k}}{{A}_{l}}B} \right].
\label{eq:106}
\end{equation}

Therefore,
\begin{equation}
\operatorname{Re}\left( {{c}_{k,l}} \right)={{A}_{k}}{{A}_{l}}B\left( 2{{\zeta }_{0}}-1 \right)={{A}_{k}}{{A}_{l}}B\left( 1-2{{\zeta }_{1}} \right)={{A}_{k}}{{A}_{l}}B\left( {{\zeta }_{0}}-{{\zeta }_{1}} \right).
\label{eq:107}
\end{equation}

Similar circuits estimate all $c_{k,l}$ in batches. For real symmetric $\mathbf{K}$, summing all blocks gives the required real quadratic form. Blocking reduces the number of circuit configurations to $O((M/B)^2)$, or one configuration when $B=M$. Larger $B$ reduces the number of configurations but increases circuit width and depth and lowers individual outcome probabilities.

\Acknowledgements{}

\makeentitle


\section*{References}
\begin{thebibliography}{99}
\bibitem{Gao2012}Q. Gao, F. Wu, H. W. Zhang, W. X. Zhong, W. P. Howson, F. W. Williams, A fast precise integration method for structural dynamics problems, Structural Engineering and Mechanics 43 (2012) 1--13. 
\bibitem{Mosby2016}M. Mosby, K. Matous, Computational homogenization at extreme scales, Extreme Mechanics Letters 6 (2016) 68--74. 
\bibitem{Turner1956}M. J. Turner, R. W. Clough, H. C. Martin, L. J. Topp, Stiffness and deflection analysis of complex structures, Journal of the Aeronautical Sciences 23 (1956) 805--823. 
\bibitem{Clough1960}R. W. Clough, The finite element method in plane stress analysis, in: Proceedings of the 2nd ASCE Conference on Electronic Computation, Pittsburgh, PA (1960).
\bibitem{Wu2013InterBelt}F. Wu, Y. Sun, W. X. Zhong, Inter-belt finite element for the analysis of incompressible material problems, Applied Mathematics and Mechanics 34 (2013) 1--9 (in Chinese). 
\bibitem{Bangerth2011}W. Bangerth, C. Burstedde, T. Heister, M. Kronbichler, Algorithms and data structures for massively parallel generic adaptive finite element codes, ACM Transactions on Mathematical Software 38 (2011) Article 14, 1--28. 
\bibitem{Koric2014}S. Koric, Q. Lu, E. Guleryuz, Evaluation of massively parallel linear sparse solvers on unstructured finite element meshes, Computers \& Structures 141 (2014) 19--25. 
\bibitem{Wu2024Discrepancy}F. Wu, Y. Zhao, Y. Yang, X. Zhang, N. Zhou, A new discrepancy for sample generation in stochastic response analyses of aerospace problems with uncertain parameters, Chinese Journal of Aeronautics 37 (2024) 192--211. 
\bibitem{Wu2023ADDP}F. Wu, D. Huang, X. Xu, K. Zhao, N. Zhou, An adaptive divided-difference perturbation method for solving stochastic problems, Structural Safety 103 (2023) 102346. 
\bibitem{Feynman1982}R. P. Feynman, Simulating physics with computers, International Journal of Theoretical Physics 21 (1982) 467--488. 
\bibitem{Lloyd1996}S. Lloyd, Universal quantum simulators, Science 273 (1996) 1073--1078. 
\bibitem{Kim2023}Y. Kim, A. Eddins, S. Anand, et al., Evidence for the utility of quantum computing before fault tolerance, Nature 618 (2023) 500--505. 
\bibitem{Preskill2018}J. Preskill, Quantum computing in the NISQ era and beyond, Quantum 2 (2018) 79. 
\bibitem{Peruzzo2014}A. Peruzzo, J. McClean, P. Shadbolt, et al., A variational eigenvalue solver on a photonic quantum processor, Nature Communications 5 (2014) 4213. 
\bibitem{Cao2019}Y. Cao, J. Romero, J. P. Olson, et al., Quantum chemistry in the age of quantum computing, Chemical Reviews 119 (2019) 10856--10915. 
\bibitem{McArdle2020}S. McArdle, S. Endo, A. Aspuru-Guzik, S. C. Benjamin, X. Yuan, Quantum computational chemistry, Reviews of Modern Physics 92 (2020) 015003. 
\bibitem{AuYeung2024}R. Au-Yeung, B. Camino, O. Rathore, V. Kendon, Quantum algorithms for scientific computing, Reports on Progress in Physics 87 (2024) 116001. 
\bibitem{Balducci2022}G. Tosti Balducci, B. Chen, M. M\"oller, M. Gerritsma, R. De Breuker, Review and perspectives in quantum computing for partial differential equations in structural mechanics, Frontiers in Mechanical Engineering 8 (2022) 914241. 
\bibitem{MengYang2023}Z. Meng, Y. Yang, Quantum computing of fluid dynamics using the hydrodynamic Schr\"odinger equation, Physical Review Research 5 (2023) 033182. 
\bibitem{MengYang2024}Z. Meng, Y. Yang, Quantum spin representation for the Navier--Stokes equation, Physical Review Research 6 (2024) 043130. 
\bibitem{Meng2024Processor}Z. Meng, J. Zhong, S. Xu, et al., Simulating unsteady flows on a superconducting quantum processor, Communications Physics 7 (2024) 349. 
\bibitem{Wu2026VBQC}F. Wu, Y. Yang, L. Zhu, C. Li, Y. Guo, X. Guo, A voxel-based quantum computing method (VBQC) for solid mechanics problem, arXiv:2606.03515 (2026). 
\bibitem{Zhang2010Heterogeneous}H.-W. Zhang, J.-K. Wu, J. L\"u, Z.-D. Fu, Extended multiscale finite element method for mechanical analysis of heterogeneous materials, Acta Mechanica Sinica 26 (2010) 899--920. 
\bibitem{XuHu2026Potential}Y. Xu, H. Hu, Potential energy minimization for structural analysis via decomposition-free quantum computing, Computers \& Structures 330 (2026) 108309. 
\bibitem{Jin2024Schrodingerization}S. Jin, N. Liu, Y. Yu, Quantum simulation of partial differential equations via Schr\"odingerization, Physical Review Letters 133 (2024) 230602. 
\bibitem{Xu2026Elasto}Y. Xu, K. Zhao, H.-T. Liu, K. Huang, Z. Liu, H. Fan, H. Hu, Hamiltonian simulation of elastodynamics on a quantum computer via energy conservation mapping, Journal of the Mechanics and Physics of Solids 213 (2026) 106639. 
\bibitem{Xu2025}Y. Xu, H. Hu, Decomposition-free variational quantum linear solver: Application in computational mechanics, Computer Methods in Applied Mechanics and Engineering 447 (2025) 118396. 
\bibitem{Raisuddin2022}O. M. Raisuddin, S. De, FEqa: Finite element computations on quantum annealers, Computer Methods in Applied Mechanics and Engineering 395 (2022) 115014. 
\bibitem{Harrow2009}A. W. Harrow, A. Hassidim, S. Lloyd, Quantum algorithm for linear systems of equations, Physical Review Letters 103 (2009) 150502. 
\bibitem{Montanaro2016}A. Montanaro, S. Pallister, Quantum algorithms and the finite element method, Physical Review A 93 (2016) 032324. 
\bibitem{Cerezo2021Review}M. Cerezo, A. Arrasmith, R. Babbush, et al., Variational quantum algorithms, Nature Reviews Physics 3 (2021) 625--644. 
\bibitem{BravoPrieto2023}C. Bravo-Prieto, R. LaRose, M. Cerezo, Y. Subasi, L. Cincio, P. J. Coles, Variational quantum linear solver, Quantum 7 (2023) 1188. 
\bibitem{Ying2023}J.-W. Ying, J.-C. Shen, L. Zhou, W. Zhong, M.-M. Du, Y.-B. Sheng, Preparing a fast Pauli decomposition for variational quantum solving linear equations, Annalen der Physik 535 (2023) 2300212. 
\bibitem{Sato2021}Y. Sato, R. Kondo, S. Koide, H. Takamatsu, N. Imoto, Variational quantum algorithm based on the minimum potential energy for solving the Poisson equation, Physical Review A 104 (2021) 052409. 
\bibitem{Patil2022}H. Patil, Y. Wang, P. S. Krsti\'c, Variational quantum linear solver with a dynamic ansatz, Physical Review A 105 (2022) 012423. 
\bibitem{PellowJarman2021}A. Pellow-Jarman, I. Sinayskiy, A. Pillay, F. Petruccione, A comparison of various classical optimizers for a variational quantum linear solver, Quantum Information Processing 20 (2021) 202. 
\bibitem{Trahan2023}C. J. Trahan, M. Loveland, N. Davis, E. Ellison, A variational quantum linear solver application to discrete finite-element methods, Entropy 25 (2023) 580. 
\bibitem{Ali2023}M. Ali, M. Kabel, Performance study of variational quantum algorithms for solving the Poisson equation on a quantum computer, Physical Review Applied 20 (2023) 014054. 
\bibitem{Arora2025}A. Arora, B. M. Ward, C. Oskay, An implementation of the finite element method in hybrid classical/quantum computers, Finite Elements in Analysis and Design 248 (2025) 104354. 
\bibitem{Liu2024}Y. Liu, J. Liu, J. R. Raney, P. Wang, Quantum computing for solid mechanics and structural engineering---A demonstration with variational quantum eigensolver, Extreme Mechanics Letters 67 (2024) 102117. 
\bibitem{McClean2018}J. R. McClean, S. Boixo, V. N. Smelyanskiy, R. Babbush, H. Neven, Barren plateaus in quantum neural network training landscapes, Nature Communications 9 (2018) 4812. 
\bibitem{Cerezo2021BP}M. Cerezo, A. Sone, T. Volkoff, L. Cincio, P. J. Coles, Cost function dependent barren plateaus in shallow parametrized quantum circuits, Nature Communications 12 (2021) 1791. 
\bibitem{Childs2012}A. M. Childs, N. Wiebe, Hamiltonian simulation using linear combinations of unitary operations, Quantum Information and Computation 12 (2012) 901--924. 
\bibitem{Wu2025Voxel}F. Wu, C. Li, Y.-X. Yang, L. Zhu, X. Guo, Quantum simulation of Hamiltonian in solid mechanics based on voxel representation, Chinese Journal of Computational Mechanics 42 (2025) 329--338 (in Chinese). 
\bibitem{Chakraborty2024}S. Chakraborty, Implementing any linear combination of unitaries on intermediate-term quantum computers, Quantum 8 (2024) 1496. 
\bibitem{Mottonen2005}M. M\"ott\"onen, J. J. Vartiainen, V. Bergholm, M. M. Salomaa, Transformation of quantum states using uniformly controlled rotations, Quantum Information and Computation 5 (2005) 467--473. 
\bibitem{Giovannetti2008}V. Giovannetti, S. Lloyd, L. Maccone, Architectures for a quantum random access memory, Physical Review A 78 (2008) 052310. 
\bibitem{Huang2021Regression}H.-Y. Huang, K. Bharti, P. Rebentrost, Near-term quantum algorithms for linear systems of equations with regression loss functions, New Journal of Physics 23 (2021) 113021. 
\bibitem{Schneider2022Voxel}M. Schneider, Voxel-based finite elements with hourglass control in fast Fourier transform-based computational homogenization, International Journal for Numerical Methods in Engineering 123 (2022) 6286--6313. 

\bibitem{Hantzko2024}L. Hantzko, L. Binkowski, S. Gupta, Tensorized Pauli decomposition algorithm, Physica Scripta 99 (2024) 085128. 

\bibitem{MSPD2026}F. Wu, C. Li, X. Wu, Y. Guo, X. Guo, A matrix sparsity-based Pauli decomposition algorithm, Acta Mechanica Sinica (2026) in press. 

\bibitem{Shende2006Multiplexor}V. V. Shende, S. S. Bullock, I. L. Markov, Synthesis of quantum-logic circuits, IEEE Transactions on Computer-Aided Design of Integrated Circuits and Systems 25 (2006) 1000--1010. 

\bibitem{Soudackov2026QFlux}A. V. Soudackov, D. G. A. Cabral, B. C. Allen, et al., QFlux: Quantum circuit implementations of molecular dynamics. Part III---State initialization and unitary decomposition, ChemRxiv (2026). 





\end{thebibliography}
\end{document}